\documentclass[aps,showpacs]{revtex4}

\usepackage{amssymb} 
\usepackage{graphicx} 
\usepackage{bm} 
\usepackage[T1]{fontenc} 
\usepackage{euscript} 
\usepackage{mathrsfs}

\begin{document} 

\newcommand{\cequ}{\begin{eqnarray}} 
\newcommand{\fequ}{\end{eqnarray}} 
\newcommand{\anticomut}[2]{\left\{#1,#2\right\}} 
\newcommand{\comut}[2]{\left[#1,#2\right]} 
\newcommand{\comutd}[2]{\left[#1,#2\right]^{*}} 

\title{\bf Hamiltonian thermodynamics of charged three-dimensional 
dilatonic black holes} 
\author{Gon\c{c}alo A. S. Dias\footnote{Email: gadias@fisica.ist.utl.pt}} 
\affiliation{Centro Multidisciplinar de Astrof\'{\i}sica - CENTRA \\ 
Departamento de F\'{\i}sica, Instituto Superior T\'ecnico - IST,\\ 
Universidade T\'ecnica de Lisboa - UTL,\\ 
Av. Rovisco Pais 1, 1049-001 Lisboa, Portugal} 
\author{Jos\'e P. S. Lemos\footnote{Email: lemos@fisica.ist.utl.pt}} 
\affiliation{Centro Multidisciplinar de {}Astrof\'{\i}sica - CENTRA \\ 
Departamento de F\'{\i}sica, Instituto Superior T\'ecnico - IST,\\ 
Universidade T\'ecnica de Lisboa - UTL,\\ 
Av. Rovisco Pais 1, 1049-001 Lisboa, Portugal} 
%\date{\today} 

\begin{abstract} 
The action for a class of three-dimensional dilaton-gravity theories, 
with an electromagnetic Maxwell field and a cosmological constant, can 
be recast in a Brans-Dicke-Maxwell type action, with its free $\omega$ 
parameter.  For a negative cosmological constant, these theories have 
static, electrically charged, and spherically symmetric black hole 
solutions. Those theories with well formulated asymptotics are studied 
through a Hamiltonian formalism, and their thermodynamical properties 
are found out. The theories studied are general relativity 
($\omega\to\pm\infty$), a dimensionally reduced cylindrical 
four-dimensional general relativity theory ($\omega=0$), and a 
theory representing a class of theories 
($\omega=-3$), all with a Maxwell term. The Hamiltonian formalism is 
setup in three dimensions through foliations on the right region of 
the Carter-Penrose diagram, with the bifurcation 1-sphere as the left 
boundary, and anti-de Sitter infinity as the right boundary.  The 
metric functions on the foliated hypersurfaces and the radial 
component of the vector potential one-form are the canonical 
coordinates. The Hamiltonian action is written, the Hamiltonian being 
a sum of constraints.  One finds a new action which yields an 
unconstrained theory with two pairs of canonical coordinates 
$\{M,P_M;\,Q,P_Q\}$, where $M$ is the mass parameter, which for 
$\omega<-\frac32$ and for $\omega=\pm\infty$ needs a careful 
renormalization, $P_M$ is the conjugate momenta of $M$, $Q$ is the 
charge parameter, and $P_Q$ is its conjugate momentum. The resulting 
Hamiltonian is a sum of boundary terms only.  A quantization of the 
theory is performed. The Schr\"odinger evolution operator is 
constructed, the trace is taken, and the partition function of the 
grand canonical ensemble is obtained, where the chemical potential is 
the scalar electric potential $\bar{\phi}$.  Like the uncharged cases 
studied previously, the charged black hole entropies differ, in 
general, from the usual quarter of the horizon area due to the 
dilaton. 
\end{abstract} 


\maketitle 

\section{Introduction} 
\label{intro} 

In a previous paper \cite{diaslemos} we have motivated and studied the 
Hamiltonian thermodynamics of three-dimensional dilatonic black 
holes. To construct a classical Hamiltonian formalism is important for 
several reasons. First, it is always elegant to write the equations of 
the theory as a pair of symmetric first order equations. Second, if 
one can put the classical theory into a Hamiltonian form, then by 
applying certain rules one can get a quantized version of the theory 
in a first approximation. Third, through Euclideanization of the 
Schr\"odinger time evolution operator $\exp(-iHt)$, where $H$ is the 
Hamiltonian, one gets the partition function, which in turn leads to a 
thermodynamical decription of the system.  Thus, the Hamiltonian 
formalism and the thermodynamical description are linked subjects. 
The Hamiltonian thermodynamics of several different black hole systems 
have been studied first by Louko and Whiting in \cite{louko1} and 
further analyzed in \cite{louko2,louko3,louko4,louko5,bose} and 
\cite{diaslemos}, (see also \cite{kunstatter1,kunstatter2}).  The 
Louko-Whiting method relies on the Hamiltonian approach of 
\cite{kuchar}, which in turn is an important ramification of the 
Arnowitt, Deser, and Misner approach, the ADM approach, and its 
major developments \cite{adm,rt}.  Other methods, like direct 
calculation from quantum fields in curved spacetime (see, e.g, 
\cite{hawking1,bekenstein1}), or path integral methods (see, e.g, 
\cite{hawking2,york1,braden,zaslavskii1,peca}), have been used to 
study the thermodynamics of black holes.  Now, the study of 
thermodynamics of black holes in any specific dimension is important, 
for instance, to understand universal properties independent of the 
dimension itself. In particular, one can single out three dimensions 
as an adequate interesting dimension, because three dimensional 
general relativity has a special black hole with simple properties 
\cite{btz}, and dilatonic extensions of this theory have black holes 
with properties similar to four-dimensional general relativistic black 
holes \cite{lemos1,sakleberlemos}. 

The main purpose of this paper is to generalize our previous work on 
Hamiltonian thermodynamics of three-dimensional dilatonic black holes 
\cite{diaslemos}, to black holes, still in three dimensions, in which 
the graviton-dilaton Brans-Dicke theory, with its $\omega$ parameter, 
is coupled to a gauge field, namely, the electromagnetic field. 
Three-dimensional Einstein-Maxwell theory with its black hole solution 
has been studied in \cite{mtz} (see also \cite{btz}), and electric 
dilatonic extensions also have charged black holes 
\cite{lemoszanchin,oscarlemos} (see also \cite{lemos1,sakleberlemos}). 
Further analyses on the properties of these black holes have been 
worked out in \cite{zaslavskii2,carlip,ads3_bh}.  In principle, 
$\omega$ can take any value, i.e., $\infty>\omega>-\infty$.  But, 
since we want to perform a canonical Hamiltonian analysis, using an 
ADM formalism supplied with proper boundary conditions, it is 
necessary to pick up only those $\omega$s which yield black hole 
solutions that fulfill boundary conditions adequate to a Hamiltonain 
method.  As in the uncharged case \cite{diaslemos}, the cases of most 
interest are then black holes for which $\omega\to\pm \infty$, 
$\omega=0$, and $\omega=-3$.  Now, unlike the uncharged case, it was 
realized in \cite{mtz,oscarlemos} (see also \cite{btz}) that the 
inclusion of electric charge creates a complication to the definition 
of the mass $M$. Indeed, for $\omega<-\frac32$, which includes the 
general relativity case (since it can be considered as the 
$\omega\rightarrow-\infty$ limit), one finds a divergent asymptotic 
behavior of the mass.  This problem can be cured by a proper mass 
renormalization \cite{mtz,oscarlemos}.  But the whole issue has never 
been confronted when one makes a full use of the Hamiltonian 
formalism, as we do here.  We solve this difficulty and show that the 
mass redefinition conforms wholly with the formalism for the 
$\omega\rightarrow\infty$ case, is not needed in the $\omega=0$ case 
and it is compatible with the $\omega=-3$ 
case.  In each case, the 
black hole system and its thermodynamics can then be studied.  We have 
chosen to study the systems in a grand canonical ensemble.  In such an 
ensemble, the boundary radius $l$, the temperature $\textbf T$ 
(or its inverse $\beta$), and 
the electric potential (i.e., the chemical potential) $\bar\phi$ at 
the boundary radius are fixed.  The electric potential is conjugate to 
the electric charge and, in this ensemble, the charge itself is not 
fixed.  Our main results are, a complete derivation from a Hamiltonian 
formalism of the mass-energy $\left\langle E\right\rangle$, the 
entropy $S$, and the charge $\left\langle Q\right\rangle$, as well as 
the establishment that the grand canonical ensemble for charged three 
dimensional black holes is locally stable for the black holes studied. 
Thus, the choice of such an ensemble is appropriate, since the 
corresponding boundary onditions lead to a well posed problem and a 
stable ensemble. 

The structure of the paper is as follows. In Sec. \ref{solucoes} we 
present the classical solutions of the three-dimensional dilatonic 
charged black holes, whose quantization through Hamiltonian methods we 
will perform.  There is a free parameter $\omega$ for which we choose 
three different values, $\omega=\infty,\,0,\,-3$, corresponding to the 
BTZ black hole, the dimensionally reduce four-dimensional cylindrical 
black hole, and a three-dimensional dilatonic black hole, 
respectively. We also introduce the spacetime foliation through which 
we will define the canonical coordinates and which will allow us to 
write the action as a sum of constraints multiplied by their 
respective Lagrange multipliers. Then, follow three sections, 
Secs. \ref{mtz}, \ref{zero}, and \ref{menostres}, where we develop the 
thermodynamic Hamiltonian formalism for the $\omega=\infty,\,0,\,-3$ 
charged black holes, respectively.  We divide each section into 
{A}-{F} subsections to better analyze the Hamiltonian thermodynamics 
of the systems. 

\section{Charged black hole solutions allowing Hamiltonian 
  description} 
\label{solucoes} 

\subsection{The 3D black hole solutions} 
\label{bhsolutions} 
Three-dimensional theories which contain black holes have been studied 
in several works. Indeed in 
\cite{btz,carlip,lemos1,sakleberlemos,ads3_bh}, a Brans-Dicke action, 
with gravitational and dilaton fields plus a cosmological constant, 
was used to study such solutions.  Now, we can couple the graviton and 
dilaton fields to an electromagnetic field.  in order to find charged 
black hole solutions in three dimensions.  For three-dimensional 
general relativity this has been done in \cite{mtz} (see also 
\cite{btz}), whereas for Brans-Dicke theory this has been done in 
\cite{oscarlemos} (see also \cite{ads3_bh}).  The Brans-Dicke action 
with a Maxwell term is given by 
\begin{equation} 
S = \frac{1}{2\pi}\int d^3x \,\sqrt{-g}\, e^{-2\phi}\,(R-4\omega 
(\partial \phi)^2 + 4\lambda^2+F^{\mu\nu}F_{\mu\nu})+\bar{B}\,, 
\label{3dchbdaction} 
\end{equation} 
where $g$ is the determinant of the three-dimensional metric 
$g_{\mu\nu}$, $R$ is the curvature scalar, $\phi$ is a scalar dilaton 
field, $\omega$ is the Brans-Dicke parameter, $\lambda$ is the 
cosmological constant, $F_{\mu\nu}=\partial_ \mu A_\nu - \partial_\nu 
A_\mu$ is the Maxwell tensor, where $A=A_\mu dx^\mu$ is the vector 
potential one-form, and $\bar{B}$ is a generic surface term. 
A general solution is a solution for the 
metric, the dilaton and the gauge field. We search for static solutions. 
The solution for the metric is then \cite{oscarlemos} 
\begin{eqnarray} 
\label{solutions} 
ds^2 &=& -\left[ (aR)^2 - \frac{b}{(aR)^{\frac{1}{\omega+1}}} 
+ \frac{k\chi^2}{(aR)^{\frac{2}{\omega+1}}} \right] 
dT^2 + \frac{dR^2}{(aR)^2 - 
\frac{b}{(aR)^{\frac{1}{\omega+1}}} + 
\frac{k\chi^2}{(aR)^{\frac{2}{\omega+1}}}}+ 
R^2 d\varphi^2, \,\,\,\, \omega\neq-2\,,-\frac{3}{2}\,,-1\,, 
\label{solutions_1}\\ 
ds^2 &=& -\left[\left(1+\frac{\chi^2}{4}\ln R \right)R^2-bR \right] dT^2+ 
\frac{dR^2}{\left(1+\frac{\chi^2}{4}\ln R \right)R^2-bR} 
+R^2 d\varphi^2 \qquad \qquad \quad \,\,\,\,\, \omega=-2\,,\\ 
ds^2 &=& \left(4\lambda^2 R^2 \ln(bR) + \chi^2 R^4\right) dT^2 
- \frac{dR^2}{4\lambda^2 R^2 \ln(bR) + \chi^2 R^4}+  R^2 d\varphi^2, 
\qquad \qquad \qquad \qquad \,  \omega=-\frac{3}{2}\,,\\ 
ds^2 &=& -dT^2 + dR^2 + d\varphi^2, \qquad \qquad \qquad \qquad 
\qquad \qquad \qquad \qquad \qquad \qquad \qquad \qquad \qquad 
\,\,\,\, \omega=-1, 
\label{mink} 
\end{eqnarray} 
where $T,R$ are Schwarzschild coordinates, $a$ is a constant related 
to the cosmological constant (see below), and $b$ is a constant of 
integration related to the mass (see below), $\chi$ is constant of 
integration related to the electric charge, and $k$ is defined as 
$k=\frac18\frac{(\omega+1)^2}{(\omega+2)}$. The general solution for 
$\phi$ is given by 
\begin{eqnarray} 
\phi &=& -\frac{1}{2(\omega+1)}\ln(a\,R)\,, \qquad \omega\neq-1\,, 
\label{phisolutions_1}\\ 
\phi &=& \textrm{constant}\,, \qquad \qquad \qquad \omega=-1\,. 
\end{eqnarray} 
The general solution for the vector potential $A$, written as 
$A=A_\mu(r)dx^\mu = A_T(r)\,dT$, is given by 
\begin{eqnarray} 
A &=& \frac14 \chi (\omega+1) \left(a\,R \right)^{-\frac{1}{\omega+1}} 
dT\,, 
\qquad \omega\neq-1\,, 
\label{vector_potential}\\ 
A &=&0\,, \qquad \qquad \qquad \omega=-1\,. 
\end{eqnarray} 
For the constant $a$ one has 
\begin{eqnarray} 
a &=& \frac{2\left|(\omega+1)\lambda\right|}{\sqrt{\left|(\omega+2) 
(2\omega+3)\right|}}\,,\qquad \omega\neq-2,-\frac{3}{2},-1\,, 
\label{a_1} \\ 
a &=& 1\,, \qquad \qquad \qquad \qquad \qquad \omega=-2,-\frac{3}{2}\,, 
\label{a_2}\\ 
a &=& 0\,, \qquad \qquad \qquad \qquad \qquad \omega=-1\,. 
\label{a_3} 
\end{eqnarray} 
For $\omega=-2$ we have $\lambda^2 = -\chi^2/16$, which means that, 
unlike for the uncharged case, the cosmological constant is not null. 
The constant $b$ is related to the ADM mass of the solutions by 
\begin{eqnarray} 
M &=& \frac{\omega+2}{\omega+1}\,b\,, \qquad \,\,\,\, 
\omega\neq-2,\,-\frac{3}{2},\,-1\,,\\ 
M &=& -b \qquad \qquad \quad \, \omega=-2\,,\\ 
M &=& -4\lambda^2 \ln{b}, \qquad \omega=-\frac{3}{2}\,,\\ 
M &=& 0, \qquad \qquad \quad \,\,\, \omega=-1\,. 
\end{eqnarray} 
The constant $\chi$ is s related to the ADM charge of the solutions by 
\begin{eqnarray} 
\chi^2 &=& Q^2\,, \qquad \omega\neq -2,-\frac32,-1\,, 
\label{chi_2}\\ 
\chi^2 &=& 0\,, \qquad\;\; \omega=-1\,. 
\end{eqnarray} 
For $\omega=-2,-\frac32$, there is no good 
definition for $Q$ in terms of $\chi$. 

Since we want to perform a canonical Hamiltonian analysis, using an 
ADM formalism supplied with proper boundary conditions, it is 
necessary to pick up only those solutions that fulfill the conditions 
we want to impose. First, we are interested only in solutions with 
horizons, so we take $b$ to be positive.  Second, apart from a measure 
zero of solutions, all solutions have a non-zero $|\lambda|$. This 
does not mean straight away that the solutions are asymptotically 
anti-de Sitter. Some have one type or another of singularities at 
infinity, which do not allow an imposition of proper boundary 
conditions.  So, from \cite{oscarlemos} with the corresponding 
Carter-Penrose diagrams we discard the following solutions: 
$\omega=-1$ which is simply the Minkowski solution, 
$-1>\omega>-\frac32$ since it has a curvature singularity at 
$R=+\infty$, which is inside the horizon, and has a null topological 
singularity at $R=0$, $\omega=-\frac32$ since all the Carter-Penrose 
boundary is singular, the same happening in the interval 
$-\frac32>\omega>-2$.  Thus, the cases of interest to be studied are 
black holes for which $\infty>\omega>-1$ and $-2>\omega>-\infty$. For 
$b$ positive these solutions have ADM mass $M$ positive.  As in 
\cite{diaslemos} we choose three typical amenable cases where an 
analytical study can be done.  The first case is the charged BTZ black 
hole \cite{mtz} (see also \cite{btz}), which can be found by taking 
appropriately the limit $\omega\to\pm\infty$.  This black hole is thus a 
solution of the Einstein-Maxwell action in three dimensions.  The 
other cases are $\omega=0$ and $\omega=-3$. The theory with $\omega=0$ 
is equivalent to cylindrical four-dimensional Einstein-Maxwell general 
relativity, and the corresponding charge black hole was found in 
\cite{lemoszanchin}, see also \cite{oscarlemos}.  The theory with 
$\omega=-3$ is just a case of 3D charged Brans-Dicke theory, with a 
black hole solution that can be analyzed in this context 
\cite{oscarlemos}. 

Now, as mentioned in the introduction, the definition of the ADM mass 
$M$ is not straightforward when there is electric charge, in the sense 
that one can face a divergent asymptotic behavior of the mass $M$, see 
\cite{mtz,oscarlemos} as well as \cite{btz}.  The problem is more 
acute when $\omega<-\frac32$, which includes the general relativity 
case (since general relativity 
can be considered the $\omega\rightarrow-\infty$), see 
\cite{oscarlemos} for a full discussion (see also \cite{btz,mtz}). 
For such $\omega$, the mass $M$ can be written as 
$M=M_{Q=0}+\textrm{Div}_M (\chi,R)$, where $M_{Q=0}$ is the mass for 
the uncharged case and $\textrm{Div}_M (\chi,R)$ is the divergent part 
when there is charge.  To treat this divergence, as explained in 
\cite{oscarlemos}, we consider a large radius $R_*$ frontier. Then we 
write the mass $M$ as 
\begin{eqnarray} 
M &=& M(R_*)+\left[\textrm{Div}_M 
(\chi,R)-\textrm{Div}_M (\chi,R_*)\right]\,, 
\label{mass_ren} 
\end{eqnarray} 
where the function $M(R_*)$ is equal to 
\begin{eqnarray} 
M(R_*) &=& M_{Q=0}+\textrm{Div}_M (\chi,R_*)\,. 
\label{mass_uncharged} 
\end{eqnarray} 
The term in square brackets in Eq. (\ref{mass_ren}) tends to zero as 
the radius $R$ tends to the frontier radius, $R\to R_*$. $M(R_*)$ is 
interpreted as the energy inside the radius $R_*$. From 
Eq. (\ref{mass_uncharged}) the term $-\textrm{Div}_M (\chi,R_*)$ is 
interpreted as the energy outside the frontier in $R_*$, apart from an 
infinite constant, which is hidden in $M(R_*)$. The total mass 
definition, $M$, is then well defined.  In practical terms, one needs 
not pay attention to the divergent factors $\textrm{Div}_M$ because 
all one needs to do is consider null the asymptotic limits, $\lim 
\textrm{Div}_M(\chi,R)=0$, ignoring the frontier at $R_*$.  For 
$\omega>-3/2$ the mass term behavior of the three-dimensional black 
holes is similar to the Reissner-Nordstr\"om case, and it offers no 
problems. Indeed, for the Reissner-Nordstr\"om black hole the metric 
function $g_{tt}$ and the mass term, following the prescription above, 
are written as $g_{tt}= 1-\frac{2M(R_*)}{R} + Q^2\left(\frac{1}{R} 
-\frac{1}{R_*}\right)$, and $M=M(R_*) + \frac{Q^2}{2R_*}$, 
respectively.  The energy inside the frontier $R_*$ is then given by 
$M(R_*)$, and the electrostatic energy outside is given by 
$\frac{Q^2}{2R_*}$. As this last term tends to zero as $R_*\to\infty$ 
there are no real divergences in this case, thus implying that 
$M(R_*)$ carries no infinity within, and so the prescription is 
trivial, it gves nothing new. This is similar to what happens for 
three-dimensional black holes with $\omega>-3/2$. On the other hand, 
for $\omega<-3/2$ one should follow the prescription carefully, i.e., 
one has to hide an infinite constant in the definition of the energy 
within the frontier at $R_*$. 

\subsection{ADM form of the metric and dilaton} 
\label{whichallowadm} 
The ansatz for the metric and dilaton fields 
with which we start our canonical analysis is given by 
\begin{equation} 
ds^2 = -N(t,r)^2 dt^2+\Lambda(t,r)^2(dr+N^r(t,r) 
dt)^2+R(t,r)^2d\varphi^2\,, 
\label{ADM_ansatz} 
\end{equation} 
\begin{equation} 
{\rm e}^{-2\phi} = \left(a\,R(t,r)\right)^{\frac{1}{\omega+1}}\,. 
\label{ADMphiansatz} 
\end{equation} 
where $\{T,R,\varphi\}$ are the Schwarzschild type time, 
radius, and angular coordinates. This is the ADM ansatz for 
spherically symmetric solutions applied to three dimensions. 
In this we follow the basic formalism developed by Kucha\v{r} 
\cite{kuchar}.  The canonical coordinates $R$ and $\Lambda$ are functions 
of $t$ and $r$, i.e., $R=R(t,r)$ and $\Lambda=\Lambda(t,r)$. Now, $r=0$ is 
generically on the horizon as analyzed in \cite{kuchar}, but for our 
purposes $r=0$ represents the horizon bifurcation point of the 
Carter-Penrose diagram \cite{louko1} (see also 
\cite{louko2}-\cite{louko5} and \cite{diaslemos}). In three spacetime 
dimensions the point actually represents a circle.  Asymptotically 
the coordinate $r$ 
tends to $\infty$, and 
$t$ is another time coordinate. The remaining functions are the lapse 
function $N=N(t,r)$, and the 
shift function $N^r=N^r(t,r)$, and they will play the role 
of Lagrange multipliers of the Hamiltonian of the theory.  The 
canonical coordinates $R=R(t,r),\,\Lambda=\Lambda(t,r)$ and the lapse 
function $N=N(t,r)$ are taken to be positive.  The angular coordinates 
are left untouched, due to spherical symmetry.  The dilaton is a 
simple function of the radial canonical coordinate, and it can be 
traded directly by it through equation (\ref{ADMphiansatz}), as will 
be done below.  The ansatz (\ref{ADM_ansatz})-(\ref{ADMphiansatz}) is 
written in order to perform the foliation of spacetime into spacelike 
hypersurfaces, and thus separates the spatial part of the spacetime 
from the temporal part.  Indeed, the canonical analysis requires the 
explicit separation of the time coordinate from the other space 
coordinates, and so in all expressions time is treated separately from 
the other coordinates. It breaks explicit, but not implicit, 
covariance of the theory, but it is necessary in order to 
perform the Hamiltonian analysis.  The metric coefficients of the 
induced metric on the hypersurfaces become the canonical variables, 
and the momenta are determined in the usual way, by replacing the time 
derivatives of the canonical variables, the velocities.  Then, using 
the Hamiltonian, one builds a time evolution operator to construct an 
appropriate thermodynamic ensemble for the geometries of a quantum 
theory of gravity.  Assuming that a quantum theory only makes sense if 
its classical form can be quantized by Hamiltonian methods, one should 
pick up only solutions which can be put consistently in a Hamiltonian 
form.  Thus, in the following we perform a Hamiltonian analysis to 
extract the entropy and other thermodynamic properties in the 
three-dimensional Brans-Dicke-Maxwell black holes mentioned above, 
those for which $\omega=\infty, 0, -3$. 

\subsection{ADM form for the vector potential $A$} 
\label{adm_vectorpotential} 
The canonical description of the vector potential one-form is given by 
\begin{eqnarray} 
A &=& \Gamma dr + \Phi dt\,, 
\label{a_ansatz} 
\end{eqnarray} 
where $\Gamma$ and $\Phi$ are functions of $t$ and $r$, i. e., 
$\Gamma(t,r)$ and $\Phi(t,r)$. 
The function $\Gamma(t,r)$ is the canonical coordinate associated with 
the electric field, and the function $\Phi(t,r)$ is the Lagrange 
multiplier associated with the electromagnetic constraint, which is 
Gauss' Law. In the same way as with the gravitational degrees of freedom 
above, when the Maxwell term in the action (\ref{3dchbdaction}) is 
written, the conjugate momentum to the coordinate $\Gamma(t,r)$ can be 
derived the usual way from the Lagrangian. 

\section{Hamiltonian Thermodynamics of the charged 
BTZ black hole ({\large $\omega=\infty$})} 
\label{mtz} 

\subsection{The metric and the vector potential one-form} 

\begin{figure} 
[htmb] 
\centerline{\includegraphics 
%[width=5.7cm,height=15cm] 
{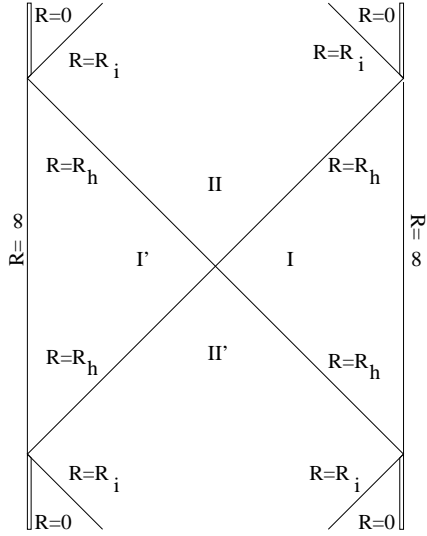}} 
\caption {{\small The Carter-Penrose diagram for 
the charged BTZ case, 
where $R_{\textrm{h}}$ is the outer horizon radius, 
$R_{\rm i}$ is the inner horizon radius, and the double 
line is the timelike singularity at $R=0$. The static 
region I, from the bifurcation point to 
spacelike infinity $R=\infty$, is the relevant region for 
our analysis. The regions II and II' are inside the outer 
horizon, beyond the foliation domain. Region I' is the 
symmetric of region I, towards left infinity.}} 
\label{omega_infinity_maxwell} 
\end{figure}

For $\omega=\infty$, the corresponding three-dimensional 
charged Brans-Dicke theory is precisely the 
one provided by three-dimensional Einstein-Maxwell theory 
\cite{btz,mtz,oscarlemos}. 
Then the general metric in Eq. (\ref{solutions}), the $\phi$ field in 
Eq. (\ref{phisolutions_1}), and the vector potential one-form 
(\ref{vector_potential}) reduce to the following 
\begin{eqnarray} 
ds^2 &=& -\left(\frac{R^2}{l^2} - M - 
\pi Q^2 \ln \left(\frac{R}{l}\right)\right)dT^2+ 
\frac{dR^2}{\frac{R^2}{l^2} - M -     \pi Q^2 
\ln \left(\frac{R}{l}\right)} 
+R^2d\varphi^2\,, 
\label{solucao_mtz}\\ 
{\rm e}^{-2\phi}&=&1\,, 
\label{dilaton__mtz}\\ 
A &=& \frac{1}{2\pi}Q \ln\left(\frac{R}{l}\right)dT\,, 
\label{vector_potential_oo} 
\end{eqnarray} 
where $l$ is the AdS length, related to the cosmological constant by 
$2\lambda^2=l^{-2}$, $M$ is the mass, and $Q$ is the 
electric charge. In this solution the scalar field $\phi$ is trivial. 
Next, in Fig. \ref{omega_infinity_maxwell}, we show the Carter-Penrose 
diagram of the three-dimensional charged black hole. 
Despite there being a large and rich structure, the outside region 
in I is the relevant one, thus our foliation is from the bifurcation 
point to 
$R=\infty$, and our boundary conditions reflect properties of the 
outer horizon at $R=R_{\rm h}$, and of the spacelike infinity 
at $R=\infty$. 

\subsection{Canonical formalism} 
Given Eqs. (\ref{ADM_ansatz})-(\ref{a_ansatz}), 
the three-dimensional action (\ref{3dchbdaction}), 
now an Einstein-Maxwell action, becomes, excluding surface terms, 
\begin{eqnarray} 
S[\Lambda,\,R,\,\Gamma,\,\dot{\Lambda},\,\dot{R},\,\dot{\Gamma}; 
\,N,\,N^r,\,\Phi]  &=& \int dt 
\int_0^{\infty} dr 
\left\{- 2N^{-1}\dot{\Lambda}\dot{R} + 2 N^{-1}N^r R'\dot{\Lambda} + 
2 N^{-1}(N^r)'\Lambda \dot{R} + 2 N^{-1}N^r \Lambda ' \dot{R}\right. 
\nonumber \\ 
&& \left. - 2N^{-1}(N^r)^2\Lambda 'R' + 2N\Lambda^{-2}\Lambda 
'R'-2N\Lambda^{-1}R'' 
+ 4N\lambda^2\Lambda R \right. \nonumber \\ 
&& \left. + \pi N^{-1}\Lambda^{-1}R\left(\dot{\Gamma}-\Phi'\right)^2 
\right\}\,, 
\label{infty} 
\end{eqnarray} 
where $2\lambda^2=l^{-2}$ and $l$ is the 
AdS length, which then yields the quantity $a$ is 
$a^2 = 2\lambda^2 =l^{-2}$. 
In Eq. (\ref{infty}) a $\dot{}$ means derivative with respect to time 
$t$ and a $'$ means derivative with respect to $r$, and where all the 
explicit functional dependences are omitted. 
{}From the action (\ref{infty}) 
one determines the canonical momenta, conjugate to $\Lambda$, $R$, and 
$\Gamma$ respectively, 
\begin{eqnarray} 
\label{p_lambda_oo} 
P_{\Lambda} &=& -2 N^{-1} \left\{\dot{R}-R'N^r\right\}\,, \\ 
P_R &=& -2 N^{-1}\left\{ \dot{\Lambda}-(\Lambda 
N^r)'\right\}\,, \label{p_r_oo}\\ 
P_{\Gamma} &=& 2\pi N^{-1}\Lambda^{-1}R\left(\dot{\Gamma}-\Phi'\right)\,. 
\label{p_gamma_oo} 
\end{eqnarray} 
By performing a Legendre transformation, we obtain 
\begin{eqnarray} \label{Hamiltonian_oo} 
\mathcal{H} &=& N\left\{-\frac12 P_R P_\Lambda 
+\frac{1}{4\pi}P_{\Gamma}^2\Lambda R^{-1}-2\Lambda^{-2}\Lambda'R' 
+2\Lambda^{-1}R''-4\lambda^2\Lambda R\right\}+N^r\left\{P_R 
R'-P_{\Lambda}'\Lambda-\Gamma P_{\Gamma}'\right\} 
+\tilde{\Phi}\left\{-P_{\Gamma}'\right\} \nonumber \\ 
&& \equiv NH+N^rH_r+\tilde{\Phi}G \,, 
\end{eqnarray} 
defining implicitly in this way 
the constraints $H$ and $H_r$, and $G$, and 
where the new Lagrange multiplier is $\tilde{\Phi}\equiv \Phi-N^r\Gamma$. 
Here additional surface terms have been ignored, as for now we are 
interested in the bulk terms only.  The action in Hamiltonian form is 
then 
\begin{equation} 
\label{haction_oo} 
S[\Lambda,\,R,\,\Gamma,\,P_{\Lambda}, 
\,P_{R},\,P_{\Gamma};\,N,\,N^r,\,\tilde{\Phi}] = 
\int dt \int_0^{\infty} dr 
\left\{P_{\Lambda}\dot{\Lambda}+P_R\dot{R}+P_{\Gamma}\dot{\Gamma} 
-NH-N^rH_r-\tilde{\Phi}G\right\}\,. 
\end{equation} 
The equations of motion are 
\begin{eqnarray} 
\dot{\Lambda} &=& -\frac12 N P_R + (N^r \Lambda)'\,, \\ 
\dot{R} &=& -\frac12 N P_\Lambda - N^r R'\,, \\ 
\dot{P_R} &=& 4\lambda^2 N \Lambda - (2N'\Lambda^{-1})' + 
(N^r P_R)'+\frac{1}{4\pi}P_{\Gamma}^2N\Lambda R^{-2}\,,\\ 
\dot{P_{\Lambda}}&=& 4\lambda^2 N R - 2N'\Lambda^{-2}R'+N^r(P_\Lambda)'- 
\frac{1}{4\pi}P_{\Gamma}^2NR^{-1}\,,\\ 
\dot{\Gamma}&=&\frac{1}{2\pi}P_{\Gamma}N\Lambda R^{-1}+ 
\left(N^r\Gamma\right)'+\tilde{\Phi}'\,,\\ 
\dot{P_{\Gamma}}&=&N^rP_{\Gamma}'\,. 
\end{eqnarray} 
In order to have a well defined variational principle, we need to 
eliminate the surface terms of the original bulk action, which render 
the original action itself ill defined when one seeks a correct 
determination of the equations of motion through variational methods. 
These surface terms are eliminated through judicious choice of extra 
surface terms which should be added to the action.  The action 
(\ref{haction_oo}) has the following extra surface terms, after 
variation 
\begin{equation} 
\label{general_surf_terms_oo} 
\textrm{Surface terms}=\left.\left(-N^rP_R\delta R+N^r\Lambda \delta 
P_\Lambda 
+2N\Lambda^{-2}R'\delta\Lambda-2N\Lambda^{-1}\delta 
R'+2N'\Lambda^{-1}\delta R+N^r\Gamma \delta P_{\Gamma}+ 
\tilde{\Phi}\delta P_{\Gamma}\right)\right|_0^\infty\,. 
\label{surface_terms_oo} 
\end{equation} 
In order to evaluate this expression, we need to know the asymptotic 
conditions of each 
of the above functions individually, which are functions of $(t,r)$. 

Starting with the limit $r\rightarrow 0$ we assume 
\begin{eqnarray} 
\label{falloff_0_i_oo} 
\Lambda(t,r) &=& \Lambda_0+O(r^2)\,,\\ 
R(t,r)&=& R_0+R_2r^2+O(r^4)\,,\\ 
P_\Lambda(t,r) &=& O(r^3)\,,\\ 
P_R(t,r) &=& O(r)\,,\\ 
N(t,r) &=& N_1(t)r+O(r^3)\,,\\ 
N^r(t,r) &=& O(r^3)\,,\\ 
\Gamma(t,r) &=& O(r)\,,\\ 
P_{\Gamma}(t,r) &=& 2\pi Q_0 + 2\pi Q_2 r^2 + O(r^4)\,,\\ 
\tilde{\Phi}(t,r)&=& \tilde{\Phi}_0 (t) + O(r^2)\,. 
 \label{falloff_0_f_oo} 
\end{eqnarray} 
Note that there are time dependences on the left hand side of the 
falloff conditions and that there are no such dependences on the right 
hand side, in the lower orders of the expansion in $r$. This apparent 
discrepancy stems from the fact that there is in fact no time 
dependence in the lower orders of the majority of the functions, but 
there may still exist such dependence for higher orders. Nevertheless, 
terms such as $R_0$ are functions, independent of $(t,r)$, thus 
constant, but undetermined. Their variation makes sense, as we may 
still vary between different values for these constant functions. 
With these conditions, we have for the surface terms at $r=0$ 
\begin{equation} \label{surface_0_oo} 
\left.\textrm{Surface terms}\right|_{r=0} = -2 N_1\Lambda_0^{-1}\delta 
R_0 - \tilde{\Phi}_0\delta\left(2\pi Q_0 \right)\,. 
\end{equation} 
For $r\rightarrow\infty$ we have 
\begin{eqnarray} \label{falloff_inf_i_oo} 
\Lambda(t,r) &=& lr^{-1}+l^3\eta(t)r^{-3} + O^\infty(r^{-5})\,,\\ 
R(t,r) &=& r + l^2\rho(t)r^{-1}+O^\infty(r^{-3})\,,\\ 
P_\Lambda(t,r) &=& O^\infty(r^{-2})\,,\\ 
P_R(t,r) &=& O^\infty(r^{-4})\,,\\ 
N(t,r) &=& R(t,r)'\Lambda(t,r)^{-1}(\tilde{N}_+(t)+O^\infty(r^{-5}))\,,\\ 
N^r(t,r) &=& O^\infty(r^{-2})\,,\\ 
\Gamma(t,r) &=& O^{\infty}(r^{-2})\,,\\ 
P_{\Gamma}(t,r) &=& 2\pi Q_+(t) + O^{\infty}(r^{-2})\,,\\ 
\tilde{\Phi}(t,r) &=& \tilde{\Phi}_+ (t)+ O^{\infty}(r^{-1})\,. 
 \label{falloff_inf_f_oo} 
\end{eqnarray} 
These conditions imply for the surface terms in the limit 
$r\rightarrow\infty$ 
\begin{equation} 
\left.\textrm{Surface terms}\right|_{r\rightarrow\infty} = 
2\delta(M_+(t))\tilde{N}_+ + \tilde{\Phi}_+ 
\delta\left(2\pi Q_+\right)\,, 
\end{equation} 
where $M_+(t)=2(\eta(t)+2\rho(t))$. 
So, the surface term added to (\ref{haction_oo}) is 
\begin{equation} \label{surface_inf_oo} 
S_{\partial\Sigma}\left[\Lambda,R,Q_0,Q_+; 
N,\tilde{\Phi}_0,\tilde{\Phi}_+\right] 
=\int dt \left(2 
 R_0 N_1 \Lambda_0^{-1} - \tilde{N}_+ M_+ + \tilde{\Phi}_0 
\left(2\pi Q_0\right) - \tilde{\Phi}_+ \left(2\pi Q_+\right)\right)\,. 
\end{equation} 
What is left after varying 
this last surface term and adding it to the varied initial action 
(see Eq. (\ref{haction_oo})) is 
\begin{equation}\label{variationofsurfaceterm_oo} 
\int dt \left(2 R_0 \delta(N_1 \Lambda_0^{-1}) - 
M_+\delta\tilde{N}_+ + 2\pi Q_0 \delta\tilde{\Phi}_0 
- 2\pi Q_+ \delta\tilde{\Phi}_+ \right)\,. 
\end{equation} 
We choose to fix $N_1 \Lambda_0^{-1}$ and $\tilde{\Phi}_0 $ 
on the horizon ($r=0$), and $\tilde{N}_+$ and $\tilde{\Phi}_0$ 
at infinity. These choices make the surface variation 
(\ref{variationofsurfaceterm_oo}) disappear. 
The term $N_1 \Lambda_0^{-1}$ is the integrand of 
\begin{equation} 
n^a(t_1)n_a(t_2) = -\cosh \left(\int_{t_1}^{t_2}\,dt\,N_1(t) 
  \Lambda_0^{-1}(t) \right)\,, 
\end{equation} 
which is the rate of the boost suffered by the future unit normal to the 
constant $t$ hypersurfaces defined at the 
bifurcation circle, i.e., at $r\rightarrow0$, due to the evolution of the 
constant $t$ hypersurfaces. By fixing the integrand we are fixing the rate 
of the boost, which allows us to control the metric singularity when 
$r\rightarrow0$ \cite{louko1}. Finally, we also fix $\tilde{\Phi}_0$ and 
$\tilde{\Phi}_+$ at $r=0$ and at infinity, respectively. 

\subsection{Reconstruction, canonical transformation, and action} 
\label{reconstruction} 
In order to reconstruct the mass and the time from the canonical data, 
which amounts to making a canonical transformation, we have to rewrite 
the form of the solutions of 
Eqs.~(\ref{solucao_mtz})-(\ref{vector_potential_oo}).  However, in 
addition to reconstructing the mass as done in \cite{diaslemos}, we 
have to consider how to reconstruct the charge from the canonical 
data.  We follow Kucha\v{r} \cite{kuchar} for this reconstruction.  We 
concentrate our analysis on the right static region of the 
Carter-Penrose diagram. 

Developing the Killing time $T$ as function of $(t,r)$ in the 
expression for the vector potential (\ref{vector_potential_oo}), and 
making use of the gauge freedom that allows us to write 
\begin{eqnarray} 
A &=& \frac{1}{2\pi}Q\ln\left(\frac{R}{l}\right)dT+d\xi\,, 
\label{vector_potential_plusgauge_oo} 
\end{eqnarray} 
where $\xi(t,r)$ is an arbitrary continuous function of $t$ and $r$, 
we can write the one-form potential as 
\begin{eqnarray} 
A &=& \left(\frac{1}{2\pi}Q\ln\left(\frac{R}{l}\right)T'+\xi'\right)dr 
+ 
\left(\frac{1}{2\pi}Q\ln\left(\frac{R}{l}\right)\dot{T}+\dot{\xi}\right)dt\,. 
\label{vector_potential_plusgauge_tr_oo} 
\end{eqnarray} 
Equating the expression (\ref{vector_potential_plusgauge_tr_oo}) 
with Eq. (\ref{a_ansatz}), and making use of the definition of 
$P_{\Gamma}$ in Eq. (\ref{p_gamma_oo}), we arrive at 
\begin{eqnarray} 
P_{\Gamma} &=& 2\pi Q\,. 
\label{p2piq} 
\end{eqnarray} 
In the static region, we define $F$ as 
\begin{equation}\label{f_1_oo} 
F(t,r) = \frac{R^2}{l^2} - M - \frac{1}{4\pi} P_{\Gamma}^2 
\ln\left(\frac{R}{R_*}\right)\,, 
\end{equation} 
where we define the $R_*$ as the (large) radius of a finite frontier, 
which 
will serve as a renormalization of the asymptotic properties of the 
future canonical coordinate $M(t,r)$ (see subsection \ref{bhsolutions}). 
This frontier will be used throughout the present section. 
We now make the following substitutions 
\begin{equation} 
T=T(t,r)\,, \qquad \qquad R=R(t,r)\,, 
\end{equation} 
into the solution  (\ref{solucao_mtz}), getting 
\begin{eqnarray} 
ds^2 &=& -(F\dot{T}^2-F^{-1}\dot{R}^2)\,dt^2\,+ 
\,2(-FT'\dot{T}+F^{-1}R'\dot{R})\,dtdr\,+\,(-F(T')^2+F^{-1}\dot{R}^2)\,dr^2\,+ 
\,R^2d\varphi^2\,. 
\end{eqnarray} 
This introduces the ADM foliation directly into the solutions. 
Comparing it with the ADM metric (\ref{ADM_ansatz}), written in 
another form as 
\begin{equation} 
ds^2 = 
-(N^2-\Lambda^2(N^r)^2)\,dt^2\,+\,2\Lambda^2N^r\,dtdr\,+ 
\,\Lambda^2dr^2\,+\,R^2\,d\varphi^2\,, 
\end{equation} 
we can write a set of three equations 
\begin{eqnarray} 
\Lambda^2 &=& -F(T')^2+F^{-1}(R')^2\,,\label{adm_1_oo}\\ 
\Lambda^2N^r &=& -FT'\dot{T}+F^{-1}R'\dot{R}\,, \label{adm_2_oo}\\ 
N^2-\Lambda^2(N^r)^2 &=& F\dot{T}^2-F^{-1}\dot{R}^2\,. \label{adm_3_oo} 
\end{eqnarray} 
The first two equations, Eqs. (\ref{adm_1_oo}) and Eq. (\ref{adm_2_oo}), 
give 
\begin{equation} \label{shift_def_oo} 
N^r = \frac{-FT'\dot{T}+F^{-1}R'\dot{R}}{-F(T')^2+F^{-1}(R')^2}\,. 
\end{equation} 
This one solution, together  with Eq. (\ref{adm_1_oo}), give 
\begin{equation} \label{lapse_def_oo} 
N = \frac{R'\dot{T}-T'\dot{R}}{\sqrt{-F(T')^2+F^{-1}(R')^2}}\,. 
\end{equation} 
One can show that $N(t,r)$ is positive (see \cite{kuchar}). 
Next, putting Eqs. (\ref{shift_def_oo})-(\ref{lapse_def_oo}), 
into the definition of the conjugate momentum of 
the canonical coordinate $\Lambda$, given in 
Eq. (\ref{p_lambda_oo}), one finds 
the spatial derivative of $T(t,r)$ as a function of the canonical 
coordinates, i.e., 
\begin{equation} 
\label{t_linha_oo} 
-T' = \frac12 F^{-1}\Lambda P_\Lambda\,. 
\end{equation} 
Later we will see that $-T'=P_M$, as it 
will be conjugate to a new canonical coordinate $M$. 
Following this procedure to the end, we may then find the form of the 
new coordinate $M(t,r)$, as a function of $t$ and $r$. First, we 
need to know the form of $F$ as a function of the canonical pair 
$\Lambda\,,\,R$. For that, we replace back into Eq. (\ref{adm_1_oo}) the 
definition, in Eq. (\ref{t_linha_oo}), of $T'$, giving 
\begin{equation}\label{f_2_oo} 
F = 
\left(\frac{R'}{\Lambda}\right)^2-\left(\frac{P_\Lambda}{2}\right)^2\,. 
\end{equation} 
Equating this form of $F$ with Eq. (\ref{f_1_oo}), we obtain 
\begin{equation} \label{m_oo} 
M = \frac{R^2}{l^2}- 
\frac{1}{4\pi}P_{\Gamma}^2\ln\left(\frac{R}{R_*}\right)-F\,, 
\end{equation} 
where $F$ is given in Eq. (\ref{f_2_oo}). 
We thus have found the form of the new canonical coordinate, $M$. It 
is now a straightforward calculation to determine the Poisson bracket 
of this variable with $P_M=-T'$ and see that they are conjugate, thus 
making Eq. (\ref{t_linha_oo}) the conjugate momentum of $M$, i.e., 
\begin{equation}\label{p_m_oo} 
P_M = \frac12 F^{-1}\Lambda P_\Lambda\,. 
\end{equation} 

The other natural transformation has been given in (\ref{p2piq}), 
namely the one that relates $P_{\Gamma}$ with $Q$ through 
$P_{\Gamma}=2\pi Q$. As $M$, $Q$ is also a natural choice for 
canonical coordinate, being another physical parameter of the solution 
(\ref{solucao_mtz}). The one coordinate which remains to be found is 
$P_Q$, the conjugate momentum to the charge.  It is also necessary to 
find out the other new canonical variable which commutes with $M$, 
$P_M$, and $Q$, and which guarantees, with its conjugate momentum, 
that the transformation from $\Lambda,\,R,\,\Gamma$, to $M$, $Q$, and 
the 
new variable is canonical. 
Immediately it is seen that $R$ commutes 
with $M$, $P_M$, and $Q$. It is then a candidate. It remains to be seen 
whether $P_R$ also commutes with $M$, $P_M$, and $Q$.  As with $R$, it is 
straightforward to see that $P_R$ does not commute with $M$ and $P_M$, 
as these contain powers of $R$ in their definitions, and 
$\left\{R(t,r),\,P_R(t,r^*) \right\}=\delta(r-r^*)$.  So, rename the 
canonical variable $R$ as $R=\textrm{R}$.  We have then to find a new 
conjugate momentum to $\textrm{R}$ which also commutes with $M$, $P_M$, 
and $Q$, making the transformation from 
$\left\{\Lambda,\,R,\,\Gamma;\,P_\Lambda,\,P_R,\,P_{\Gamma} 
\,\right\}\rightarrow\, 
\left\{M,\,\textrm{R},\,Q;\,P_M,\,P_{\textrm{R}},\,P_Q\,\right\}$ 
a canonical 
one. 
The way to proceed is to look at the constraint $H_r$, which is called 
in this formalism the super-momentum. This is the constraint which 
generates spatial diffeomorphisms in all variables. Its form, in the 
initial canonical coordinates, is 
$H_r=-\Lambda\,P_\Lambda'+P_R\,R'-\Gamma P_{\Gamma}'$. 
In this formulation, $\Lambda$ is a spatial density, $R$ is a 
spatial scalar, and $\Gamma$ is also a spatial density. 
As the new variables, $M$, $\textrm{R}$, and $Q$, are 
spatial scalars, the generator of spatial diffeomorphisms is written 
as $H_r=P_M M'+P_{\textrm{R}} \textrm{R}'+P_Q Q'$, regardless of the 
particular form of the canonical coordinate transformation. It is thus 
equating these two expressions of the super-momentum $H_r$, with $M$, 
$P_M$, and $Q$  written as functions of $\Lambda,\,R,\,\Gamma$ 
and their respective 
momenta, that gives us one equation for the new $P_{\textrm{R}}$ and 
$P_Q$. This means that we have two unknowns, $P_{\textrm{R}}$ and 
$P_Q$, for one equation only. This sugests that we should make 
the coefficients of $R'$ and $P_{\Gamma}'$ equal to zero independently. 
This results in 
\begin{eqnarray} 
P_{\textrm{R}} &=& P_R - 2\lambda^2 F^{-1}\Lambda P_\Lambda R + 
F^{-1}\Lambda^{-1} P_\Lambda R''- F^{-1}\Lambda^{-2}P_\Lambda \Lambda'R' 
\nonumber \\ 
&& - F^{-1}\Lambda^{-1} P_\Lambda' R' + \frac{1}{8\pi}P_{\Gamma}^2 R^{-1} 
F^{-1}\Lambda P_{\Lambda}\,, 
\label{p_rnew_oo}\\ 
P_Q &=& -2\pi\Gamma+\frac12 P_{\Gamma} 
\ln\left(\frac{R}{R_*}\right)F^{-1}\Lambda P_{\Lambda}\,. 
\label{p_q_oo} 
\end{eqnarray} 
It can be shown that $P_Q$ commutes with the new $P_{\textrm{R}}$ 
and with the rest of the new coordinates, except with $Q$. 
We have now all the canonical variables of the new set determined. For 
completeness and future use, we write the inverse transformation for 
$\Lambda$ and $P_\Lambda$, 
\begin{eqnarray} \label{inversetrans_1_oo} 
\Lambda &=& \left((\textrm{R}')^2F^{-1}-P_M^2F\right)^{\frac12}\,, \\ 
P_\Lambda &=& 2 F P_M 
\left((\textrm{R}')^2F^{-1}-P_M^2F\right)^{-\frac12}\,. 
\label{inversetrans_2_oo} 
\end{eqnarray} 
In summary, the canonical transformations are the following, 
\begin{eqnarray} \label{setcanonicaltrans_oo} 
\textrm{R} &=& R\,,\nonumber \\ 
M &=& \frac{R^2}{l^2} - 
\frac{1}{4\pi}P_{\Gamma}^2\ln\left(\frac{R}{R_*}\right)- F\,, 
\nonumber \\ 
Q &=& \frac{P_{\Gamma}}{2\pi}\,, \nonumber \\ 
P_{\textrm{R}} &=& P_R - 2\lambda^2 F^{-1}\Lambda P_\Lambda R + 
F^{-1}\Lambda^{-1} P_\Lambda R''-F^{-1}\Lambda^{-2}P_\Lambda \Lambda'R' 
\nonumber \\ 
&& - F^{-1}\Lambda^{-1} P_\Lambda' R' + 
\frac{1}{8\pi}P_{\Gamma}^2 R^{-1} F^{-1}\Lambda P_{\Lambda}\,, 
\nonumber \\ 
P_M &=& \frac12 F^{-1}\Lambda P_\Lambda\,, \nonumber \\ 
P_Q &=& -2\pi\Gamma+\frac12 P_{\Gamma} 
\ln\left(\frac{R}{R_*}\right)F^{-1}\Lambda P_{\Lambda}\,. 
\end{eqnarray} 
It remains to be seen that this set of transformations 
is in fact canonical. 
In order to prove that the set of equalities in expression 
(\ref{setcanonicaltrans_oo}) is canonical we start with the equality 
\begin{eqnarray} \label{oddidentity_oo} 
P_\Lambda \delta\Lambda + P_R \delta R +P_\Gamma \delta\Gamma 
- P_M \delta M - P_{\textrm{R}}\delta\textrm{R}-P_Q\delta Q 
&=& \left( \delta R \ln 
\left|\frac{2 R'+\Lambda P_\Lambda}{2 R'- \Lambda 
P_\Lambda}\right|\right)'+ \nonumber \\ 
&& +\, \delta\left(\Gamma P_{\Gamma}+\Lambda P_\Lambda + 
R' \ln \left|\frac{2 R'-\Lambda P_\Lambda}{2 R'+ \Lambda 
P_\Lambda}\right|\right)\,. 
\end{eqnarray} 
We now integrate expression (\ref{oddidentity_oo}) in $r$, in 
the interval from $r=0$ to $r=\infty$. The first term on the right 
hand side of Eq. (\ref{oddidentity_oo}) vanishes due to the falloff 
conditions (see Eqs. (\ref{falloff_0_i_oo})-(\ref{falloff_0_f_oo}) and 
Eqs. (\ref{falloff_inf_i_oo})-(\ref{falloff_inf_f_oo})). 
We then obtain the following expression 
\begin{eqnarray} \label{int_oddidentity_oo} 
\int_0^\infty\,dr\,\left(P_\Lambda \delta\Lambda + P_R \delta 
R + P_\Gamma \delta\Gamma \right) 
-\int_0^\infty\,dr\,\left(P_M \delta M + 
P_{\textrm{R}}\delta\textrm{R} + P_Q \delta Q \right) &=& 
\delta\omega\,\left[\Lambda,\,R,\,\Gamma,\,P_\Lambda,\,P_\Gamma\right]\,, 
\end{eqnarray} 
where 
$\delta\omega\,\left[\Lambda,\,R,\,\Gamma,\,P_\Lambda,\,P_\Gamma\right]$ 
is a well defined functional, which is also an exact form. This equality 
shows 
that the difference between the Liouville form of 
$\left\{R,\,\Lambda,\,\Gamma;\,P_R,\,P_\Lambda,\,P_\Gamma\right\}$ 
and the Liouville form 
of $\left\{\textrm{R},\,M,\,Q;\,P_{\textrm{R}},\,P_M,\,P_Q\right\}$ is an 
exact 
form, which implies that the transformation of variables given by the 
set of equations (\ref{setcanonicaltrans_oo}) is canonical. 

Armed with the certainty of the canonicity of the new variables, we 
can write the asymptotic form of the canonical variables and of the 
metric function $F(t,r)$. These are, for $r\rightarrow 0$ 
\begin{eqnarray} \label{newfalloff_0_i_oo} 
F(t,r) &=& 4 R_2^2 \Lambda_0^{-2} r^2 + O(r^4)\,, \\ 
\textrm{R}(t,r) &=& R_0+R_2\,r^2+O(r^4)\,, \\ 
M(t,r) &=& M_0 + M_2 \, r^2+O(r^4)\,, \\ 
Q(t,r) &=& Q_0 + Q_2 \, r^2+O(r^4)\,, \\ 
P_\textrm{R}(t,r) &=& O(r)\,, \\ 
P_M(t,r) &=& O(r)\,,\\ 
P_Q(t,r) &=& O(r)\,, 
\label{newfalloff_0_f_oo} 
\end{eqnarray} 
with 
\begin{eqnarray} 
    M_0 &=& l^{-2}R_0^2-\pi Q_0^2 \ln 
\left(\frac{R_0}{R_*}\right)\,,\\ 
    M_2 &=& 2l^{-2}R_0R_2 - 2\pi\ln 
\left(\frac{R_0}{R_*}\right)Q_0Q_2- 
    \pi Q_0^2R_2R_0^{-1}-4\Lambda_0^{-2}R_2^{2}\,. 
\end{eqnarray} 
For $r\rightarrow \infty$, we have 
\begin{eqnarray} \label{newfalloff_inf_i_oo} 
F(t,r) &=& 2\lambda^2\,r^2 - 2(\eta(t)+2\rho(t))+O^{\infty}(r^{-2})\,, \\ 
\textrm{R}(t,r) &=& r + (2\lambda^2)^{-1}\rho(t)\,r^{-1}+ 
O^{\infty}(r^{-3})\,, \\ 
M(t,r) &=& M_+(t) + O^{\infty}(r^{-2})\,, \\ 
Q(t,r) &=& Q_+(t) + O^{\infty}(r^{-2})\,, \\ 
P_\textrm{R}(t,r) &=& O^\infty(r^{-4})\,, \\ 
P_M(t,r) &=& O^\infty(r^{-5})\,, \\ 
P_Q(t,r) &=& O^{\infty}(r^{-2})\,, 
\label{newfalloff_inf_f_oo} 
\end{eqnarray} 
where $M_+(t)=2(\eta(t)+2\rho(t))$, as seen before in the 
surface terms (see Eq. (\ref{surface_inf_oo})). Note here 
that without the treatment of the divergence in subsection 
\ref{bhsolutions} the function $M(t,r)$ would diverge. 

We are now almost ready to write the action with the new canonical 
variables. It is now necessary to determine the new Lagrange 
multipliers. In order to write the new constraints with the new 
Lagrange multipliers, we can use the identity given by the space 
derivative of $M$, 
\begin{equation} 
M' = -\Lambda^{-1}\left( R' H +\frac12 P_\Lambda \left(H_r-\Gamma G\right) 
-\frac{1}{2\pi}P_\Gamma G \ln\left(\frac{R}{R_*}\right)\Lambda\right)\,. 
\end{equation} 
Solving for $H$ and making use of the inverse transformations of 
$\Lambda$ and $P_\Lambda$, in Eqs. (\ref{inversetrans_1_oo}) and 
(\ref{inversetrans_2_oo}), we get 
\begin{eqnarray} \label{oldconstraintsinnewvariables_1_oo} 
H &=& - \frac{M'F^{-1}\textrm{R}'+F P_M 
P_{\textrm{R}}+2\pi QQ'R'F^{-1}\ln\left(\frac{R}{R_*}\right)} 
{\left(F^{-1}(\textrm{R}')^2-FP_M^2\right)^{\frac12}}\,, \\ 
H_r &=& P_M M' + P_{\textrm{R}} 
\textrm{R}'+P_Q Q'\,, \\ 
G &=& -2\pi Q'\,. \label{oldconstraintsinnewvariables_2_oo} 
\end{eqnarray} 
Following Kucha\v{r} \cite{kuchar}, the new set of constraints, totally 
equivalent to the old set $H(t,r)=0$, $H_r(t,r)=0$, and $G=0$, outside the 
horizon points, is $M'(t,r)=0$, $P_{\textrm{R}}(t,r)=0$, and $Q'(t,r)=0$. 
By continuity, this also applies on the horizon, where $F(t,r)=0$. 
So we can say that the equivalence is valid everywhere. 
The new Hamiltonian, which is the total sum of the constraints, 
can now be written as 
\begin{equation} \label{newHamiltonian_oo} 
NH+N^rH_r+\tilde{\Phi}G= N^M M' + N^{\textrm{R}} P_{\textrm{R}}+N^QQ'\,. 
\end{equation} 
In order to determine the new Lagrange multipliers, one has to write 
the left hand side of the previous equation, 
Eq. (\ref{newHamiltonian_oo}), and replace the constraints on that side 
by their expressions as functions of the new canonical coordinates, 
spelt out in Eqs. 
(\ref{oldconstraintsinnewvariables_1_oo})-(\ref{oldconstraintsinnewvariables_2_oo}). 
After manipulation, one gets 
\begin{eqnarray} \label{new_mult_1_oo} 
N^M &=& - \frac{N F^{-1} R'}{\left(F^{-1}(\textrm{R}')^2-F 
    P_M^2\right)^{\frac12}}+ N^r P_M \,, \\ 
N^{\textrm{R}} &=& - \frac{N F P_M}{\left(F^{-1}(\textrm{R}')^2-F 
    P_M^2\right)^{\frac12}}+ N^r R'\,, \label{new_mult_2_oo}\\ 
N^Q &=& \frac{2\pi NQR'F^{-1}\ln\left(\frac{R}{R_*}\right)} 
{\left(F^{-1}(\textrm{R}')^2-FP_M^2\right)^{\frac12}}+N^r P_Q - 
2\pi\tilde{\Phi}\,. 
\label{new_mult_3_oo} 
\end{eqnarray} 
Using the inverse transformations 
Eqs. (\ref{inversetrans_1_oo})-(\ref{inversetrans_2_oo}), and the identity 
$R=\textrm{R}$, with $P_\Gamma=2\pi Q$, 
we can write the new multipliers as functions of the old variables 
\begin{eqnarray} \label{mult_trans_1_oo} 
N^M &=& - NF^{-1}R'\Lambda^{-1}+\frac12 N^r F^{-1} \Lambda P_\Lambda\,, \\ 
N^{\textrm{R}} &=& - \frac12 N P_\Lambda + N^r 
R'\,,\label{mult_trans_2_oo}\\ 
N^Q &=& NP_\Gamma R'F^{-1}\Lambda^{-1}\ln\left(\frac{R}{R_*}\right) 
- 2\pi N^r\Gamma + \frac12 N^r P_\Gamma 
F^{-1}\ln\left(\frac{R}{R_*}\right) 
\Lambda P_\Lambda - 2\pi \tilde{\Phi}\,, 
\label{mult_trans_3_oo} 
\end{eqnarray} 
allowing us to determine its asymptotic conditions from the original 
conditions given above. These transformations are non-singular for $r>0$. 
As before, for $r\rightarrow 0$, 
\begin{eqnarray} \label{mult_newfalloff_0_i_oo} 
N^M(t,r) &=& -\frac12 N_1(t) \Lambda_0 R_2^{-1} + O(r^2)\,,\\ 
N^{\textrm{R}}(t,r) &=& O(r^4)\,,\\ 
N^Q(t,r) &=& \frac12 R_2^{-1}\ln\left(\frac{R_0}{R_*}\right) 
Q_0\Lambda_0N_1(t)-2\pi\tilde{\Phi}_0(t)+O(r^2)\,, 
\label{mult_newfalloff_0_f_oo} 
\end{eqnarray} 
and for $r\rightarrow\infty$ we have 
\begin{eqnarray} \label{mult_newfalloff_inf_i_oo} 
N^M(t,r) &=& -\tilde{N}_+(t) +  O^\infty(r^{-2})\,,\\ 
N^{\textrm{R}}(t,r) &=& O^\infty(r^{-1})\,,\\ 
N^Q(t,r)&=&-2\pi\tilde{\Phi}_+(t)+O^\infty(r^{-1})\,. 
\label{mult_newfalloff_inf_f_oo} 
\end{eqnarray} 
The conditions 
(\ref{mult_newfalloff_0_i_oo})-(\ref{mult_newfalloff_inf_f_oo}) 
show that the transformations in 
Eqs. (\ref{mult_trans_1_oo})-(\ref{mult_trans_2_oo}) are satisfactory in 
the case of $r\rightarrow\infty$, but not for $r\rightarrow 0$. This 
is due to fact that in order to fix the Lagrange multipliers for 
$r\rightarrow\infty$, as we are free to do, we fix $\tilde{N}_+(t)$, 
which we already do when adding the surface term 
\begin{equation} 
- \int\, dt \, \tilde{N}_+ M_+ 
\end{equation} 
to the action, in order to obtain the equations of motion in the bulk, 
without surface terms. 
The same is true for $\tilde{\Phi}_+$. 
However, at $r=0$, we see that fixing the multiplier $N^M$ to values 
independent of the canonical variables is not the same as fixing $N_1 
\Lambda_0^{-1}$ to values independent of the canonical variables. 
The same is true of the fixation of $N^Q$ with respect to 
$\tilde{\Phi}_0$. 
We need to rewrite the multipliers $N^M$ and $N^Q$ for the asymptotic 
regime 
$r\rightarrow 0$ without affecting their behavior for 
$r\rightarrow\infty$. 
In order to proceed we have to make one assumption, which is that the 
expression given in the asymptotic condition of $M(t,r)$, as $r\rightarrow 
0$, for the term of order zero, $M_0\equiv 
l^{-2}R_0^2-\pi Q_0^2 \ln \left(\frac{R_0}{R_*}\right)$, 
defines $R_0$ as a function of $M_0$ and $Q_0$, 
and $R_0$ is the horizon radius function, $R_0\equiv 
R_{\textrm{h}}(M_0,Q_0)$. Also, we assume that $M_0> 
M_{\textrm{\tiny{crit}}}(Q_0)$, 
where 
\begin{eqnarray} 
M_{\textrm{\tiny{crit}}}(Q_0) &=& \frac{\pi Q_0^2}{2} 
\left(1-\ln\left(\frac{\pi Q_0^2}{2}\right)\right)\,. 
\label{critical_m_oo} 
\end{eqnarray} 
With these assumptions, we are working in the domain of the classical 
solutions. 
We can immediately obtain that the variation of 
$R_0$ is given in relation to the variations of $M_0$ and $Q_0$ as 
\begin{equation} \label{var_r_m_oo} 
\delta R_0 = \left(2l^{-2}R_0-\pi Q_0^2 R_0^{-1}\right)^{-1} 
\left(\delta M_0 +2\pi Q_0 \ln\left(\frac{R_0}{R_*}\right)\delta 
Q_0\right)\,, 
\end{equation} 
This expression will be used when we derive the equations of 
motion from the new action. 
We now define the new multipliers $\tilde{N}^M$ and $\tilde{N}^Q$ as 
\begin{eqnarray} \label{new_n_m_oo} 
\tilde{N}^M &=& - N^M 
\left[(1-g)+2\,g\,\left(2l^{-2}R_0-\pi Q_0^2 R_0^{-1}\right)^{-1} 
\right]^{-1}\,,\\ 
\tilde{N}^Q &=& \tilde{N}^M 2\,g\,Q_0 \ln\left(\frac{R_0}{R_*}\right) 
\left(2l^{-2}R_0-\pi Q_0^2 R_0^{-1}\right)^{-1} - N^Q\,, 
\label{new_n_q_oo} 
\end{eqnarray} 
where $g(r)=1+O(r^2)$ for $r\rightarrow 0$ and $g(r)=O^\infty(r^{-5})$ 
for $r\rightarrow\infty$. The new multipliers, functions of the old 
multipliers $N^M$ and $N^Q$, have as their properties for 
$r\rightarrow0$ 
\begin{eqnarray} 
\tilde{N}^M(t,r) &=& \tilde{N}_0^M(t) + O(r^{2})\,,\\ 
\tilde{N}^Q(t,r) &=& 2\pi\tilde{\Phi}_0(t) + O(r^{2})\,, 
\end{eqnarray} 
and as their properties for 
$r\rightarrow \infty$ 
\begin{eqnarray} 
\tilde{N}^M(t,r) &=& \tilde{N}_+(t) + O^\infty(r^{-2})\,,\\ 
\tilde{N}^Q(t,r) &=& 2\pi\tilde{\Phi}_+(t) + O^\infty(r^{-1})\,. 
\end{eqnarray} 
When the constraints $M'=0=Q'$ hold, $\tilde{N}_0^M$ is given by 
\begin{equation} 
\tilde{N}_0^M =  N_1 \Lambda_0^{-1} \,. 
\end{equation} 
With this new constraint $\tilde{N}^M$, fixing $N_1 \Lambda_0^{-1}$ at 
$r=0$ or fixing $\tilde{N}_0^M$ is equivalent, there being no problems 
with $N^{\textrm{R}}$, which is left as determined in 
Eq. (\ref{new_mult_2_oo}). With respect to $\tilde{N}^Q$ the same 
happens, i. e., fixing the zero order term of the expansion of $N^Q$ 
for $r\to0$ is the same as fixing $\tilde{\Phi}_0$. At infinity there 
were no initial problems with the definitions of both $\tilde{N}^M$ 
and $\tilde{N}^Q$. 

The new action is now written as the sum of $S_\Sigma$, the bulk action, 
and 
$S_{\partial\Sigma}$, the surface action, 
\begin{eqnarray} 
S\left[M, \textrm{R}, Q, P_M, P_{\textrm{R}}, P_Q; \tilde{N}^M, 
  N^{\textrm{R}}, \tilde{N}^Q\right] &=& \int \,dt\, \int_0^\infty \, dr 
\, 
\left( P_M\dot{M} + P_\textrm{R} \dot{\textrm{R}} + P_Q \dot{Q} 
\right. \nonumber \\ 
&& \left. +\tilde{N}^QQ' - N^{\textrm{R}}P_{\textrm{R}} + \tilde{N}^M 
(1-g)\,M' \right. \nonumber \\ 
&& + \tilde{N}^M 2\,g\, \left(2l^{-2}R_0-\pi Q_0^2 R_0^{-1}\right)^{-1} 
\left(M'-Q'Q_0\ln\left(\frac{R_0}{R_*}\right)\right) 
\nonumber \\ 
&&  + \int \, dt \, \left\{\left(2 R_0 \tilde{N}_0^M - 
  \tilde{N}_+ M_+ \right) + 
 2\pi\left(\tilde{\Phi}_0Q_0-\tilde{\Phi}_+Q_+\right) 
  \right\}\,. \label{newaction_oo} 
\end{eqnarray} 
The new equations of motion are now 
\begin{eqnarray} \label{new_eom_1_oo} 
\dot{M} &=& 0\,, \\ 
\dot{\textrm{R}} &=& N^{\textrm{R}}\,, \\ 
\dot{Q} &=& 0\,,\\ 
\dot{P}_M &=& (N^M)'\,, \\ 
\dot{P}_{\textrm{R}} &=& 0\,, \\ 
\dot{P}_Q &=& (N^Q)'\,,\\ 
M' &=& 0\,, \\ 
P_{\textrm{R}} &=& 0\,, \\ 
Q'&=&0\,.\label{new_eom_9_oo} 
\end{eqnarray} 
where we understood $N^M$ to be a function of the new constraint, 
defined through Eq. (\ref{new_n_m_oo}) and $N^Q$ as the function of the 
new constraint defined through Eq. (\ref{new_n_q_oo}). 
The resulting boundary terms of 
the variation of this new action, Eq. (\ref{newaction_oo}), are, 
first, terms proportional to $\delta M$, $\delta \textrm{R}$, 
and $\delta Q$ on the initial and final hypersurfaces, 
and, second, 
\begin{eqnarray}\label{surface_varied_terms_final} 
          \int \, dt \, \left(2 R_0 \delta\tilde{N}_0^M - 
      M_+ \delta\tilde{N}_+ \right) + 
2\pi\left(Q_0\delta\tilde{\Phi}_0- 
Q_+\delta\tilde{\Phi}_+\right) \,. 
\end{eqnarray} 
To arrive at (\ref{surface_varied_terms_final}) we have used 
the expression in Eq. (\ref{var_r_m_oo}). 
The action in 
Eq. (\ref{newaction_oo}) yields the equations of motion, 
Eqs. (\ref{new_eom_1_oo})-(\ref{new_eom_9_oo}), provided that we fix the 
initial and final values of the new canonical variables and that we 
also fix the values of $\tilde{N}^M_0$ and of $\tilde{N}_+$, and of 
$\tilde{\Phi}_0$ and $\tilde{\Phi}_+$. 
Thanks to 
the redefinition of the Lagrange multiplier, from $N^M$ to 
$\tilde{N}^M$, the fixation of those quantities, $\tilde{N}^M_0$ and 
$\tilde{N}_+$, has the same meaning it had before the 
canonical transformations and the redefinition of $N^M$. 
The same happens with $\tilde{\Phi}_0$ and $\tilde{\Phi}_+$. 
This keeping of meaning is guaranteed through the use of our gauge freedom 
to 
choose the multipliers, and at the same time not fixing the boundary 
variations independently of the choice of Lagrange multipliers, which 
in turn allow us to have a well defined variational principle for the 
action. 

\subsection{Hamiltonian reduction} 
\label{h_red_oo} 

We now solve the constraints in order to reduce to the true dynamical 
degrees of freedom. The equations of motion 
(\ref{new_eom_1_oo})-(\ref{new_eom_9_oo}) allow us to write $M$ and $Q$ 
as independent functions of space, $r$, 
\begin{eqnarray} \label{m_t_oo} 
M(t,r) &=& \textbf{m}(t)\,,\\ 
Q(t,r) &=& \textbf{q}(t)\,. 
\label{q_t_oo} 
\end{eqnarray} 
The reduced action, with the constraints taken into account, is then 
\begin{equation} \label{red_action_oo} 
S 
\left[\textbf{m},\textbf{p}_{\textbf{m}},\textbf{q},\textbf{p}_{\textbf{q}} 
;\tilde{N}_0^M,\tilde{N}_+,\tilde{\Phi}_0,\tilde{\Phi}_+\right] 
= \int 
dt\,\,\textbf{p}_{\textbf{m}} \dot{\bf{m}}+\textbf{p}_{\textbf{q}} 
\dot{\bf{q}}-\textbf{h}\,, 
\end{equation} 
where 
\begin{eqnarray} \label{new_p_m_oo} 
\textbf{p}_{\textbf{m}} &=& \int_0^\infty dr\,P_M\,,\\ 
\textbf{p}_{\textbf{q}} &=& \int_0^\infty dr\,P_Q\,, 
\label{new_p_q_oo} 
\end{eqnarray} 
and the reduced Hamiltonian, $\textbf{h}$, is now written as 
\begin{equation} \label{red_Hamiltonian_oo} 
\textbf{h}(\textbf{m},\,\textbf{q};t)= 
-2 R_{\textrm{h}} \tilde{N}_0^M + 
\tilde{N}_+ \textbf{m}+2\pi\textbf{q} 
\left(\tilde{\Phi}_+-\tilde{\Phi}_0\right)\,, 
\end{equation} 
with $R_{\textrm{h}}$ being the horizon radius. 
We also have 
that $\textbf{m}>M_{\textrm{\tiny{crit}}}(\textbf{q})$, 
according to the assumptions made in the previous 
subsection. Thanks to the functions $\tilde{N}_0^M(t)$, 
$\tilde{N}_+(t)$,  $\tilde{\Phi}_0(t)$, and $\tilde{\Phi}_+(t)$ 
the Hamiltonian $\textbf{h}$ is an explicitly time 
dependent function. The variational principle associated with the 
reduced action, Eq. (\ref{red_action_oo}), will fix the values of 
$\textbf{m}$ and $\textbf{q}$ on the initial and final hypersurfaces, 
or in the spirit of the classical analytical mechanics, 
the Hamiltonian principle fixes 
the initial and final values of the canonical coordinates. 
The equations of motion are 
\begin{eqnarray} \label{red_eom_1_oo} 
\dot{\textbf{m}} &=& 0\,, \\ 
\dot{\textbf{q}} &=& 0\,, \label{red_eom_2_oo} \\ 
\dot{\textbf{p}}_{\textbf{m}} &=& 2 \tilde{N}_0^M 
(2l^{-2}R_{\textrm{h}}-\pi Q_0^2R_\textrm{h}^{-1})^{-1} - \tilde{N}_+\,, 
\label{red_eom_3_oo}\\ 
\dot{\textbf{p}}_{\textbf{q}} &=& 2\pi\left\{ 2\,\textbf{q} 
\ln\left(\frac{R_\textrm{h}}{R_*}\right) 
(2l^{-2}R_{\textrm{h}}-\pi Q_0^2R_\textrm{h}^{-1})^{-1}\tilde{N}_0^M 
 + \tilde{\Phi}_0-\tilde{\Phi}_+\right\}\,. 
\label{red_eom_4_oo} 
\end{eqnarray} 
The equation of motion for $\textbf{m}$, Eq. (\ref{red_eom_1_oo}), is 
understood as saying that $\textbf{m}$ is, on a classical solution, 
equal to the mass parameter $M$ of the solution, Eq. (\ref{solucao_mtz}). 
The same goes for the function $\textbf{q}$, where Eq. 
(\ref{red_eom_2_oo}) 
implies that $\textbf{q}$ is equal to the charge parameter $Q$ on a 
classical solution, Eq. (\ref{solucao_mtz}). 
In order to interpret the other equation of motion, 
Eq. (\ref{red_eom_3_oo}), we have to recall that from 
Eq. (\ref{p_m_oo}) one has $P_M=-T'$, 
where $T$ is the Killing time. This, together with 
the definition of $\textbf{p}_{\textbf{m}}$, given in 
Eq. (\ref{new_p_m_oo}), yields 
\begin{equation} 
\textbf{p}_{\textbf{m}} = T_0 - T_+\,, 
\end{equation} 
where $T_0$ is the value of the Killing time at the left end of the 
hypersurface of a certain $t$, and $T_+$ is the Killing time at 
spatial infinity, the right end of the same hypersurface of $t$. As 
the hypersurface evolves in the spacetime of the black hole solution, 
the right hand side of Eq. (\ref{red_eom_2_oo}) is equal 
to $\dot{T}_0-\dot{T}_+$. Finally, after the definition 
\begin{eqnarray} 
    \textbf{p}_{\textbf{q}} = \xi_0 - \xi_+\,, 
\end{eqnarray} 
obtained from Eqs. (\ref{vector_potential_plusgauge_tr_oo}), 
(\ref{p_q_oo}), and (\ref{new_p_q_oo}), Eq. (\ref{red_eom_4_oo}) 
gives in the right hand side $\dot{\xi}_0 - \dot{\xi}_+$, 
which is the diference of the time derivatives of the 
electromagnetic gauge $\xi(t,r)$ at $r=0$ and at infinity. 

\subsection{Quantum theory and partition function} 

The next step is to quantize the reduced Hamiltonian theory, by 
building the time evolution operator quantum mechanically and then 
obtaining a partition function through the analytic continuation of 
the same operator \cite{diaslemos}-\cite{louko5}. 
The variables $\textbf{m}$ and $\textbf{q}$ are regarded here 
as configuration variables. These variables satisfy the inequality 
$\textbf{m}>M_{\textrm{\tiny{crit}}}(\textbf{q})$. The 
wave functions will be of the form $\psi(\textbf{m},\textbf{q})$, 
with the inner product given by 
\begin{equation} 
\left(\psi,\chi \right) = \int_A \mu d\textbf{m}d\textbf{q} 
\, \bar{\psi}\chi\,, 
\end{equation} 
where $A$ is the domain of integration defined by 
$\textbf{m}>M_{\textrm{\tiny{crit}}}(\textbf{q})$ and 
$\mu(\textbf{m},\textbf{q})$ is a smooth and positive weight 
factor for the integration measure. It is assumed that $\mu$ 
is a slow varying function, otherwise arbitrary. We are thus 
working in the Hilbert space defined as $\mathscr{H}:= 
L^2(A;\mu d\textbf{m}d\textbf{q})$. 

The Hamiltonian operator, written as $\hat{\textbf{h}}(t)$, acts 
through pointwise multiplication by the function 
$\textbf{h}(\textbf{m},\textbf{q};t)$, 
which on a function of our working Hilbert space reads 
\begin{equation} 
\hat{\textbf{h}}(t)\psi(\textbf{m},\textbf{q})= 
\textbf{h}(\textbf{m},\textbf{q};t) 
\psi(\textbf{m},\textbf{q})\,. 
\end{equation} 
This Hamiltonian operator is an unbounded essentially self-adjoint 
operator. The corresponding time evolution operator in the same 
Hilbert space, which is unitary due to the fact that the Hamiltonian 
operator is self-adjoint, is 
\begin{equation} 
\hat{K}(t_2;t_1) = \exp 
\left[-i\int_{t_1}^{t_2}dt'\,\hat{\textbf{h}}(t') \right]\,. 
\label{K} 
\end{equation} 
This operator acts also by pointwise multiplication in the Hilbert 
space. 
We now define 
\begin{eqnarray} \label{t_oo} 
\mathcal{T} &:=& \int_{t_1}^{t_2}dt\,\tilde{N}_+(t)\,,\\ 
\Theta &:=& \int_{t_1}^{t_2}dt\,\tilde{N}^{M}_0 (t)\,, 
\label{theta_oo}\\ 
\Xi_0 &:=& \int_{t_1}^{t_2}dt\,\tilde{\Phi}_0(t)\,, 
\label{csi_0_oo}\\ 
\Xi_+ &:=& \int_{t_1}^{t_2}dt\,\tilde{\Phi}_+(t)\,. 
\label{csi_+_oo} 
\end{eqnarray} 
Using (\ref{red_Hamiltonian_oo}), (\ref{K})-(\ref{csi_+_oo}) 
we write the function $K$, which is in fact the 
action of the operator in the Hilbert space, as 
\begin{equation} 
K\left(\textbf{m};\mathcal{T},\Theta,\Xi_0,\Xi_+\right) = \exp 
\left[-i\textbf{m}\mathcal{T}+ 2\,i\,R_{\textrm{h}}\Theta- 
2\pi\,i\textbf{q}(\Xi_+-\Xi_0) 
\right]\,. 
\label{K2} 
\end{equation} 
This expression indicates that $\hat{K}(t_2;t_1)$ depends 
on $t_1$ and $t_2$ only through the functions $\mathcal{T}$, 
$\Theta$, $\Xi_0$, and $\Xi_+$. Thus, the 
operator corresponding to the function $K$ can now be written as 
$\hat{K}(\mathcal{T},\Theta,\Xi_0,\Xi_+)$. The composition law in time 
$\hat{K}(t_3;t_2)\hat{K}(t_2;t_1)=\hat{K}(t_3;t_1)$ can be regarded as 
a sum of the parameters $\mathcal T$, $\Theta$, $\Xi_0$, and $\Xi_+$ 
inside the operator $\hat{K}(\mathcal{T},\Theta,\Xi_0,\Xi_+)$. 
These parameters are evolutions parameters defined by the 
boundary conditions, i.e., $\mathcal{T}$ is the Killing time 
elapsed at right spatial infinity and $\Theta$ is the boost 
parameter elapsed at the bifurcation circle; $\Xi_0$ and $\Xi_+$ 
are line integrals along timelike curves of constant $r$, and 
constant angular variables, at $r=0$ and at infinity. 

\subsection{Thermodynamics} 

We can now build the partition function for this system. The path to 
follow is to continue the operator to imaginary time and take the 
trace over a complete orthogonal basis. 
Our classical thermodynamic situation consists of a three-dimensional 
spherically symmetric charged black hole, asymptotically anti-de Sitter, 
in 
thermal equilibrium with a bath of Hawking radiation. Ignoring 
back reaction from the radiation, the geometry is described by the 
solutions in Eqs. (\ref{solucao_mtz})-(\ref{vector_potential_oo}). 
Thus, we consider a thermodynamic ensemble in which the temperature, 
or more appropriately here, the inverse temperature 
$\beta$, and the electric potential $\bar\phi$ 
are fixed. This characterizes a grand canonical ensemble, 
and the partition function 
$\mathcal{Z}(\beta,\bar\phi)$ arises naturally in such an ensemble. 
To analytically continue the Lorentzian solution 
we put 
$\mathcal{T}=-i\beta$, and $\Theta-2\pi i$, this latter choice 
based on the regularity of the classical Euclidean solution. 
We also choose $\Xi_0=0$ and $\Xi_+=i\beta\bar\phi$. 

We arrive then at the following expression for the partition function 
\begin{equation} \label{partition_function_1_oo} 
\mathcal{Z}(\beta,\bar\phi) = \textrm{Tr} \left[\hat{K} 
(-i\beta,-2\pi i,0,i\beta\bar\phi)\right]\,. 
\end{equation} 
{}From Eq. (\ref{K2}) this is realized as 
\begin{equation} \label{partition_function_2_oo} 
\mathcal{Z}(\beta.\bar\phi) = \int_A \mu\,d\textbf{m}d\textbf{q}\, 
\exp\left[-\beta (\textbf{m}-2\pi\textbf{q}\bar\phi)+ 4\pi 
R_{\textrm{h}}\right]\left\langle 
\textbf{m}|\textbf{m}\right\rangle\,. 
\label{Z1} 
\end{equation} 
Since $\left\langle \textbf{m}|\textbf{m}\right\rangle$ is equal to 
$\delta(0)$, one has to regularize Eq. (\ref{Z1}).  Following the 
procedure developed in the Louko-Whiting approach 
\cite{louko1}, this means regularizing and normalizing 
the operator $\hat{K}$ beforehand.  This leads to 
\begin{equation} \label{partition_function_3_oo} 
\mathcal{Z}_{\textrm{ren}}(\beta,\bar\phi) = \mathcal{N} 
\int_{A} \mu\,d\textbf{m}d\textbf{q}\, 
\exp\left[-\beta (\textbf{m}-2\pi\textbf{q}\bar\phi) 
+4\pi R_{\textrm{h}}\right]\,, 
\end{equation} 
where $\mathcal{N}$ is a normalization factor 
and ${A}$ is the domain of integration. 
Provided the weight 
factor $\mu$ is slowly varying compared to the 
exponential in Eq. (\ref{partition_function_3_oo}), and using the 
fact that the horizon radius $R_{\textrm{h}}$ is a function of 
$\textbf{m}$ and $\textbf{q}$, the integral in 
Eq. (\ref{partition_function_3_oo}) is convergent. 
Changing integration variables, from $\textbf{m}$ to $R_{\textrm{h}}$, 
keeping $\textbf{q}$, where 
\begin{equation} \label{m_r_h_oo} 
\textbf{m} = l^{-2} R_{\textrm{h}}^2 - \pi \textbf{q}^2 
\ln\left(\frac{R_{\textrm{h}}}{R_*}\right) \,, 
\end{equation} 
the integral Eq. (\ref{partition_function_3_oo}) becomes 
\begin{equation} \label{partition_function_ren_oo} 
\mathcal{Z}_{\textrm{ren}}(\beta,\bar\phi) = \mathcal{N} \int_{A'} 
\widetilde{\mu}\,dR_{\textrm{h}}d\textbf{q}\,\exp(-I_*)\,, 
\end{equation} 
where ${A'}$ is new the domain of integration after changing 
variables, and 
the function $I_*(R_{\textrm{h}})$,  a kind of an effective 
action (see \cite{york1}), is written as 
\begin{equation} \label{eff_action_oo} 
I_*(R_{\textrm{h}}):= \beta\left(l^{-2} R_{\textrm{h}}^2- 
\pi \textbf{q}^2 \ln\left(\frac{R_{\textrm{h}}}{R_*}\right) 
-2\pi\textbf{q}\bar\phi\right)-4\pi R_{\textrm{h}}\,. 
\end{equation} 
The new domain of integration, $A'$, is defined by the inequalities 
$0 \leq R_{\textrm{h}}$ and 
$\textbf{q}^2 \leq 2R_{\textrm{h}}^2 \pi^{-1}l^{-2}$. 
The new weight factor $\widetilde{\mu}$ includes the Jacobian of the 
transformation. 
Since the weight factor $\widetilde{\mu}$ 
is slowly varying, we can estimate 
the integral of $\mathcal{Z}_{\textrm{ren}}(\beta,\bar\phi)$ by 
the saddle point approximation. 
For that we have to calculate the critical points. 
Firstly one finds the value of $\textbf{q}$ for which 
\begin{eqnarray} \label{extrema_q_oo} 
\frac{\partial I_*(R_{\textrm{h}},\textbf{q})}{\partial\textbf{q}}=0\,. 
\end{eqnarray} 
The value is 
\begin{eqnarray} \label{q_star} 
    \textbf{q}^* &=& -\bar\phi 
    \left[\ln\left(\frac{R_\textrm{h}}{R_*}\right)\right]^{-1}\,. 
\end{eqnarray} 
Replacing this value in (\ref{eff_action_oo}), we obtain 
\begin{eqnarray} \label{eff_action_at_q_star_oo} 
    I_*(R_{\textrm{h}},\textbf{q}^*(R_{\textrm{h}})) &=& 
    \beta\left(l^{-2} R_{\textrm{h}}^2+ 
\pi \bar\phi^2 \left[\ln\left(\frac{R_{\textrm{h}}}{R_*}\right) 
\right]^{-1}\right) 
-4\pi R_{\textrm{h}}\,. 
\end{eqnarray} 
Deriving expression (\ref{eff_action_at_q_star_oo}) with respect to 
$R_{\textrm{h}}$ and making it zero, we obtain 
\begin{eqnarray} \label{1st_der_r_eff_action_oo} 
\frac{\partial I_*(R_{\textrm{h}})}{\partial R_{\textrm{h}}}&=& 
2\beta l^{-2}R_{\textrm{h}} - 
\beta\bar\phi^2\left[\ln\left(\frac{R_{\textrm{h}}} 
{R_*}\right)\right]^{-2} 
R_{\textrm{h}}^{-1}-4\pi=0\,, 
\end{eqnarray} 
which implies 
\begin{eqnarray}\label{solution_poly_oo} 
    2\beta l^{-2}R_{\textrm{h}}^2 - 
        \beta\bar\phi^2\left[\ln\left(\frac{R_{\textrm{h}}} 
        {R_*}\right)\right]^{-2} 
    -4\pi R_{\textrm{h}}&=&0\,. 
\end{eqnarray} 
The solutions $R_{\textrm{h}}^*$ to the equation (\ref{solution_poly_oo}) 
gives us the pair $(R_{\textrm{h}}^*,\textbf{q}^*)$ of critical points, 
with $\textbf{q}^*$ given in Eq. (\ref{q_star}), where 
$R_{\textrm{h}}=R_{\textrm{h}}^*$. 
Of these, the classical solution comes from the one critical pair 
that is also the global minimum of the effective action $I_*$, 
denoted by $(R_{\textrm{h}}^+,\textbf{q}^+)$, depending on the 
interval of values to which $\bar\phi$ belongs. 
The renormalized partition function is then 
\begin{equation} 
\label{saddle_point_partition_function_P_oo} 
\mathcal{Z}_{\textrm{ren}}(\beta,\bar\phi) = \textrm{P} \exp 
\left[-\beta\left(l^{-2}(R_{\textrm{h}}^+)^2 - \pi(\textbf{q}^+)^2 
\ln\left(\frac{R_{\textrm{h}}^+}{R_*}\right)- 
2\pi\,\textbf{q}^+\bar\phi\right) 
+4\pi R_{\textrm{h}}^+\right]\,, 
\end{equation} 
where $\textrm{P}$ is a slowly varying prefactor and 
$(R_{\textrm{h}}^+,\textbf{q}^+)$ 
is the global minimum of the effective action 
(\ref{eff_action_oo}). 
In the domain of integration the 
dominating contribution comes from the vicinity of 
$R_{\textrm{h}}=R_{\textrm{h}}^+$. 
We now write the logarithm of 
$\mathcal{Z}_{\textrm{ren}}$ as 
\begin{equation} \label{log_partition_function_oo} 
\ln(\mathcal{Z}_{\textrm{ren}}) = \ln \textrm{P}- 
\beta\left(l^{-2}(R_{\textrm{h}}^+)^2 - \pi(\textbf{q}^+)^2 
\ln\left(\frac{R_{\textrm{h}}^+}{R_*}\right)- 
2\pi\,\textbf{q}^+\bar\phi\right) +4\pi R_{\textrm{h}}^+\,. 
\end{equation} 
By ignoring the prefactor's logarithm, which closer to 
$R_{\textrm{h}}^+$ is less relevant, we are able to determine the 
value of $\textbf{m}$ at the critical point, where we find that it 
corresponds to the value of the mass of the classical solutions of the 
black holes (see Eq. (\ref{solucao_mtz})). 
Thus, when the critical point dominates the partition function, we have 
that the mean energy $\left\langle E\right\rangle$ is given by 
\begin{equation} 
\left\langle E\right\rangle = -\frac{\partial}{\partial\beta}\ln 
\mathcal{Z}_{\textrm{ren}} \approx 2\lambda^{2} 
(R_{\textrm{h}}^+)^2 - \pi (\textbf{q}^+)^2 
\ln\left(\frac{R_{\textrm{h}}^+}{R_*}\right) = \textbf{m}^+\,, 
\end{equation} 
where $\textbf{m}^+$ is obtained from Eq. (\ref{m_r_h_oo}) evaluated at 
$R_{\textrm{h}}^+$. The thermal expectation value of the charge is 
\begin{equation} 
\left\langle Q\right\rangle = \beta^{-1}\frac{\partial}{\partial\bar\phi} 
\ln \mathcal{Z}_{\textrm{ren}}\approx 2\pi\,\textbf{q}^+\,. 
\end{equation} 
The temperature of the black hole, $\textbf{T}\equiv 
\beta^{-1}$, is 
\begin{equation} 
\textbf{T}=\frac{1}{4\pi}\left(4\lambda^{2} 
(R_{\textrm{h}}^+)-\pi(\textbf{q}^+)^2 
(R_{\textrm{h}}^+)^{-1}\right)\,, 
\label{temperature_oo} 
\end{equation} 
where $\textbf{q}^+$ is the function in (\ref{q_star}) evaluated at 
$R_{\textrm{h}}^+$. If the maximum value of the charge $\textbf{q}^+$ 
is chosen, i.e., $\textbf{q}^2=2R_{\textrm{h}}^2 \pi^{-1}l^{-2}$, 
which is the value of the charge of the  extreme solution of the 
black hole, and is replaced into (\ref{temperature_oo}), then the 
temperature is null, as expected from an extreme solution. 
It can be shown that 
$\partial\textbf{m}^+/\partial\beta<0$, which through 
the constant $\bar\phi$ heat capacity 
$C_{\bar\phi}=-\beta^2(\partial \left\langle 
E\right\rangle/\partial\beta)$ 
tells us that the system is thermodynamically stable. 
The entropy is given by 
\begin{equation} 
S = 
\left(1-\beta\frac{\partial}{\partial\beta}\right) 
(\ln\mathcal{Z}_{\textrm{ren}}) 
\approx 4 \pi R_{\textrm{h}}^+\,. 
\end{equation} 
This is the entropy of the charged BTZ black hole \cite{btz,mtz} (see 
also \cite{ads3_bh}). This entropy includes the extreme solution, as 
the specific value of the charge is irrelevant for the determination 
of the entropy, provided we are in the domain of validity of the 
approximation and $\textbf{q}^2\leq2R_{\textrm{h}}^2 \pi^{-1}l^{-2}$. 

\section{Hamiltonian Thermodynamics of the general 
relativistic cylindrical dimensionally reduced 
charged black hole ({\large $\omega=0$})} 
\label{zero} 

\subsection{The metric, the scalar $\phi$, and the vector potential 
one-form} 

\begin{figure} 
[htmb] 
\centerline{\includegraphics 
%[width=5.7cm,height=15cm] 
{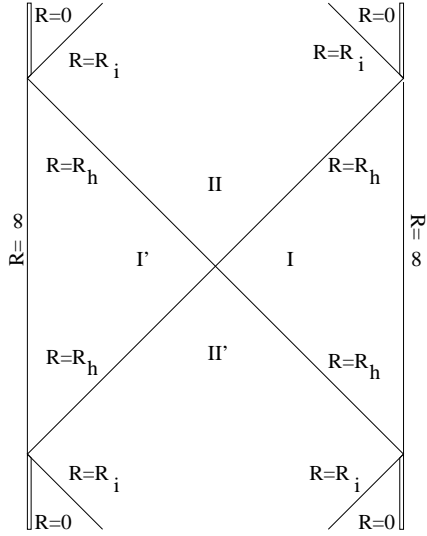}} 
\caption {{\small The Carter-Penrose diagram for 
the charged $\omega=0$ case, 
where $R_{\textrm{h}}$ is the outer horizon radius, 
$R_{\rm i}$ is the inner horizon radius, and the double 
line is the timelike singularity. 
}} 
\label{omega_0} 
\end{figure} 

For $\omega=0$, the corresponding three-dimensional 
charged Brans-Dicke theory is obtained from the 
cylindrical dimensionally reduced black hole of four-dimensional 
general relativity, with a Maxwell term 
\cite{lemos1,sakleberlemos,oscarlemos}. 
Then general metric in Eq. (\ref{solutions}), the $\phi$ field in 
Eq. (\ref{phisolutions_1}), and the vector potential one-form 
(\ref{vector_potential}) reduce to the following 
\begin{eqnarray} 
ds^2 &=& - \left[ (a\,R)^2 - \frac{M}{2(a\,R)}+ 
\frac{Q^2}{16(a\,R)^2}\right]\,dT^2 + 
\frac{dR^2}{(a\,R)^2 - \frac{M}{2(a\,R)}+ 
\frac{Q^2}{16(a\,R)^2}}+R^2\,d\varphi^2\,, 
\label{metric_zero}\\ 
e^{-2\phi} &=& a\,R\,,\label{phi_zero}\\ 
A &=& \alpha^{-1}\frac{Q}{R}\,dT \,, 
\label{vector_potential_zero} 
\end{eqnarray} 
with $M=2b$, $a=\sqrt{\frac23}|\lambda|=l^{-1}$, and 
$\alpha=4a$, where the $l$ is the AdS length. 
The Schwarzschild coordinates 
are again $\{T,R,\varphi\}$. 
Unlike the charged BTZ solution, 
this solution (\ref{metric_zero})-(\ref{vector_potential_zero}) 
has a metric function whose mass term 
depends on $R$. This behavior is 
similar to the Reissner-N\"{o}rdstrom black hole metric function 
\cite{lemos1}. 
In Fig. \ref{omega_0} we show the Carter-Penrose diagram of the black 
hole solution for $\omega=0$, where again $R_{\textrm{h}}$ 
is the horizon radius, given by the larger positive real root of 
\begin{equation} 
(a\,R_{\textrm{h}})^4-\frac{1}{2}M(a\,R_{\textrm{h}}) 
+\left(\frac{Q}{4}\right)^2=0\,, 
\label{radius_0} 
\end{equation} 
$R=0$ is the radius of the timelike curvature 
singularity, and $R=\infty$ is the spatial infinity. 

\subsection{Canonical formalism} 
The action with $\omega=0$ becomes, following 
Eqs. (\ref{ADM_ansatz})-(\ref{a_ansatz}), 
and excluding surface terms, 
\begin{eqnarray} 
S[\Lambda,\,R,\,\Gamma,\,\dot{\Lambda},\,\dot{R},\,\dot{\Gamma}; 
\,N,\,N^r,\,\Phi] &=& \int dt 
\int_0^{\infty} 
dr \,\alpha \left\{\lambda^2\Lambda N R^2 
-N^{-1}R\dot{R}\dot{\Lambda}-\frac12 N^{-1}\Lambda\dot{R}^2+N^{-1} 
R\dot{R} 
(\Lambda N^r)'\right. \nonumber\\ && 
+N^{-1}N^r R R'\dot{\Lambda} - \left. N^{-1}N^r\Lambda R R'(N^r)' 
-N^{-1}(N^r)^2RR'\Lambda'+N^{-1}N^r\Lambda\dot{R}R' \right. 
\nonumber\\ && \left. -\frac12 N^{-1}(N^r)^2\Lambda(R')^2 
-(\Lambda^{-1})'RR'N 
-\frac12 \Lambda^{-1}(R')^2N-\Lambda^{-1}RR''N\right. \nonumber \\ 
&& \left. + \frac12 \alpha N^{-1}\Lambda^{-1}R^2\left(\dot{\Gamma}- 
\Phi'\right)^2\right\}\,, 
\end{eqnarray} 
where $\dot{}$ means 
derivative with respect to time $t$ and $'$ is the derivative with 
respect to $r$, and where all the explicit functional dependences are 
omitted. Depending on the situation we use four different 
letters containing the same information but with slightly different 
numerical values. Thus, $a$, $\lambda$, $l$ and $\alpha$ are 
related by $a^2=\frac23\lambda^2 = l^{-2}$, and $\alpha=4a$, 
where $l$ is the AdS length. 
{}From this action one determines the canonical momenta, 
conjugate to $\Lambda$, $R$, and $\Gamma$ respectively 
\begin{eqnarray} \label{p_lambda} 
P_{\Lambda} &=& -\alpha N^{-1} R \left\{\dot{R}-R'N^r\right\}\,, \\ 
P_R &=& -\alpha N^{-1}\left\{ R[\dot{\Lambda}-(\Lambda 
  N^r)']+\Lambda[\dot{R}-N^rR'] \right\}\,, \label{p_r}\\ 
P_\Gamma &=& \alpha N^{-1}\Lambda^{-1}R^2\left(\dot{\Gamma}- 
\Phi'\right)\,. \label{p_gamma} 
\end{eqnarray} 
By performing a Legendre transformation, we obtain 
\begin{eqnarray} 
\mathcal{H} &=& N\left\{-\alpha^{-1}R^{-1}P_{\Lambda}P_{R}+\frac12 
  \alpha^{-1} \Lambda R^{-2}(P_{\Lambda}^2+P_\Gamma^2)+ 
\alpha \Lambda^{-1}RR''-\alpha\Lambda^{-2}RR'\Lambda'+\frac12 
\Lambda^{-1} (R')^2- 
\lambda^2\Lambda R^2\right\} \nonumber \\ 
&& + N^r\left\{P_R R'-P_{\Lambda}'\Lambda-P_\Gamma'\Gamma\right\} 
 + \tilde{\Phi}\left\{-P_\Gamma'\right\}\equiv NH+N^rH_r+\tilde{\Phi}G\,, 
\end{eqnarray} 
with the new Lagrange multiplier being $\tilde{\Phi}:=\Phi-N^r\Gamma$. 
The action in Hamiltonian form is then 
\begin{equation} \label{haction} 
S[\Lambda,\,R,\,\Gamma,\,P_{\Lambda},\,P_{R},\,P_\Gamma;\,N,\,N^r,\, 
\tilde{\Phi}] 
 = \int dt \int_0^{\infty} dr 
 \left\{ P_{\Lambda}\dot{\Lambda}+P_R\dot{R}+P_\Gamma\dot{\Gamma} 
 -NH-N^rH_r-\tilde{\Phi}G \right\}\,. 
\end{equation} 
The equations of motion are 
\begin{eqnarray} 
\dot{\Lambda} &=& -N\alpha^{-1}R^{-1}P_R+N\alpha^{-1}\Lambda 
R^{-2}P_{\Lambda}+N^r\Lambda\,, \\ 
\dot{R} &=& -N\alpha^{-1}P_{\Lambda}R^{-1}+N^rR'\,, \\ 
\dot{\Gamma}&=& \alpha^{-1} N \Lambda R^{-2} P_{\Gamma} + 
(N^r\Gamma)'\tilde{\Phi}'\,,\\ 
\dot{P_R} &=& -N\alpha^{-1}P_{\Lambda}P_RR^{-2}+N\alpha^{-1}\Lambda 
(P_{\Lambda}^2+P_\Gamma^2)R^{-3}-\left((N\alpha)'\Lambda^{-1}R\right)' 
-(N\alpha)(\Lambda^{-1}R')'\nonumber \\ 
&& + 2N\alpha \lambda^2\Lambda R+(N^rP_R)'\,,\\ 
\dot{P_{\Lambda}}&=& 
-\frac12N\alpha^{-1}R^{-2}(P_{\Lambda}^2+P_\Gamma^2)-(N\alpha)' 
RR'\Lambda^{-2}-\frac12 N\alpha 
(R')^2\Lambda^{-2}+N\alpha\lambda^2R^2+N^rP_{\Lambda}'\,,\\ 
\dot{P}_\Gamma &=& N^r P_\Gamma'\,. 
\end{eqnarray} 
In order to have a well defined variational principle, we need to 
eliminate the surface terms of the original bulk action, which render 
the original action itself ill defined for a correct determination of the 
equations of motion through variational methods. 
Through the choice of added surface terms one can achieve this 
elimination. 
The action 
(\ref{haction}) 
has the following extra surface terms, after variation 
\begin{eqnarray} 
\textrm{Surface terms}&=& \left. 
\left\{\alpha\left(-N\Lambda^{-1}R\delta 
R'+N'\Lambda^{-1}R\delta R - 
N^rP_R\delta R+N^r\Lambda \delta 
P_\Lambda+NRR'\Lambda^{-2}\delta\Lambda\right)\right.\right.\nonumber\\ 
&& \left.\left. +N^r\Gamma\delta P_\Gamma + \tilde{\Phi}\delta P_\Gamma 
\right\}\right|_0^\infty\,. 
\end{eqnarray} 
In order to evaluate this expression, we need to know the asymptotic 
conditions of each of the functions of $(t,r)$. 

Starting with the limit $r\rightarrow 0$ we assume 
\begin{eqnarray} \label{falloff_0_i} 
\Lambda(t,r) &=& \Lambda_0(t)+O(r^2)\,,\\ 
R(t,r)&=& R_0(t)+R_2(t)r^2+O(r^4)\,,\\ 
\Gamma(t,r) &=& O(r)\,,\\ 
P_\Lambda(t,r) &=& O(r^3)\,,\\ 
P_R(t,r) &=& O(r)\,,\\ 
P_\Gamma(t,r) &=& Q_0(t)+Q_2(t)r^2+O(r^4)\,,\\ 
N(t,r) &=& N_1(t)r+O(r^3)\,,\\ 
N^r(t,r) &=& N^r_1(t)r+O(r^3)\,,\\ 
\tilde{\Phi}(t,r) &=& \tilde{\Phi}_0(t)+O(r^2)\,. 
 \label{falloff_0_f} 
\end{eqnarray} 
With these conditions, we have for the surface terms at $r=0$, 
\begin{equation}\label{surface_0} 
\left.\textrm{Surface terms}\right|_{r=0} = -\alpha 
N_1R_0\Lambda_0^{-1}\delta R_0 - \tilde{\Phi}_0\delta Q_0\,. 
\end{equation} 
In the same way, for $r\rightarrow\infty$, 
\begin{eqnarray} \label{falloff_inf_i} 
\Lambda(t,r) &=& lr^{-1}+l^3\eta(t)r^{-4} 
+ O^\infty(r^{-5})\,,\\ 
R(t,r) &=& r + l^2\rho(t)r^{-2}+O^\infty(r^{-3})\,,\\ 
\Gamma(t,r) &=& O^\infty(r^{-2})\,,\\ 
P_\Lambda(t,r) &=& O^\infty(r^{-2})\,,\\ 
P_R(t,r) &=& O^\infty(r^{-4})\,,\\ 
P_\Gamma(t,r) &=& Q_+(t)+O^\infty(r^{-1})\,,\\ 
N(t,r) &=& R(t,r)'\Lambda(t,r)^{-1}(\tilde{N}_+(t) 
+O^\infty(r^{-5}))\,,\\ 
N^r(t,r) &=& O^\infty(r^{-2})\,,\\ 
\tilde{\Phi}(t,r) &=& \tilde{\Phi}_+(t) + O^\infty(r^{-1})\,. 
\label{falloff_inf_f} 
\end{eqnarray} 
These conditions imply for the surface terms in the limit 
$r\rightarrow\infty$, 
\begin{equation} 
\left.\textrm{Surface terms}\right|_{r\rightarrow\infty} = 
\alpha\delta(M_+)\tilde{N}_+ + \tilde{\Phi}_+\delta Q_+\,, 
\end{equation} 
where $M_+(t)=\alpha(\eta(t)+3\rho(t))$. 
So, the surface term to be added to (\ref{haction}) is 
\begin{equation} \label{surface_inf} 
S_{\partial\Sigma}\left[\Lambda,R,Q_0,Q_+; 
N,\tilde{\Phi}_0,\tilde{\Phi}_+\right]= 
\int dt \left(\frac12 
\alpha R_0^2 N_1 \Lambda_0^{-1} - \tilde{N}_+ M_+ 
+ \tilde{\Phi}_0 Q_0 - \tilde{\Phi}_+ Q_+ \right)\,. 
\end{equation} 
What is left after varying 
this last surface term and adding it to the varied initial action 
(see Eq. (\ref{haction})) is 
\begin{equation}\label{variationofsurfaceterm} 
\int dt \left(\frac12 \alpha R_0^2 \delta(N_1 \Lambda_0^{-1}) 
 - M_+ \delta\tilde{N}_+ + Q_0\delta\tilde{\Phi}_0 
 - Q_+\delta\tilde{\Phi}_+ \right)\,. 
\end{equation} 
We choose to fix $N_1 \Lambda_0^{-1}$ and $\tilde{\Phi}_0$ on 
the horizon, and $\tilde{N}_+$ and $\tilde{\Phi}_+$ 
at infinity, which makes the surface variation 
(\ref{variationofsurfaceterm}) disappear. 

\subsection{Reconstruction, canonical transformation, and action} 

In order to reconstruct the mass and the time from the canonical data, 
which amounts to making a canonical transformation, we have to rewrite 
the general form of the solutions in Eqs. 
(\ref{metric_zero})-(\ref{vector_potential_zero}). 
However, in 
addition to reconstructing the mass as done in \cite{diaslemos}, we 
have to consider how to reconstruct the charge from the canonical 
data.  We follow Kucha\v{r} \cite{kuchar} for this reconstruction.  We 
concentrate our analysis on the right static region of the 
Carter-Penrose diagram. 

Developing the Killing time $T$ as function of $(t,r)$ in the expression 
for the vector potential (\ref{vector_potential_zero}), and making use of 
the gauge freedom that allows us to write 
\begin{eqnarray} 
    A &=& \alpha^{-1}\frac{Q}{R}dT+d\xi\,, 
    \label{vector_potential_plusgauge_zero} 
\end{eqnarray} 
 where $\xi(t,r)$ is an arbitrary continuous function of $t$ and $r$, 
 we can write the one-form potential as 
\begin{eqnarray} 
    A &=& \left(\alpha^{-1}\frac{Q}{R}T'+\xi'\right)dr + 
\left(\alpha^{-1}\frac{Q}{R}\dot{T}+\dot{\xi}\right)dt\,. 
    \label{vector_potential_plusgauge_tr_zero} 
\end{eqnarray} 
Equating the expression (\ref{vector_potential_plusgauge_tr_zero}) with 
Eq. (\ref{a_ansatz}), and making use of the definition of $P_{\Gamma}$ in 
Eq. (\ref{p_gamma}), we arrive at 
\begin{eqnarray} 
    P_{\Gamma} &=& Q\,. 
    \label{p_q_zero} 
\end{eqnarray} 
In the right static region we define $F$ as 
\begin{equation}\label{f_1} 
F(R(t,r))=(aR(t,r))^2 - \frac{M}{2aR(t,r)} + 
\frac{P_\Gamma^2}{16(aR(t,r))^2}\,. 
\end{equation} 
Note that here, $R_*$, 
the large radius of a finite frontier, which can be used to 
renormalize the asymptotic properties of the 
future canonical coordinate $M(t,r)$ (see subsections \ref{bhsolutions} 
and \ref{reconstruction}), is not needed; in fact bringing to the 
$\omega=0$ solution introduces complications at the $r=0$ frontier. 
Thus, in this section we do not mention $R_*$. 
Then, making the following substitutions 
\begin{equation} 
T=T(t,r)\,, \qquad \qquad R=R(t,r)\,, 
\end{equation} 
in the solutions above, Eqs. (\ref{metric_zero})-(\ref{phi_zero}), 
one has 
\begin{eqnarray} 
ds^2 &=& -(F\dot{T}^2-F^{-1}\dot{R}^2)\,dt^2\,+ 
\,2(-FT'\dot{T}+F^{-1}R'\dot{R})\,dtdr\,+\,(-F(T')^2+F^{-1}\dot{R}^2)\,dr^2\,+ 
\,R^2d\varphi^2\,. 
\end{eqnarray} 
This introduces the ADM foliation directly into the solutions. 
Comparing it with the ADM metric (\ref{ADM_ansatz}), written in 
another form as 
\begin{equation} 
ds^2 = 
-(N^2-\Lambda^2(N^r)^2)\,dt^2\,+\,2\Lambda^2N^r\,dtdr\,+ 
\,\Lambda^2dr^2\,+\,R^2\,d\varphi^2\,, 
\end{equation} 
we can write a set of three equations 
\begin{eqnarray} 
\Lambda^2 &=& -F(T')^2+F^{-1}(R')^2\,,\label{adm_1}\\ 
\Lambda^2N^r &=& -FT'\dot{T}+F^{-1}R'\dot{R}\,, \label{adm_2}\\ 
N^2-\Lambda^2(N^r)^2 &=& F\dot{T}^2-F^{-1}\dot{R}^2\,. \label{adm_3} 
\end{eqnarray} 
The first two equations, Eq. (\ref{adm_1}) and Eq. (\ref{adm_2}), give 
\begin{equation} \label{shift_def} 
N^r = \frac{-FT'\dot{T}+F^{-1}R'\dot{R}}{-F(T')^2+F^{-1}(R')^2}\,. 
\end{equation} 
This one solution, together  with Eq. (\ref{adm_1}), give 
\begin{equation} \label{lapse_def} 
N = \frac{R'\dot{T}-T'\dot{R}}{\sqrt{-F(T')^2+F^{-1}(R')^2}}\,. 
\end{equation} 
One can show that $N(t,r)$ is positive. 
Next, putting Eq. (\ref{shift_def}) and 
Eq. (\ref{lapse_def}) into the definition of the conjugate momentum of 
the canonical coordinate $\Lambda$, given in 
Eq. (\ref{p_lambda}), one finds the spatial derivative of $T(t,r)$ as a 
function of the canonical coordinates, i.e., 
\begin{equation}\label{t_linha} 
-T' = \alpha^{-1}R^{-1}F^{-1}\Lambda P_\Lambda\,. 
\end{equation} 
Later we will see that $-T'=P_M$, as it will be conjugate to a new 
canonical coordinate $M$. 
Following this procedure to the end, we may then find the form of the 
new coordinate $M(t,r)$, also as a function of $t$ and $r$. First, we 
need to know the form of $F$ as a function of the canonical pair 
$\Lambda\,,\,R$. For that, we replace back into Eq. (\ref{adm_1}) the 
definition of $T'$, giving 
\begin{equation}\label{f_2} 
F = \left(\frac{R'}{\Lambda}\right)^2-\left(\frac{P_\Lambda}{\alpha 
R}\right)^2\,. 
\end{equation} 
Equating this form of $F$ with Eq. (\ref{f_1}), we obtain 
\begin{equation} \label{m} 
M = \frac12 \alpha R \left(\frac{\alpha^2}{16}R^2+ 
\frac{\alpha^2\,R^2}{16}-F\right)\,, 
\end{equation} 
where $F$ is given in Eq. (\ref{f_2}). 
We thus have found the form of the new canonical coordinate, $M$. It 
is now a straightforward calculation to determine the Poisson bracket 
of this variable with $P_M=-T'$ and see that they are conjugate, thus 
making Eq. (\ref{t_linha}) the conjugate momentum of $M$, i.e., 
\begin{equation}\label{p_m} 
P_M = \alpha^{-1}R^{-1}F^{-1}\Lambda P_\Lambda\,. 
\end{equation} 

The other natural transformation has been given, namely the one that 
relates $P_{\Gamma}$ with $Q$ through $P_{\Gamma}=Q$. As $M$, $Q$ is also 
a natural choice for canonical coordinate, being another physical 
parameter of the solution (\ref{solucao_mtz}). The one coordinate which 
remains to be found is $P_Q$, the conjugate momentum to the charge. 
It is also necessary to find out the other new canonical variable which 
commutes with $M$, $P_M$, and $Q$, and which guarantees, with its 
conjugate 
momentum, that the transformation from $\Lambda,\,R,\,\Gamma$, to $M$, 
$Q$, and the 
new variable is canonical. 
Immediately is it seen that $R$ commutes 
with $M$, $P_M$, and $Q$. It is then a candidate. It remains to be seen 
whether $P_R$ also commutes with $M$, $P_M$, and $Q$.  As with $R$, it is 
straightforward to see that $P_R$ does not commute with $M$ and $P_M$, 
as these contain powers of $R$ in their definitions, and 
$\left\{R(t,r),\,P_R(t,r^*) \right\}=\delta(r-r^*)$.  So rename the 
canonical variable $R$ as $R=\textrm{R}$.  We have then to find a new 
conjugate momentum to $\textrm{R}$ which also commutes with $M$, $P_M$, 
and $Q$, making the transformation from 
$\left\{\Lambda,\,R,\,\Gamma;\,P_\Lambda,\,P_R,\,P_{\Gamma}\,\right\} 
\rightarrow\, 
\left\{M,\,\textrm{R},\,Q;\,P_M,\,P_{\textrm{R}},\,P_Q\,\right\}$ a 
canonical 
one. 
The way to proceed is to look at the constraint $H_r$, which is called 
in this formalism the super-momentum. This is the constraint which 
generates spatial diffeomorphisms in all variables. Its form, in the 
initial canonical coordinates, is 
$H_r=-\Lambda\,P_\Lambda'+P_R\,R'-\Gamma P_{\Gamma}'$. 
In this formulation, $\Lambda$ is a spatial density, $R$ is a 
spatial scalar, and $\Gamma$ is also a spatial density. 
As the new variables, $M$, $\textrm{R}$, and $Q$, are 
spatial scalars, the generator of spatial diffeomorphisms is written 
as $H_r=P_M M'+P_{\textrm{R}} \textrm{R}'+P_Q Q'$, regardless of the 
particular form of the canonical coordinate transformation. It is thus 
equating these two expressions of the super-momentum $H_r$, with $M$, 
$P_M$, and $Q$  written as functions of $\Lambda,\,R,\,\Gamma$ 
and their respective 
momenta, that gives us one equation for the new $P_{\textrm{R}}$ and 
$P_Q$. This means that we have two unknowns, sc. $P_{\textrm{R}}$ and 
$P_Q$, for one equation only. This sugests that we should make 
the coefficients of $R'$ and $P_{\Gamma}'$ equal to zero independently. 
This results in 
\begin{eqnarray} 
P_{\textrm{R}} &=& P_R - \frac{3\alpha^2}{32}F^{-1}\Lambda P_\Lambda R 
- \frac12 R^{-1} \Lambda P_\Lambda + F^{-1} P_\Lambda R''\Lambda^{-1} 
- F^{-1} \Lambda^{-2} P_\Lambda \Lambda' R'+ (R')^2 F^{-1} 
\Lambda^{-1} P_\Lambda R^{-1} \nonumber \\ 
&& - F^{-1}\Lambda^{-1} P_\Lambda' R'+\frac12\alpha^{-2}R^{-3} 
P_\Gamma^2F^{-1}\Lambda P_\Lambda \,,\label{p_rnew} \\ 
P_Q &=& -\Gamma-\alpha^{-2}R^{-2}F^{-1}\Lambda P_\Lambda P_\Gamma\,. 
\label{p_q} 
\end{eqnarray} 
We have now all the canonical variables of the new set determined. For 
completeness and future use, we write the inverse transformation for 
$\Lambda$ and $P_\Lambda$, 
\begin{eqnarray} \label{inversetrans_1} 
\Lambda &=& \left((\textrm{R}')^2F^{-1}-P_M^2F\right)^{\frac12}\,, 
\\ 
P_\Lambda &=& \alpha \textrm{R} F P_M 
\left((\textrm{R}')^2F^{-1}-P_M^2F\right)^{-\frac12}\,. 
\label{inversetrans_2} 
\end{eqnarray} 
In summary, the canonical transformations are 
\begin{eqnarray} \label{setcanonicaltrans} 
R &=& \textrm{R}\,,\nonumber \\ 
M &=& \frac12 \alpha R \left(\frac{\alpha^2}{16}R^2-F\right)\,, 
\nonumber \\ 
Q &=& P_\Gamma\,, \nonumber \\ 
P_{\textrm{R}} &=& P_R - \frac{3\alpha^2}{32}F^{-1}\Lambda P_\Lambda R 
- \frac12 R^{-1} \Lambda P_\Lambda + F^{-1} P_\Lambda R''\Lambda^{-1} 
- F^{-1} \Lambda^{-2} P_\Lambda \Lambda' R'+ (R')^2 F^{-1} 
\Lambda^{-1} P_\Lambda R^{-1} \nonumber \\ 
&& - F^{-1}\Lambda^{-1} P_\Lambda' R'+\frac12\alpha^{-2}R^{-3} 
P_\Gamma^2F^{-1}\Lambda P_\Lambda\,, \nonumber \\ 
P_M &=& \alpha^{-1}R^{-1}F^{-1}\Lambda P_\Lambda\,,\nonumber\\ 
P_Q &=& -\Gamma-\alpha^{-2}R^{-2}F^{-1}\Lambda P_\Lambda P_\Gamma\,. 
\end{eqnarray} 

In order to prove that the set of equalities in expression 
(\ref{setcanonicaltrans}) is canonical we start with the equality 
\begin{eqnarray} \label{oddidentity} 
P_\Lambda \delta\Lambda + P_R \delta R + P_\Gamma\delta\Gamma 
- P_M \delta M - P_{\textrm{R}}\delta\textrm{R} - P_Q \delta Q 
&=& 
\left( \frac12 \alpha R \delta R \ln 
\left|\frac{\alpha R R'+\Lambda P_\Lambda}{\alpha R R'- \Lambda 
P_\Lambda}\right|\right)'+ \nonumber \\ 
&& +\, \delta\left(\Gamma P_\Gamma + \Lambda P_\Lambda 
+ \frac12 \alpha R R' \ln \left|\frac{\alpha R R' 
-\Lambda P_\Lambda}{\alpha R R'+ \Lambda P_\Lambda} 
\right|\right)\,. 
\end{eqnarray} 
We now integrate expression (\ref{oddidentity}) in $r$, in 
the interval from $r=0$ to $r=\infty$. The first term on the right 
hand side of Eq. (\ref{oddidentity}) vanishes due to the falloff 
conditions (see Eqs. (\ref{falloff_0_i})-(\ref{falloff_0_f}) and 
Eqs. (\ref{falloff_inf_i})-(\ref{falloff_inf_f})). 
We then obtain the following expression 
\begin{eqnarray} \label{int_oddidentity} 
\int_0^\infty\,dr\,\left(P_\Lambda \delta\Lambda + P_R \delta 
  R + P_\Gamma \delta\Gamma \right) 
  -\int_0^\infty\,dr\,\left(P_M \delta M + 
P_{\textrm{R}}\delta\textrm{R} + P_Q \delta Q \right) 
&=& \delta\omega\,\left[\Lambda,\,R,\,\Gamma,\,P_\Lambda, 
\,P_\Gamma\right]\,, 
\end{eqnarray} 
where $\delta\omega\,\left[\Lambda,\,R,\,\Gamma,\,P_\Lambda, 
\,P_\Gamma\right]$ is a well 
defined functional, which is also an exact form. This equality shows 
that the difference between the Liouville form of 
$\left\{R,\,\Lambda,\,\Gamma;\,P_R,\,P_\Lambda,\,P_\Gamma\right\}$ 
and the Liouville form 
of $\left\{\textrm{R},\,M,\,Q;\,P_{\textrm{R}},\,P_M,\,P_Q\right\}$ 
is an exact form, which implies that the transformation of 
variables given by the set of equations (\ref{setcanonicaltrans}) 
is canonical. 

Armed with the certainty of the canonicity of the new variables, we 
can write the asymptotic form of the canonical variables and of the 
metric function $F(t,r)$. These are, for $r\rightarrow 0$ 
\begin{eqnarray} \label{newfalloff_0_i} 
F(t,r) &=& 4 R_2(t) \Lambda_0(t)^{-2} r^2 + O(r^4)\,, \\ 
\textrm{R}(t,r) &=& R_0(t)+R_2(t)\,r^2+O(r^4)\,, \\ 
M(t,r) &=& M_0(t) + M_2(t)\,r^2 + O(r^4)\,, \\ 
Q(t,r) &=& Q_0(t)+Q_2(t)\,r^2+O(r^{4})\,, \\ 
P_\textrm{R}(t,r) &=& O(r)\,, \\ 
P_M(t,r) &=& O(r)\,,\\ 
P_Q(t,r) &=& O(r)\,, 
\label{newfalloff_0_f} 
\end{eqnarray} 
with 
\begin{eqnarray} 
    M_0(t) &=& \frac12 \alpha^{-1}R_0^{-1}(t)Q_0^2(t) 
    +\frac{1}{32}\alpha^3R_0^3(t)\,,\label{m_0_zero}\\ 
    M_2(t) &=& \frac{1}{32}\alpha R_0(t)R_2(t)\left(3\alpha^2 
  R_0(t)-64R_2(t)\Lambda_0(t)^{-2}\right) 
  + \alpha^{-1}Q_0(t)R_0^{-1}(t)\left(Q_2(t)- 
  \frac12Q_0(t)R_0^{-1}(t)R_2(t)\right)\,. 
  \label{m_2_zero} 
\end{eqnarray} 
For $r\rightarrow \infty$, we have 
\begin{eqnarray} \label{newfalloff_inf_i} 
F(t,r) &=& \frac{\alpha^2}{16}\,r^2 - 
2(\eta(t)+2\rho(t))\,r^{-1}+O^{\infty}(r^{-2})\,, \\ 
\textrm{R}(t,r) &=& r + 16\rho(t)\alpha^{-2}r^{-2} 
+ O^{\infty}(r^{-3})\,, \\ 
M(t,r) &=& M_+(t) + O^{\infty}(r^{-1}) \,, \\ 
Q(t,r) &=& Q_+(t) + O^{\infty}(r^{-1}) \,, \\ 
P_\textrm{R}(t,r) &=& O^\infty(r^{-4})\,, \\ 
P_M(t,r) &=& O^\infty(r^{-6})\,, \\ 
P_Q(t,r) &=& O^\infty(r^{-2})\,, 
\label{newfalloff_inf_f} 
\end{eqnarray} 
where $M_+(t)=\alpha(\eta(t)+3\rho(t))$, as seen before in the 
surface terms (see Eq. (\ref{surface_inf})). 

We are now almost ready to write the action with the new canonical 
variables. It is now necessary to determine the new Lagrange 
multipliers. In order to write the new constraints with the new 
Lagrange multipliers, we can use the identity given by the space 
derivative of $M$, 
\begin{equation} 
M' = -\Lambda^{-1}\left( R' H +\alpha^{-1} R^{-1} 
P_\Lambda\left(H_r-\Gamma G\right) + \alpha^{-1}\Lambda R^{-1} 
P_\Gamma G \right)\,. 
\end{equation} 
Solving for $H$ and making use of the inverse transformations of 
$\Lambda$ and $P_\Lambda$, in Eqs. (\ref{inversetrans_1}) and 
(\ref{inversetrans_2}), we get 
\begin{eqnarray} \label{oldconstraintsinnewvariables_1} 
H &=& - \frac{M'F^{-1}\textrm{R}'+F P_M 
P_{\textrm{R}}+\alpha^{-1}R^{-1}F^{-1}R'QG} 
{\left(F^{-1}(\textrm{R}')^2-F 
    P_M^2\right)^{\frac12}}\,, 
    \label{oldconstraintsinnewvariables_2}\\ 
H_r &=& P_M M' + P_{\textrm{R}} \textrm{R}' + P_Q Q'\,, \\ 
G &=& -Q'\,. 
 \label{oldconstraintsinnewvariables_3} 
\end{eqnarray} 
Following Kucha\v{r} \cite{kuchar}, the new set of constraints, totally 
equivalent to the old set $H(t,r)=0$, $H_r(t,r)=0$, and $G(t,r)=0$ 
outside the horizon, is $M'(t,r)=0$, $P_{\textrm{R}}(t,r)=0$, 
and $Q'(t,r)=0$. 
By continuity, this also applies on the horizon, where $F(t,r)=0$. 
So we can say that the equivalence is valid everywhere. 
So, the new Hamiltonian, the total 
sum of the constraints, can now be written as 
\begin{equation} \label{newHamiltonian} 
NH+N^rH_r+\tilde{\Phi}G = N^M M' + N^{\textrm{R}} P_{\textrm{R}} 
+ N^Q Q'\,. 
\end{equation} 
In order to determine the new Lagrange multipliers, one has to write 
the left hand side of the previous equation, 
Eq. (\ref{newHamiltonian}), and replace the constraints on that side 
by their expressions as functions of the new canonical coordinates, 
spelt out in Eqs. 
(\ref{oldconstraintsinnewvariables_1}) 
-(\ref{oldconstraintsinnewvariables_3}). 
After manipulation, one gets 
\begin{eqnarray} \label{new_mult_1} 
N^M &=& - \frac{N F^{-1} R'}{\left(F^{-1}(\textrm{R}')^2-F 
    P_M^2\right)^{\frac12}}+ N^r P_M \,, \\ 
N^{\textrm{R}} &=& - \frac{N F P_M}{\left(F^{-1}(\textrm{R}')^2-F 
    P_M^2\right)^{\frac12}}+ N^r R'\,, \label{new_mult_2} \\ 
N^Q &=& \frac{\alpha^{-1}NR^{-1}F^{-1}R'Q}{\left(F^{-1}(\textrm{R}')^2-F 
    P_M^2\right)^{\frac12}}+N^rP_Q-\tilde{\Phi}\,. \label{new_mult_3} 
\end{eqnarray} 
Using the inverse transformations 
Eqs. (\ref{inversetrans_1})-(\ref{inversetrans_2}), and the identity 
$R=\textrm{R}$, we can write the new multipliers as functions of the 
old variables 
\begin{eqnarray} \label{mult_trans_1} 
N^M &=& - NF^{-1}R'\Lambda^{-1}+\alpha^{-1} N^r F^{-1} R^{-1} \Lambda 
P_\Lambda\,, \\ 
N^{\textrm{R}} &=& - \alpha^{-1} N R^{-1} P_\Lambda + N^r 
R'\,,\label{mult_trans_2} \\ 
N^Q &=& \alpha^{-1}N\Lambda^{-1}R^{-1}F^{-1}R'Q- 
N^r\Gamma-\alpha^{-2}N^rR^{-2}F^{-1}\Lambda 
P_\Lambda P_\Gamma-\tilde{\Phi}\,, \label{mult_trans_3} 
\end{eqnarray} 
allowing us to determine its asymptotic conditions from the original 
conditions given above. These transformations are non-singular for $r>0$. 
As before, for $r\rightarrow 0$, 
\begin{eqnarray} \label{mult_newfalloff_0_i} 
N^M(t,r) &=& -\frac12 N_1(t) \Lambda_0(t) R_2(t)^{-1} + O(r^2)\,,\\ 
N^{\textrm{R}}(t,r) &=& O(r^2)\,,\\ 
N^Q (t,r)&=& \frac12 \alpha^{-1} N_1(t) Q_0(t) \Lambda_0(t) R_0^{-1}(t) 
R_2^{-1}(t)-\tilde{\Phi}_0 + O(r^2)\,, \label{mult_newfalloff_0_f} 
\end{eqnarray} 
and for $r\rightarrow\infty$ we have 
\begin{eqnarray} \label{mult_newfalloff_inf_i} 
N^M(t,r) &=& -\tilde{N}_+(t) +  O^\infty(r^{-4})\,,\\ 
N^{\textrm{R}}(t,r) &=& O^\infty(r^{-2})\,, \\ 
N^Q(t,r)&=&-\tilde{\Phi}_+(t)+O^\infty(r^{-1})\,. 
\label{mult_newfalloff_inf_f} 
\end{eqnarray} 
These conditions 
(\ref{mult_newfalloff_0_i})-(\ref{mult_newfalloff_inf_f}) 
show that the transformations in 
Eqs. (\ref{mult_trans_1})-(\ref{mult_trans_2}) are satisfactory in 
the case of $r\rightarrow\infty$, but not for $r\rightarrow 0$. This 
is due to fact that in order to fix the Lagrange multipliers for 
$r\rightarrow\infty$, as we are free to do, we fix $\tilde{N}_+(t)$, 
which we already do when adding the surface term 
\begin{equation} 
- \int\, dt \, \tilde{N}_+ M_+ 
\end{equation} 
to the action, in order to obtain the equations of motion in the bulk, 
without surface terms. The same is true for $\tilde{\Phi}_+$. 
However, at $r=0$, we see that fixing the multiplier $N^M$ to values 
independent of the canonical variables is not the same as fixing $N_1 
\Lambda_0^{-1}$ to values independent of the canonical variables. 
The same is true of the fixation of $N^Q$ with respect to 
$\tilde{\Phi}_0$. 
We need to rewrite the multipliers $N^M$ and $N^Q$ for the asymptotic 
regime 
$r\rightarrow 0$ without affecting their behavior for 
$r\rightarrow\infty$. 
In order to proceed we have to make one assumption, which is that the 
expression given in asymptotic condition of $M(t,r)$, as $r\rightarrow 
0$, for the term of order zero, $M_0\equiv 
\frac12\alpha^{-1}Q_0(t)^2R_0^{-1}(t) 
+\frac{1}{32}\alpha^3R_0(t)^3$, 
defines $R_0$ as a function of $M_0$ and $Q_0$, 
where $R_0$ is the horizon radius function, $R_0\equiv 
R_{\textrm{h}}(M_0,Q_0)$. Also, we assume that 
$M_0>M_{\textrm{\tiny{crit}}}(Q_0)=\frac{Q^{\frac32}}{3^{\frac34}}$. 
With these assumptions, we are working in the domain 
of the classical solutions. 
We can immediately obtain that the variation of 
$R_0$ is given in relation to the variations of $M_0$ 
and of $Q_0$ as 
\begin{equation} \label{var_r_m} 
\delta R_0 = \left(\frac{32}{3}\alpha^{3}R_0^{2}- 
\frac{Q_0^2}{2\alpha R_0^2}\right)^{-1} 
\left(\delta M_0-\frac{Q_0\delta Q_0}{\alpha R_0}\right)\,. 
\end{equation} 
This 
expression will be used when we derive the equations of motion from 
the new action. 
We now define the new multipliers $\tilde{N}^M$ and $\tilde{N}^Q$ as 
\begin{eqnarray} \label{new_n_m} 
\tilde{N}^M &=& - N^M 
\left[(1-g)+ g (\alpha R_0) \left(\frac{3\alpha^3}{32} 
R_0^2 - \frac{Q_0^2}{2\alpha R_0^2}\right)^{-1}\right]^{-1}\,, \\ 
\tilde{N}^Q &=& \tilde{N}^M\,g\,Q_0 \left(\frac{3\alpha^3}{32} 
R_0^2 - \frac{Q_0^2}{2\alpha R_0^2}\right)^{-1}-N^Q\,, 
\label{new_n_q} 
\end{eqnarray} 
where $g(r)=1+O(r^2)$ for $r\rightarrow 0$ and $g(r)=O^\infty(r^{-5})$ 
for $r\rightarrow\infty$. These new multipliers, functions of the old 
multipliers $N^M$ and $N^Q$, have as their properties for 
$r\rightarrow \infty$ 
\begin{eqnarray} 
    \tilde{N}^M(t,r) &=& \tilde{N}_+(t) + O^\infty(r^{-5})\,, \\ 
    \tilde{N}^Q(t,r) &=& \tilde{\Phi}_+(t) + O^\infty(r^{-1})\,, 
\end{eqnarray} 
and for $r\rightarrow0$ 
\begin{eqnarray} 
\tilde{N}^M(t,r) &=& \tilde{N}_0^M(t) + O(r^{2})\,, \\ 
\tilde{N}^M(t,r) &=& \tilde{\Phi}_0(t) + O(r^2)\,, 
\end{eqnarray} 
where $\tilde{N}_0^M$ is given by 
\begin{equation} 
\tilde{N}_0^M = \frac12 \alpha^{-1}N_1R_0^{-1}R_2^{-1}\Lambda_0 
\left(-\frac12Q_0^2\alpha^{-1}R_0^{-2}+\frac{3}{32}\alpha^3 
R_0^2\right)\,. 
\end{equation} 
When the constraints $M'=0=Q'$ hold, the last expression is 
\begin{equation} 
\tilde{N}_0^M =  N_1 \Lambda_0^{-1} \,. 
\end{equation} 
With this new constraint $\tilde{N}^M$, fixing $N_1 \Lambda_0^{-1}$ at 
$r=0$ or fixing $\tilde{N}^M_0$ is equivalent. The same happens with 
fixing $\tilde{\Phi}_0$ or $\tilde{N}^Q$ at $r\to0$. There are no problems 
with $N^{\textrm{R}}$, which is left as determined in 
Eq. (\ref{new_mult_2}). 
The new action is now written as the sum of $S_\Sigma$, the bulk action, 
and 
$S_{\partial\Sigma}$, the surface action, 
\begin{eqnarray} 
S\left[M, \textrm{R},\Gamma, P_M, P_{\textrm{R}},P_\Gamma; 
\tilde{N}^M, N^{\textrm{R}},\tilde{N}^Q\right] 
&=& \int dt\,\int_0^\infty dr \, 
\left\{ P_M\dot{M} + P_\textrm{R} \dot{\textrm{R}}+P_Q\dot{Q} 
+\tilde{N}^QQ'-N^{\textrm{R}}P_{\textrm{R}}\right. \nonumber\\ 
&+& \left.\tilde{N}^M 
\left[(1-g)M'+g 
\left(\frac{3\alpha^2}{16}R_0^2 
-\frac{Q_0^2}{2\alpha R_0^2}\right)^{-1} 
\left(\alpha R_0M'-Q_0Q'\right)\right]\right\} 
\nonumber \\ 
&+&\int dt \, \left(\frac12 \alpha R_0^2 \tilde{N}_0^M - 
  \tilde{N}_+ M_+ +\tilde{\Phi}_0Q_0-\tilde{\Phi}_+Q_+ 
  \right)\,. \label{newaction} 
\end{eqnarray} 
The new equations of motion are now 
\begin{eqnarray} \label{new_eom_1} 
\dot{M} &=& 0\,, \\ 
\dot{\textrm{R}} &=& N^{\textrm{R}}\,, \\ 
\dot{Q}&=&0\,,\\ 
\dot{P}_M &=& (N^M)'\,, \\ 
\dot{P}_{\textrm{R}} &=& 0\,, \\ 
\dot{P}_Q &=&(N^Q)'\,,\\ 
M' &=& 0\,, \\ 
P_{\textrm{R}} &=& 0\,,\\ 
Q' &=& 0\,. \label{new_eom_9} 
\end{eqnarray} 

Here we understood $N^M$ to be a function of the new constraint 
$\tilde{N}^M$, defined through Eq. (\ref{new_n_m}), and $N^Q$ to 
be a function of the new constraint $\tilde{N}^Q$, defined through 
Eq. (\ref{new_n_q}). 
The resulting boundary terms of 
the variation of this new action, Eq. (\ref{newaction}), are, first, 
terms proportional to $\delta M$, $\delta \textrm{R}$, and $\delta Q$ 
on the initial and final hypersurfaces, and, second, the term 
\begin{eqnarray} 
          \int \, dt \, \left(\frac12 \alpha R_0^2 
\delta\tilde{N}_0^M - 
      M_+ \delta\tilde{N}_+  + Q_0\delta\tilde{\Phi}_0 - 
      Q_+\delta\tilde{\Phi}_+\right) 
\end{eqnarray} 
Here we have used the expression in Eq. (\ref{var_r_m}). The action in 
Eq. (\ref{newaction}) yields the equations of motion, 
Eqs. (\ref{new_eom_1})-(\ref{new_eom_9}), provided that we fix the 
initial and final values of the new canonical variables and that we 
also fix the values of $\tilde{N}^M_0$, $\tilde{N}_+$, $\tilde{\Phi}_0$, 
and $\tilde{\Phi}_+$. Thanks to the redefinition of both the Lagrange 
multipliers, from $N^M$ to $\tilde{N}^M$ and from $N^Q$ to $\tilde{N}^Q$, 
the fixation of those quantities, $\tilde{N}^M_0$, $\tilde{N}_+$, 
$\tilde{\Phi}_0$, and $\tilde{\Phi}_+$ has the same meaning it had 
before the canonical transformations and the redefinition of $N^M$ and 
$N^Q$. This same meaning is guaranteed through the use of our gauge 
freedom to choose the multipliers, and at the same time not fixing 
the boundary variations independently of the choice of Lagrange 
multipliers, which in turn allow us to have a well defined 
variational principle for the action. 

\subsection{Hamiltonian reduction} 

We now solve the constraints in order to reduce to the true dynamical 
degrees of freedom. The equations of motion 
(\ref{new_eom_1})-(\ref{new_eom_9}) 
allow us to write $M$ and $Q$ as independent functions of the radial 
coordinate $r$, 
\begin{eqnarray} \label{m_t} 
M(t,r) &=& \textbf{m}(t)\,,\\ 
Q(t,r) &=& \textbf{q}(t)\,. 
\label{q_t} 
\end{eqnarray} 
The reduced action, with the constraints and  Eqs. (\ref{m_t})-(\ref{q_t}) 
taken into account, is 
\begin{equation} \label{red_action} 
S 
\left[\textbf{m},\textbf{p}_{\textbf{m}},\textbf{q},\textbf{p}_{\textbf{q}}; 
\tilde{N}_0^M,\tilde{N}_+,\tilde{\Phi}_0,\tilde{\Phi}_+\right] 
= \int 
dt\,\,\textbf{p}_{\textbf{m}}\dot{\bf{m}}+ 
\textbf{p}_{\textbf{q}}\dot{\bf{q}} -\textbf{h}\,, 
\end{equation} 
where 
\begin{eqnarray} \label{new_p_m} 
\textbf{p}_{\textbf{m}} &=& \int_0^\infty dr\,P_M\,, \\ 
\textbf{p}_{\textbf{q}} &=& \int_0^\infty dr\,P_Q\,, 
\label{new_p_q} 
\end{eqnarray} 
and the reduced Hamiltonian, $\textbf{h}$, is now written as 
\begin{equation} \label{red_Hamiltonian} 
\textbf{h}(\textbf{m},\,\textbf{q};t) = -\frac12 \alpha 
R_{\textrm{h}}^2 \tilde{N}_0^M 
+ \tilde{N}_+ \textbf{m}+ 
\left(\tilde{\Phi}_+-\tilde{\Phi}_0\right) 
\textbf{q}\,, 
\end{equation} 
with $R_{\textrm{h}}$ being the horizon radius. We also have 
that $\textbf{m}>M_{\textrm{\tiny{crit}}}(\textbf{q})= 
\frac{\textbf{q}^\frac32}{3^\frac34}$, 
according to the assumptions made in the previous 
subsection. Thanks to the functions $\tilde{N}_0^M(t)$, 
$\tilde{N}_+(t)$, $\tilde{\Phi}_0(t)$, and $\tilde{\Phi}_+(t)$ 
the Hamiltonian $\textbf{h}$ is an explicitly time 
dependent function. The variational principle associated with the 
reduced action, Eq. (\ref{red_action}), will fix the value of 
$\textbf{m}$ and $\textbf{q}$ on the initial and final hypersurfaces, 
or in the spirit of the classical analytical mechanics, 
the Hamiltonian principle fixes the initial and final values of 
the canonical coordinate. The equations of motion are 
\begin{eqnarray} \label{red_eom_1} 
\dot{\textbf{m}} &=& 0\,, \\ 
\dot{\textbf{q}} &=& 0\,,  \label{red_eom_2}\\ 
\dot{\textbf{p}}_{\textbf{m}} &=& \alpha R_{\textrm{h}} 
\left(\frac{3}{32}\alpha^3R_{\textrm{h}}^2- 
\frac{\textbf{q}^2}{2\alpha R_{\textrm{h}}^2}\right)^{-1} 
\tilde{N}_0^M-\tilde{N}_+\,, \label{red_eom_3}\\ 
\dot{\textbf{p}}_{\textbf{q}} &=& 
-\textbf{q}\left(\frac{3}{32}\alpha^3R_{\textrm{h}}^2- 
\frac{\textbf{q}^2}{2\alpha R_{\textrm{h}}^2}\right)^{-1} 
\tilde{N}_0^M + \tilde{\Phi}_0-\tilde{\Phi}_+\,. 
\label{red_eom_4} 
\end{eqnarray} 
The equation of motion for $\textbf{m}$, Eq. (\ref{red_eom_1}), is 
understood as saying that $\textbf{m}$ is, on a classical solution, 
equal to the mass parameter $M$ of the solutions in Eq. 
(\ref{metric_zero}). The same holds for $\textbf{q}$ in Eq. 
(\ref{red_eom_2}), 
i.e., $\textbf{q}$ is the charge parameter in Eqs. 
(\ref{metric_zero})-(\ref{vector_potential_zero}). 
In order to interpret equation of motion for $\textbf{p}_{\textbf{m}}$, 
Eq. (\ref{red_eom_3}), we have to recall that from Eq. (\ref{p_m}) 
one has $P_M=-T'$, where $T$ is the Killing time. This, together with 
the definition of $\textbf{p}_{\textbf{m}}$, given in 
Eq. (\ref{new_p_m}), yields 
\begin{equation} 
\textbf{p}_{\textbf{m}} = T_0 - T_+\,, 
\end{equation} 
where $T_0$ is the value of the Killing time at the left end of the 
hypersurface of a certain $t$, and $T_+$ is the Killing time at 
spatial infinity, the right end of the same hypersurface of $t$. As 
the hypersurface evolves in the spacetime of the black hole solutions, 
the right hand side of Eq. (\ref{red_eom_2}) is equal to 
$\dot{T}_0 - \dot{T}_+$. Finally, after the definition 
\begin{eqnarray} 
    \textbf{p}_{\textbf{q}} = \xi_0 - \xi_+\,, 
\end{eqnarray} 
obtained from to Eqs. (\ref{vector_potential_plusgauge_tr_zero}), 
(\ref{p_q}), and (\ref{new_p_q}), Eq. (\ref{red_eom_4}) 
gives in the right hand side $\dot{\xi}_0 - \dot{\xi}_+$, 
which is the diference of the time derivatives of the 
electromagnetic gauge $\xi(t,r)$ at $r=0$ and at infinity. 

\subsection{Quantum theory and partition function} 

Following Sec. \ref{mtz} we need not repeat here the steps taken 
there in order to build the time evolution operator. 

Thus the operator $K$ can now be written as 
\begin{equation} 
K\left(\textbf{m},\textbf{q};\mathcal{T},\Theta,\Xi_0,\Xi_+\right) = \exp 
\left[-i\textbf{m}\mathcal{T}+\frac{i}{2}\alpha R^2_{\textrm{h}}\Theta 
-i\textbf{q}\left(\Xi_+-\Xi_0\right)\right]\,. 
\label{kappa_function_zero} 
\end{equation} 
The same composition law as in Sec. \ref{mtz} holds here as well. 
Here again $\mathcal{T}$ is the Killing time elapsed at right spatial 
infinity and $\Theta$ is the boost parameter elapsed at the 
bifurcation circle; $\Xi_0$ and $\Xi_+$ are line integrals along 
timelike curves of constant $r$ and constant angular variables, at 
$r=0$ and at infinity. 

\subsection{Thermodynamics} 

We can now build the partition function for this system. 
The path to follow is to continue the operator to imaginary 
time and take the trace over a complete orthogonal basis. 
Our classical thermodynamic situation consists of a three-dimensional 
spherically symmetric charged black hole, asymptotically anti-de Sitter, 
in thermal equilibrium with a bath of Hawking radiation. Ignoring 
back reaction from the radiation, the geometry is described by the 
solutions in Eq. (\ref{metric_zero}) and (\ref{vector_potential_zero}). 
Thus, we consider a thermodynamic ensemble in which the temperature, 
or more appropriately here, the inverse temperature 
$\beta$ is fixed. This characterizes a grand canonical ensemble, 
and the partition function 
$\mathcal{Z}(\beta,\bar{\phi})$ arises naturally in such an ensemble. 
To analytically continue the Lorentzian solution 
we put $\mathcal{T}=-i\beta$, and $\Theta-2\pi i$, this latter choice 
based on the regularity of the classical Euclidean solution. 
We also choose $\Xi_0=0$ and $\Xi_+=i\beta\bar{\phi}$. 

We arrive then at the following expression for the 
partition function 
\begin{equation} \label{partition_function_1} 
\mathcal{Z}(\beta,\bar{\phi}) = \textrm{Tr} \left[\hat{K} 
(-i\beta,-2\pi i,0,i\beta\bar{\phi})\right]\,. 
\end{equation} 
{}From Eq. (\ref{kappa_function_zero}) this is realized as 
\begin{equation} \label{partition_function_2} 
\mathcal{Z}(\beta,\bar{\phi}) = \int_A \mu\,d\textbf{m}d\textbf{q}\, 
\exp\left[-\beta (\textbf{m}-\textbf{q}\bar{\phi})+\pi 
\alpha R_{\textrm{h}}^2\right]\left\langle 
\textbf{m}|\textbf{m} 
\right\rangle\,. 
\end{equation} 
Since $\left\langle \textbf{m}|\textbf{m}\right\rangle$ is equal to 
$\delta (0)$, one has to regularize (\ref{partition_function_2}). 
Again, following the Louko-Whiting procedure \cite{louko1}-\cite{louko5}, 
we have to regularize and normalize the operator 
$\hat{K}$ beforehand. This leads to 
\begin{equation} \label{partition_function_3} 
\mathcal{Z}_{\textrm{ren}}(\beta,\bar{\phi}) = \mathcal{N} \int_{A} 
\mu\,d\textbf{m}d\textbf{q}\,\exp\left[-\beta 
(\textbf{m}-\textbf{q}\bar{\phi})+ 
\pi \alpha R_{\textrm{h}}^2\right]\,, 
\end{equation} 
where $\mathcal{N}$ is a normalization factor and $A$ is the domain 
of integration.  Provided the weight 
factor $\mu$ is slowly varying compared to the 
exponential in Eq. (\ref{partition_function_3}), and using the 
fact that the horizon radius $R_{\textrm{h}}$ is function of 
$\textbf{m}$ and $\textbf{q}$, 
the integral in Eq. (\ref{partition_function_3}) is 
convergent. Changing integration variables, from 
$\textbf{m}$ to $R_{\textrm{h}}$, where 
\begin{equation} \label{m_r_h} 
\textbf{m} = \frac{1}{32}\alpha^3 R_{\textrm{h}}^3 
+\frac{\textbf{q}^2}{2\alpha R_{\textrm{h}}}\,, 
\end{equation} 
the integral Eq. (\ref{partition_function_3}) becomes 
\begin{equation} \label{partition_function_ren} 
\mathcal{Z}_{\textrm{ren}}(\beta,\bar{\phi}) = \mathcal{N} \int_{A'} 
\widetilde{\mu}\,dR_{\textrm{h}}d\textbf{q}\,\exp(-I_*)\,, 
\end{equation} 
where $A'$ is the new domain of integration after 
changing variables, and the function 
$I_*(R_{\textrm{h}},\textbf{q})$, a kind of an 
effective action (see \cite{york1}), is written as 
\begin{equation} \label{eff_action} 
I_*(R_{\textrm{h}},\textbf{q}):= \frac{1}{32}\beta\, 
\alpha^3 R_{\textrm{h}}^3 
+\frac{\beta}{2\alpha}\frac{\textbf{q}^2}{R_{\textrm{h}}}- 
\beta\bar{\phi}\textbf{q}-\pi\alpha R_{\textrm{h}}^2\,. 
\end{equation} 
The domain of integration, $A'$, is defined by the inequalities 
$0\leq R_{\textrm{h}}$ and 
$\textbf{q}^2\leq3\alpha^4R_{\textrm{h}}^4/16$. 
The new weight factor $\widetilde{\mu}$ includes the Jacobian of the 
change of variables. Since the weight factor is slowly varying, 
we can estimate the integral of 
$\mathcal{Z}_{\textrm{ren}}(\beta,\bar{\phi})$ 
by the saddle point approximation. 
The critical points are given as a pair of values 
$(R_{\textrm{h}},\textbf{q})$ 
\begin{eqnarray} 
R_{\textrm{h}}^\pm &=& \frac{32\pi}{3\beta\alpha^2} 
\left(1\pm\sqrt{1+\frac{3}{64}\beta^2\alpha^2\bar{\phi}^2 
\pi^{-2}}\right)\,, \label{crit_r_zero}\\ 
\textbf{q}^\pm&=&\alpha R_{\textrm{h}}^\pm\bar{\phi}\,. 
\label{crit_q_zero} 
\end{eqnarray} 
The critical points belong to the domain of 
the effective action (\ref{eff_action}), and so the effective 
action has a null derivative at these points 
in Eqs. (\ref{crit_r_zero})-(\ref{crit_q_zero}). 
As $\bar{\phi}^2>-\frac{3}{64}\beta^{-2}\alpha^{-2}\bar{\phi}^{-2}\pi^{2}$ 
there are always two critical points, given by 
(\ref{crit_q_zero})-(\ref{crit_r_zero}). 
The critical point at $(R_{\textrm{h}}^-,\textbf{q}^-)$ 
is outside the domain of physical interest as $R_{\textrm{h}}^-<0$. 
The critical point at $(R_{\textrm{h}}^+,\textbf{q}^+)$ is an 
global minimum. 
The expression (\ref{crit_q_zero}) implies that the critical 
points are on the straight line defined by itself, on 
the plane of the arguments of the effective action 
(\ref{eff_action}). Then, if we replace the value of 
the charge as a function of 
$R_{\textrm{h}}$, i. e., $\textbf{q}(R_{\textrm{h}})$ 
as given in (\ref{crit_q_zero}) into the effective action 
we obtain a one variable function of $R_{\textrm{h}}$, 
written as 
\begin{eqnarray} 
I_*(R_{\textrm{h}})=\frac{1}{32}\beta\alpha^3R_{\textrm{h}}^3 
-\frac12\beta\alpha\bar{\phi}^2R_{\textrm{h}}- 
\alpha\pi R_{\textrm{h}}^2\,, \label{eff_action_r_zero} 
\end{eqnarray} 
which is a function of $R_{\textrm{h}}$ on the straight line 
defined by $\textbf{q}=\alpha R_{\textrm{h}}\bar{\phi}$. 
The zeros of this effective action on the straight 
line are given by 
\begin{eqnarray} 
 R_{\textrm{h}}^0 &=& 0\,,\\ \label{zero_eff_action_r_zero} 
 {R_{\textrm{h}}^0}^\pm &=& 16\beta^{-1}\alpha^{-2}\pi 
 \left(1\pm\sqrt{1+\frac{1}{16}\beta^2\alpha^2\bar{\phi}^2\pi^{-2}} 
 \right)\,. \label{zeros_eff_action_r_zero} 
\end{eqnarray} 
Here $R_{\textrm{h}}^-<0$ is outside the physical domain of 
interest. As 
$\bar{\phi}^2>-\frac{1}{16}\beta^{-2}\alpha^{-2}\bar{\phi}^{-2}\pi^{2}$ 
there are always three zeros of the function 
(\ref{eff_action_r_zero}), which means that there are no 
phase transitions, as the zeros of the effective action 
maintain their respective hierarchy whatever the value of 
$\bar{\phi}$, which is the order parameter of the system. 
So the classical solution is found in the global minimum 
of the critical points $(R_{\textrm{h}}^+,\textbf{q}^+)$ 
of the effective action $I_*(R_{\textrm{h}},\textbf{q})$ 
(Eq. (\ref{eff_action})). 
The partition function $\mathcal{Z}_{\textrm{ren}}(\beta,\bar{\phi})$ 
can now be written, through the saddle-point approximation 
\begin{eqnarray} \label{z_approx} 
    \mathcal{Z}_{\textrm{ren}}(\beta,\bar{\phi})&\approx&\textrm{P} 
    \exp\left[-I_*(R_{\textrm{h}}^+,\textbf{q}^+)\right]\,,\\ 
    \ln\mathcal{Z}_{\textrm{ren}}(\beta,\bar{\phi}) &\approx& 
\ln\textrm{P} 
    -I_*(R_{\textrm{h}}^+,\textbf{q}^+)\,. 
    \label{log_z_approx} 
\end{eqnarray} 
where $\textrm{P}$ is a slowly varying prefactor, and 
$I_*(R_{\textrm{h}}^+,\textbf{q}^+)$ is the effective action 
evaluated at the global minimum critical point. 
By ignoring the prefactor's logarithm, which closer to 
$(R_{\textrm{h}}^+,\textbf{q}^+)$ is less relevant, 
we are able to determine the 
value of $\textbf{m}$ at the critical point, where we find that it 
corresponds to the value of the mass of the classical solution of the 
black hole given in Eq. (\ref{metric_zero}). Thus, 
when the critical point dominates the partition function, we have that 
the mean energy $\left\langle E \right\rangle$ is given by 
\begin{equation} 
\left\langle E\right\rangle = -\frac{\partial}{\partial\beta}\ln 
\mathcal{Z}_{\textrm{ren}} \approx \frac{1}{32} \alpha^3 
(R_{\textrm{h}}^+)^3-\frac12\alpha\bar{\phi}^2R_{\textrm{h}}^+ = 
\textbf{m}^+\,, 
\label{m_plus_zero} 
\end{equation} 
where $\textbf{m}^+$ is obtained from Eq. (\ref{m_r_h}) evaluated at 
$(R_{\textrm{h}}^+,\textbf{q}^+)$. 
The thermal expectation value of the charge is 
\begin{equation} 
    \left\langle Q\right\rangle = 
\beta^{-1}\frac{\partial}{\partial\bar{\phi}} 
    \ln \mathcal{Z}_{\textrm{ren}}\approx \textbf{q}^+\,, 
    \label{q_plus_zero} 
\end{equation} 
as given in Eq. (\ref{crit_q_zero}). 
We now can write the temperature $\textbf{T}\equiv\beta^{-1}$ as 
a function of the critical point given by $R_{\textrm{h}}^+$ and 
$\textbf{q}^+$ 
\begin{eqnarray} 
    \textbf{T} &=& \frac{1}{4\pi} 
    \left(3a\,R_{\textrm{h}}^+-\frac{1}{16} 
        \left(\textbf{q}^+\right)^2\left(a\,R_{\textrm{h}}^+\right)^{-3} 
        \right)\,, \label{temp_zero} 
\end{eqnarray} 
with $R_{\textrm{h}}^+$ explicitly written in (\ref{crit_r_zero}), 
and where $\textbf{q}^+$ is given 
in Eq. (\ref{crit_q_zero}). If we choose the maximum charge of the 
domain given by $\textbf{q}^2\leq3\alpha^4R_{\textrm{h}}^4/16$, 
then we find ourselves with the extreme solution of the black hole, 
whose temperature, if $\textbf{q}^2=3\alpha^4R_{\textrm{h}}^4/16$ 
is replaced into (\ref{temp_zero}), is equal to zero. 
It can be shown that 
$\partial\textbf{m}^+/\partial\beta<0$, which through 
the constant $\bar{\phi}$ heat capacity 
$C_{\bar{\phi}}=-\beta^2(\partial \left\langle 
E\right\rangle/\partial\beta)$ 
tells us that the system is thermodynamically stable. 
Finally there is the entropy of the black hole, $S$, 
which is given by 
\begin{equation} 
S = 
\left(1-\beta\frac{\partial}{\partial\beta}\right) 
(\ln\mathcal{Z}_{\textrm{ren}}) 
\approx 4\pi a 
{R_{\textrm{h}}^+}^2\,. 
\label{entropy_zero} 
\end{equation} 
Here we see that despite the fact that the extreme solution has null 
temperature, it has an entropy given in (\ref{entropy_zero}), which is 
independent of the value of the charge in the domain of validity 
of the approximation. 

\section{Hamiltonian Thermodynamics of a representative 
charged dilatonic black hole ({\large $\omega=-3$})} 
\label{menostres} 

\subsection{The metric} 
For $\omega=-3$, the general metric in Eq. (\ref{solutions}), the 
$\phi$ field in Eq. (\ref{phisolutions_1}), and the vector potential 
(\ref{vector_potential}) reduce to the following, respectively, 
\begin{eqnarray} 
ds^2 &=& - \left[ (a\,R)^2 - 2(a\,R)^\frac12 \, M 
-\frac12(a\,R)Q^2\right]\,dT^2 + 
\frac{dR^2}{(a\,R)^2 - 2(a\,R)^\frac12 \,M-\frac12(a\,R)Q^2} 
+R^2\,d\varphi^2\,, 
\label{metric_dilatonic}\\ 
e^{-2\phi} &=& \frac{1}{(a\,R)^\frac12}\,, 
\label{phi_dilatonic}\\ 
A &=& -\frac12(a\,R)^{\frac12}Q dT\,, 
\label{vector_potential_dilatonic} 
\end{eqnarray} 
with $2M=b$ and 
$a=\frac{4\sqrt{3}}{3}|\lambda|=\frac{2\sqrt{6}}{3}l^{-1}$. 
\begin{figure} 
[htmb] 
\centerline{\includegraphics 
%[width=6cm,height=6cm] 
{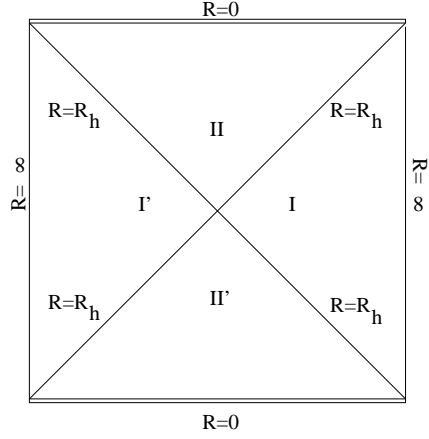}} 
\caption {{\small The Carter-Penrose Diagram for the $\omega=-3$ case, 
with $M>0$.}} 
\label{omega_-3} 
\end{figure} 
In Fig. \ref{omega_-3} 
the Carter-Penrose diagram of the black hole 
solution for the case $\omega=-3$ is shown, which is analogous to 
the case $\omega=0$, where again the singularity at $R=0$ is a 
curvature singularity, and $R=\infty$ is spatial infinity. 
The horizon radius is given by the largest real solution of 
the following equation 
\begin{eqnarray} 
    (a\,R_\textrm{h})^2 - 2(a\,R_\textrm{h})^\frac12\,M 
-\frac12(a\,R_\textrm{h})Q^2 &=& 0\,. 
\label{horizon_radius_dilatonic} 
\end{eqnarray} 
Note that this diagram, Fig. \ref{omega_-3}, 
holds for static black holes with positive 
mass $M>0$. Here I and I' are the right and left static 
outer regions of the Carter-Penrose diagram, respectively, and II 
and II' are the future and past inner dynamical regions of the 
Carter-Penrose diagram. $R_\textrm{h}$ is the horizon radius given 
in Eq. (\ref{horizon_radius_dilatonic}) and the spacelike infinity 
is asymptotically anti-de Sitter, both in the right and in the left 
regions, denoted by I and I', respectively. There are no extreme 
black hole solutions for $\omega=-3$ \cite{oscarlemos}. 

Since we are now familiar with the whole formalism, we will be 
briefer in this section omitting several of the details. 

\subsection{Canonical formalism } 
The action becomes, following 
Eqs. (\ref{ADM_ansatz})-(\ref{a_ansatz}), 
and up to surface terms, 
\begin{eqnarray} \label{action_-3} 
S[\Lambda,\,R,\,\dot{\Lambda},\,\dot{R};\,N,\,N^r] &=& 
\int dt \int_0^{\infty} dr\, 
\left\{4\,(a\,R)^{-\frac12}N\lambda^2\Lambda R- 
(a\,R)^{-\frac12}N^{-1}\dot{\Lambda}\dot{R}+ 
\frac14 (a\,R)^{-\frac12} N^{-1} \Lambda R^{-1} \dot{R}^2\right. 
\nonumber \\ && + 
(a\,R)^{-\frac12} N^{-1} \dot{R} (N^r \Lambda)' + 
(a\,R)^{-\frac12} N^{-1}N^r\dot{\Lambda}R' 
-(a\,R)^{-\frac12} N^{-1}N^r(N^r)'\Lambda R' 
\nonumber \\ && 
-(a\,R)^{-\frac12} N^{-1}(N^r)^2 \Lambda' R' 
-\frac12 (a\,R)^{-\frac12} N^{-1}N^r \Lambda R^{-1}\dot{R}R' 
+\frac14 (a\,R)^{-\frac12} N^{-1}(N^r)^2\Lambda R^{-1}(R')^2 
\nonumber \\ && \left. 
+\frac14 (a\,R)^{-\frac12} N \Lambda^{-1}R^{-1}(R')^2 - 
(a\,R)^{-\frac12} N (\Lambda^{-1})'R' 
 -(a\,R)^{-\frac12}N\Lambda^{-1}R''\right. \nonumber \\ 
&& +\left. 2\,N^{-1}\Lambda^{-1}(a\,R)^{-\frac12}R 
\left(\dot{\Gamma}-\Phi'\right)^2\right\}\,. 
\end{eqnarray} 
Depending on the situation we use three different letters containing 
the same information but with slightly different numerical 
values. Thus, $a$, $\lambda$, and $l$ are related by 
$a^2=\frac{16}{3}\lambda^2 = \frac83 l^{-2}$, where $l$ is defined 
now as the AdS length.  From the action above, 
we obtain the conjugate momenta 
\begin{eqnarray} 
P_\Lambda &=& -N^{-1}(a\,R)^{-\frac12}\left(\dot{R}-R'N^r\right)\,,\\ 
P_R &=& -N^{-1}(a\,R)^{-\frac12} \left\{\dot{\Lambda}- 
(\Lambda N^r)'-\frac12 \Lambda R^{-1}(\dot{R}-R'N^r)\right\}\,,\\ 
P_\Gamma &=& 4\,N^{-1}\Lambda^{-1}(a\,R)^{-\frac12}R 
\left(\dot{\Gamma}-\Phi'\right)\,. 
\end{eqnarray} 
By performing a Legendre transformation we obtain the Hamiltonian, 
which is a sum of constraints, i.e., 
\begin{eqnarray} \label{Hamiltonian_-3} 
\mathcal{H} &=& N\left\{ -(a\,R)^{\frac12}P_\Lambda P_R + 
\frac14 a^{\frac12}R^{-\frac12}\Lambda P_\Lambda^2 + 
(a\,R)^{-\frac12}\left[(\Lambda^{-1})'R'+ 
\Lambda^{-1}R''-\frac14 \Lambda^{-1}(R')^2R^{-1}\right] 
\right. \nonumber \\ 
&& \left. -4(a\,R)^{-\frac12}\lambda^2\Lambda R 
+\frac18 N\Lambda R^{-1}(a\,R)^{\frac12}P_\Gamma^2 \right\} 
\nonumber \\ 
&& +\, N^r\left\{P_R R' - \Lambda P_\Lambda' -\Gamma P_\Gamma' 
\right\} + \tilde{\Phi}\left\{-P_\Gamma'\right\}\, 
\equiv\,NH+N^r H_r+\tilde{\Phi}G\,. 
\end{eqnarray} 
We can now write the action in Hamiltonian form, which reads 
\begin{equation} \label{haction_-3} 
S[\Lambda,\,R,\,\Gamma,\,P_{\Lambda},\,P_{R},\,P_\Gamma; 
\,N,\,N^r,\,\tilde{\Phi}] = \int dt 
\int_0^{\infty} dr \left\{ P_{\Lambda}\dot{\Lambda}+P_R\dot{R} 
+P_\Gamma\dot{\Gamma}-NH-N^rH_r-\tilde{\Phi}G \right\}\,, 
\end{equation} 
with the constraints defined in Eq. (\ref{Hamiltonian_-3}). 
{}From here we derive the equations of motion for the canonical variables 
and respective canonical momenta 
\begin{eqnarray} \label{eom_-3_1} 
\dot{\Lambda} &=& -N (a\,R)^{\frac12} P_R + 
\frac12 N a^{\frac12}R^{-\frac12} 
\Lambda P_\Lambda + (N^r\Lambda)'\,,\\ \label{eom_-3_2} 
\dot{R} &=& -N (a\,R)^{\frac12} P_\Lambda + N^r R'\,, \\ \label{eom_-3_3} 
\dot{\Gamma} &=& \frac14 N(a\,R)^\frac12\Lambda R^{-1}P_\Gamma 
+(N^rP_\Gamma)'+\tilde{\Phi}'\,,\\ \label{eom_-3_4} 
\dot{P}_\Lambda &=& -\frac14 N a^{\frac12}R^{-\frac12} P_\Lambda^2 - 
\left(N(a\,R)^{-\frac12}\right)'R'\Lambda^{-2} - 
\frac14 N (a\,R)^{-\frac12} (R')^2R^{-1}\Lambda^{-2} + 
4N(a\,R)^{-\frac12}\lambda^2R + N^rP_\Lambda'\nonumber \\ 
&&-\frac18N(a\,R)^\frac12R^{-1}P_\Gamma^2\,, \\ 
\dot{P}_R &=& \frac12 N a^{\frac12}R^{-\frac12} P_\Lambda P_R + 
\frac18 N a^{\frac12}R^{-\frac32} \Lambda P_\Lambda^2 - 
N a^{-\frac12} (\Lambda^{-1})'(R^{-\frac12})' \nonumber \\ 
&& + \frac14 N (a\,R)^{-\frac12} \Lambda^{-1} R^{-1} R'' 
- \left(\left(N a^{-\frac12}\right)'\Lambda^{-1} R^{-\frac12} 
\right)' 
- \frac12 N a^{-\frac12} \Lambda^{-1}(R^{-\frac12})'' \nonumber \\ 
\label{eom_-3_5} 
&& + 2 N a^{-\frac12} \lambda^2 \Lambda R^{-\frac12} + (N^rP_R)' 
+\frac{1}{16}N\Lambda(a\,R)^\frac12R^{-2}P_\Gamma^2\,,\\ 
\dot{P_\Gamma} &=& N^rP_\Gamma'\,. 
\label{eom_-3_6} 
\end{eqnarray} 
For a correct variational principle to be applied, we have to find out 
what surface terms are left over from the variation performed, with the 
purpose of deriving the equations of motion in 
Eqs. (\ref{eom_-3_1})-(\ref{eom_-3_6}). These surface 
terms are, 
\begin{eqnarray} \label{surf_terms_-3} 
\textrm{Surface terms} &=& N (a\,R)^{-\frac12} R' \Lambda^{-2} 
\delta\Lambda -  N (a\,R)^{-\frac12} \Lambda^{-1} \delta R' + 
\left((a\,R)^{-\frac12}\right)'\Lambda^{-1} \delta R + 
\frac12(a\,R)^{-\frac12}\Lambda^{-1}R^{-1}R'\delta R \nonumber \\ 
&& - \left. N^rP_R\delta R + N^r\Lambda\delta P_\Lambda+ 
N^r\Gamma\delta P_\Gamma+\tilde{\Phi}\delta P_\Gamma 
\right|_0^{\infty}\,. 
\end{eqnarray} 
In order to know the form of (\ref{surf_terms_-3}) for $r \to 0$, we 
assume 
\begin{eqnarray} \label{falloff_0_i_-3} 
\Lambda(t,r) &=& \Lambda_0+O(r^2)\,,\\ 
R(t,r)&=& R_0+R_2r^2+O(r^4)\,,\\ 
P_\Lambda(t,r) &=& O(r^3)\,\\ 
P_R(t,r) &=& O(r)\,,\\ 
N(t,r) &=& N_1(t)r+O(r^3)\,\\ 
N^r(t,r) &=& O(r^3)\,,\\ 
\Gamma(t,r) &=& O(r)\,,\\ 
P_\Gamma(t,r) &=& Q_0 + Q_2 r^2 + O(r^4)\,,\\ 
\tilde{\Phi}(t,r)&=&\tilde{\Phi}_0(t)+O(r^2)\,. 
\label{falloff_0_f_-3} 
\end{eqnarray} 
The surface terms become 
\begin{equation} \label{surf_terms_0_-3} 
\left. \textrm{Surface terms}\right|_{r=0} = - N_1(a\,R)^{-\frac12} 
\Lambda_0^{-1} \delta R_0-\tilde{\Phi}_0\delta Q_0\,. 
\end{equation} 
The asymptotic conditions for $r\rightarrow\infty$ are assumed as 
\begin{eqnarray} \label{falloff_inf_i_-3} 
\Lambda(t,r) &=& \frac{\sqrt{6}}{4}\,  l r^{-1} + 
2^{\frac54} l^{\frac52}\eta(t)r^{-\frac52} + O^\infty(r^{-3})\,,\\ 
R(t,r) &=& r + 2^{\frac34} l^{\frac32} 
\rho(t)r^{-\frac12}+O^\infty(r^{-1})\,,\\ 
P_\Lambda(t,r) &=& O^\infty(r^{-2})\,,\\ 
P_R(t,r) &=& O^\infty(r^{-4})\,,\\ 
N(t,r) &=& R(t,r)'\Lambda(t,r)^{-1}(\tilde{N}_+(t)+O^\infty(r^{-5}))\,,\\ 
N^r(t,r) &=& O^\infty(r^{-2})\,,\\ 
\Gamma(t,r) &=& O^\infty(r^{-2})\,,\\ 
P_\Gamma(t,r) &=& Q_+(t) + O^\infty(r^{-1})\,,\\ 
\tilde{\Phi}(t,r) &=& \tilde{\Phi}_+(t)+O^\infty(r^{-1})\,. 
\label{falloff_inf_f_-3} 
\end{eqnarray} 
where $l$ is the AdS length. 
The surface terms, Eq. (\ref{surf_terms_-3}), for 
$r\rightarrow\infty$, are written as 
\begin{equation} \label{surf_terms_inf_-3} 
\left. \textrm{Surface terms}\right|_{r\rightarrow\infty} = 
\tilde{N}_+\delta M_++\tilde{\Phi}_+\delta Q_+ \,, 
\end{equation} 
where 
\begin{equation} \label{m_+_-3} 
M_+(t) = 2^53^{-\frac54}\,\eta(t)+2^23^{\frac12}\,\rho(t)\,. 
\end{equation} 
Here we pay attention to the fact that we can define the above 
quantity $M_+$, in Eq. (\ref{m_+_-3}), because we do as explained in 
subsection \ref{bhsolutions} when $\omega<-\frac32$ 
and absorb the divergence in the definition 
of $M$. This means that the $M_+$ above in Eq. (\ref{m_+_-3}) is 
in fact related to the mass plus the divergence, 
defined above as $M(R_*)$, see Eq. (\ref{mass_uncharged}) and below 
in subsection \ref{rec_cantrans_action_-3}. 
The surface term added to (\ref{haction_-3}) is then 
\begin{equation} \label{variationofsurfaceterm_-3} 
S_{\partial\Sigma}\left[\Lambda,R,Q_0,Q_+; 
N,\tilde{\Phi}_0,\tilde{\Phi}_+\right]= 
\int dt \left(2a^{-1}(a\,R_0)^{\frac12}N_1 \Lambda_0^{-1} 
-\tilde{N}_+M_++\tilde{\Phi}_0Q_0-\tilde{\Phi}_+Q_+\right)\,. 
\end{equation} 
With this surface term added we obtain a well defined variational 
principle. What remains after variation of the total action, 
Eq. (\ref{haction_-3}) and Eq. (\ref{variationofsurfaceterm_-3}), is 
\begin{equation} 
\int\,dt\,\left(2a^{-1}(a\,R_0)^{\frac12}\delta(N_1\Lambda_0^{-1}) 
-M_+\delta\tilde{N}_++Q_0\delta\tilde{\Phi}_0- 
Q_+\delta\tilde{\Phi}_+\right)\,. 
\end{equation} 
The surface terms coming from the variation of the total action 
disappear as the result of the fixation of 
$N_1\Lambda_0^{-1}$ on the horizon, $r=0$, and of 
$\tilde{N}_+(t)$ at infinity, $r \to \infty$, with the same 
happening respectively to $\tilde{\Phi}_0$ at $r=0$ 
and to $\tilde{\Phi}_+$ at $r\to\infty$. 

\subsection{Reconstruction, canonical transformation, and action} 
\label{rec_cantrans_action_-3} 

Repeating the steps of the two previous sections, we now give the main 
results concerning $\omega=-3$. 
The vector potential is 
\begin{eqnarray} 
    A &=& -\frac12 (a\,R)^\frac12 Q \,dT + \xi\,, 
\label{vector_potential_-3} 
\end{eqnarray} 
where there is the function $\xi=\xi(t,r)$. Expanding for $(t,r)$ we find 
\begin{eqnarray} 
    A &=& \left(-\frac12 (a\,R)^\frac12 Q \,T' + \xi'\right)dr + 
    \left(-\frac12 (a\,R)^\frac12 Q \,\dot{T} + \dot{\xi}\right)dt 
    \,, \label{vector_potential_expanded_-3} 
\end{eqnarray} 
out of which we infer that 
\begin{eqnarray} 
 P_\Gamma &=& Q\,. 
 \label{p_gamma_equals_q_-3} 
\end{eqnarray} 
The metric function is given by the expression 
\begin{equation} 
F=(a\,R)^2 - 2(a\,R)^\frac12\,M-\frac12 (a\,R)P_\Gamma^2\,. 
\label{metric_function_-3} 
\end{equation} 
The Killing time $T=T(t,r)$ is a function of $(t,r)$, where we find that 
\begin{equation} 
-T' = F^{-1} \Lambda P_\Lambda (a\,R)^{\frac12}\,, 
\label{tlinha_-3} 
\end{equation} 
which is equal to minus the conjugate momentum of the new variable $M$, 
or $P_M=P_M(t,r)\equiv -T'(t,r)$. 
With the help of Eq. (\ref{tlinha_-3}) we find $F=F(t,r)$ as a function 
of the canonical variables, 
\begin{equation} 
F=(R')^2\Lambda^{-2}-aRP_\Lambda^2\,. 
\end{equation} 
Summing up, the canonical transformations are 
\begin{eqnarray} \label{setcanonicaltrans_-3} 
R &=& \textrm{R}\,, \nonumber\\ 
M &=& \frac12 (a\,R)^{-\frac12} \left((a\,R)^2-\frac12(a\,R)P_\Gamma^2 
-F\right)\,, \nonumber \\ 
Q &=& P_\Gamma\,, \nonumber \\ 
P_\textrm{R} &=& P_R - F^{-1}R^{-1}\Lambda^{-2}\left[RR'\Lambda 
  P_\Lambda' + \frac12 (R')^2 P_\Lambda \Lambda - RR''\Lambda 
  P_\Lambda + RR'\Lambda'P_\Lambda\right] \nonumber \\ 
&& +\frac14 R^{-1}\Lambda P_\Lambda - 4\lambda^2F^{-1}P_\Lambda 
R\Lambda+\frac18\,a\,F^{-1}\Lambda P_\Lambda P_\Gamma^2 
\,, \nonumber \\ 
P_M &=& F^{-1} \Lambda P_\Lambda (a\,R)^{\frac12}\,, \nonumber \\ 
P_Q &=& -\Gamma + \frac12 (a\,R)\,\Lambda P_\Lambda P_\Gamma\,. 
\end{eqnarray} 
In addition, the relevant inverse transformations back to the old variable 
$\Lambda$ and respective conjugate momentum $P_\Lambda$ are 
\begin{eqnarray} \label{inversetrans_l_-3} 
\Lambda &=& \left(F^{-1}(R')^2-FP_M^2\right)^{\frac12}\,, \\ 
P_\Lambda &=& (a\,R)^{-\frac12} FP_M 
\left(F^{-1}(R')^2-FP_M^2\right)^{-\frac12}\,. \label{inversetrans_pl_-3} 
\end{eqnarray} 
We have to show that this transformation is canonical. This requires 
using the identity 
\begin{eqnarray} \label{oddidentity_-3} 
P_\Lambda \delta\Lambda + P_R \delta R +P_\Gamma\delta\Gamma 
- P_M \delta M - P_{\textrm{R}}\delta\textrm{R}-P_Q\delta Q 
&=& \left( \frac12(a\,R)^{-\frac12}\delta 
R\ln 
  \left|\frac{(a\,R)^{-\frac12}R'+\Lambda P_\Lambda}{(a\,R)^{-\frac12}R'- 
\Lambda 
      P_\Lambda}\right|\right)'+ \nonumber \\ 
&+&\delta\left(\Gamma P_\Gamma + 
\Lambda P_\Lambda + \frac12 (a\,R)^{-\frac12} R' \ln 
 \left|\frac{(a\,R)^{-\frac12} R'-\Lambda P_\Lambda} 
 {(a\,R)^{-\frac12} R'+ \Lambda 
 P_\Lambda}\right|\right)\,. 
\end{eqnarray} 
We now integrate Eq. (\ref{oddidentity_-3}), in $r$, 
from $r=0$ to $r=\infty$. The first term on the right 
hand side of Eq. (\ref{oddidentity_-3}) vanishes due to the falloff 
conditions, Eqs. (\ref{falloff_0_i_-3})-(\ref{falloff_0_f_-3}) and 
Eqs. (\ref{falloff_inf_i_-3})-(\ref{falloff_inf_f_-3}). 
We obtain then the following expression 
\begin{eqnarray} \label{int_oddidentity_-3} 
\int_0^\infty\,dr\,\left(P_\Lambda \delta\Lambda + P_R \delta 
  R + P_\Gamma \delta\Gamma \right)- 
  \int_0^\infty\,dr\,\left(P_M \delta M + 
P_{\textrm{R}}\delta\textrm{R} + P_Q \delta Q \right) 
&=& 
\delta\omega\,\left[\Lambda,\,R,\,\Gamma,\,P_\Lambda, 
\,P_\Gamma\right]\,, 
\end{eqnarray} 
where $\delta\omega\,\left[\Lambda,\,R,\,\Gamma,\,P_\Lambda, 
\,P_\Gamma\right]$ is a well 
defined functional, which is also an exact form. As above, this 
equality shows that the difference between the Liouville form of 
$\left\{R,\,\Lambda,\,\Gamma;\,P_R,\,P_\Lambda,\,P_\Gamma\right\}$ 
and the Liouville form 
of $\left\{\textrm{R},\,M,\,Q;\,P_{\textrm{R}},\,P_M,\,P_Q\right\}$ 
is an exact form, which implies that the set of transformations 
(\ref{setcanonicaltrans_-3}) is canonical. 

With this result, we write the asymptotic conditions of the new 
canonical coordinates for $r\rightarrow 0$ 
\begin{eqnarray} \label{newfalloff_0_i_-3} 
F(t,r) &=& 4 R_2^2 \Lambda_0^{-2} r^2 + O(r^4)\,, \\ 
\textrm{R}(t,r) &=& R_0+R_2\,r^2+O(r^4)\,, \\ 
M(t,r) &=& M_0 + M_2 \,r^2 + O(r^4)\,, \\ 
Q(t,r) &=& Q_0+Q_2r^2+O(r^4)\,,\\ 
P_\textrm{R}(t,r) &=& O(r)\,, \\ 
P_M(t,r) &=& O(r)\,,\\ 
P_Q(t,r) &=& O(r)\,. \label{newfalloff_0_f_-3} 
\end{eqnarray} 
where we have 
\begin{eqnarray} \label{m_0_-3} 
M_0 &=& \frac12 (a\,R_0)^{-\frac12}\left((a\,R_0)^2- 
\frac12(a\,R_0)Q_0^2\right)\,, \\ 
M_2 &=& -\frac{\left[-\frac12(a\,R_0)Q_0^2+(a\,R_0)^2\right]R_2} 
{4R_0(a\,R_0)^\frac12}+\frac{2a^2\,R_0^2R_2^2-\frac12 
\left[2(a\,R_0)Q_0Q_2+a\,Q_0^2R_2\right]-4R_2^2\Lambda_0^{-2}} 
{2(a\,R_0)^\frac12}\,,\label{m_2_-3} 
\end{eqnarray} 
and $a=l^{-1}\sqrt{8/3}$. 
For $r\rightarrow \infty$, we have 
\begin{eqnarray} \label{newfalloff_inf_i_-3} 
F(t,r) &=& \frac{4}{3} l^{-2} \, r^2 
-\frac{16}{3}\left(\frac{\sqrt{2}\,l^{-1}}{6}\right)^{\frac12} 
\left(8\eta(t)+3^{\frac12}\rho(t)\right)\,r^{\frac12} 
+ O^{\infty}(r^{0})\,, \\ 
\textrm{R}(t,r) &=& r + 
\left(\frac{\sqrt{2}}{2}\,l^{-1}\right)^{-\frac32}\rho(t)\,r^{-\frac12}+ 
O^{\infty}(r^{-1})\,, \\ 
Q(t,r) &=& Q_+(t) + O^\infty(r^{-1})\,, \\ 
M(t,r) &=& M_+(t) + O^{\infty}(r^{-\frac12}) \,, \\ 
P_\textrm{R}(t,r) &=& O^\infty(r^{-4})\,, \\ 
P_M(t,r) &=& O^\infty(r^{-\frac95})\,, \\ 
P_Q(t,r) &=& O^\infty(r^{-2})\,. 
\label{newfalloff_inf_f_-3} 
\end{eqnarray} 
where $ M_+(t)$ is defined in Eq. (\ref{m_+_-3}). 

We now write the future constraint 
$M'$ as a function of the older constraints 
\begin{equation} 
M' = -\Lambda^{-1}\left( R' H +(a\,R)^{\frac12} P_\Lambda 
\left(H_r-\Gamma G\right)-\frac12 (a\,R)^\frac12\Lambda 
P_\Gamma G \right)\,. 
\end{equation} 
Using the inverse transformations of $\Lambda$ and $P_\Lambda$ in 
Eqs. (\ref{inversetrans_l_-3}) and (\ref{inversetrans_pl_-3}), we 
obtain the same form for the old constraints as functions of the new 
variables 
\begin{eqnarray} \label{oldconstraintsinnewvariables_1_-3} 
H &=& - \frac{M'F^{-1}\textrm{R}'+F P_M 
P_{\textrm{R}}-\frac12(a\,R)^\frac12QQ'F^{-1}R'} 
{\left(F^{-1}(\textrm{R}')^2-F 
    P_M^2\right)^{\frac12}}\,, \\ 
H_r &=& P_M M' + P_{\textrm{R}} 
\textrm{R}'+P_QQ'\,, 
\label{oldconstraintsinnewvariables_2_-3} \\ 
G &=& -Q'\,. 
\label{oldconstraintsinnewvariables_3_-3} 
\end{eqnarray} 
The new Hamiltonian, the total sum of the constraints, can now be 
written as 
\begin{equation} \label{newHamiltonian_-3} 
NH+N^rH_r+\tilde{\Phi}G = N^MM'+N^{\textrm{R}}P_{\textrm{R}} 
+N^QQ'\,. 
\end{equation} 
The new multipliers are, using Eqs. 
(\ref{oldconstraintsinnewvariables_1_-3})-(\ref{newHamiltonian_-3}), 
\begin{eqnarray} \label{new_mult_1_-3} 
N^M &=& - \frac{N F^{-1} R'}{\left(F^{-1}(\textrm{R}')^2-F 
    P_M^2\right)^{\frac12}}+ N^r P_M \,, \\ 
N^{\textrm{R}} &=& - \frac{N F P_M}{\left(F^{-1}(\textrm{R}')^2-F 
    P_M^2\right)^{\frac12}}+ N^r R'\,, \label{new_mult_2_-3} \\ 
N^Q &=& -\frac12 \frac{N (a\,R)^\frac12 F^{-1}R'Q} 
{\left(F^{-1}(\textrm{R}')^2-FP_M^2\right)^{\frac12}}+ 
N^rP_Q-\tilde{\Phi}\,. 
\label{new_mult_3_-3} 
\end{eqnarray} 
Using the inverse transformations 
Eqs. (\ref{inversetrans_l_-3})-(\ref{inversetrans_pl_-3}), and the 
identity $R=\textrm{R}$, we can write the new multipliers as functions 
of the old variables 
\begin{eqnarray} \label{mult_trans_1_-3} 
N^M &=& - NF^{-1}R'\Lambda^{-1}+ N^r F^{-1} \Lambda P_\Lambda 
(a\,R)^{\frac12} \,, \\ 
N^{\textrm{R}} &=& - N P_\Lambda (a\,R)^{\frac12} + N^r R'\,, 
\label{mult_trans_2_-3} \\ 
N^Q &=& -\frac12 N (a\,R)^\frac12 F^{-1}R'\Lambda^{-1}P_\Gamma 
-N^r\Gamma + \frac12 N^r(a\,R)\,\Lambda P_\Lambda P_\Gamma- 
\tilde{\Phi}\,. 
\label{mult_trans_3_-3} 
\end{eqnarray} 
For $r\rightarrow 0$ we have, 
\begin{eqnarray} \label{mult_newfalloff_0_i_-3} 
N^M(t,r) &=& -\frac12  N_1(t) \Lambda_0 R_2^{-1} + O(r^2)\,,\\ 
N^{\textrm{R}}(t,r) &=& O(r^4)\,,\\ 
N^Q(t,r)&=&-\frac14Q_0R_2^{-1}(a\,R_0)^\frac12N_1\Lambda_0 
-\tilde{\Phi}_0+O(r^2)\,. 
\label{mult_newfalloff_0_f_-3} 
\end{eqnarray} 
and for $r\rightarrow\infty$ we have 
\begin{eqnarray} \label{mult_newfalloff_inf_i_-3} 
N^M(t,r) &=& -\tilde{N}_+(t) +O^\infty(r^{-2})\,,\\ 
N^{\textrm{R}}(t,r) &=& O^\infty(r^{-\frac12})\,,\\ 
N^Q(t,r) &=& -\tilde{\Phi}_+(t)+O^\infty(r^{-1})\,. 
\label{mult_newfalloff_inf_f_-3} 
\end{eqnarray} 
Again, for $r\rightarrow0$, fixing $N^M(t,r)$, which means fixing 
$N_1\Lambda_0 R_2^{-1}$, is not 
equivalent to fixing $N_1 \Lambda_0^{-1}$. 
It is thus necessary to rewrite $N^M$ for $r\rightarrow0$. 
The same is true of fixing $N^Q(t,r)$ when $r\to0$. In this case 
it is not the same if one fixes $-\frac14Q_0R_2^{-1}(a\,R_0)^\frac12 
N_1\Lambda_0-\tilde{\Phi}_0$ or just $\tilde{\Phi}_0$. 
So, assuming $M_0$ as a function of $R_0$ and $Q_0$ 
allows one to define the horizon radius 
$R_0\equiv R_{\textrm{h}}(M_0,Q_0)$. 
We are thus working in the domain where 
$M_0>M_{\textrm{\tiny{crit}}}(Q_0)= 
(a^\frac32-1)\frac{Q_0^3}{12\sqrt{6}}$, the 
domain of the classical black hole solution. 
The variation of $R_0$ 
is given in terms of the variations of $M_0$ and of $Q_0$ 
in the expression 
\begin{equation} \label{deltar_0_-3} 
\delta R_0 = \left(\frac34\,a\,(a\,R)^{\frac12}- 
\frac18a\,(a\,R_0)^{-\frac12}Q_0^2\right)^{-1} 
\left(\delta M_0+\frac12(a\,R_0)^{\frac12}Q_0\delta 
Q_0\right)\,. 
\end{equation} 
The new multiplier $\tilde{N}^M$ is obtained from the old $N^M$ as 
\begin{eqnarray} \label{new_n_m_-3} 
\tilde{N}^M &=& - N^M 
\left[(1-g) + (a\,R_0)^{-\frac12}\,g 
\left(\frac34\,a\,(a\,R_0)^{\frac12}- 
\frac18a\,(a\,R_0)^{-\frac12}Q_0^2\right)^{-1}\right]^{-1}\,, \\ 
\tilde{N}^Q &=& -\frac12\tilde{N}^M\,g\,Q_0 
\left(\frac34\,a\,(a\,R_0)^{\frac12}- 
\frac18a\,(a\,R_0)^{-\frac12}Q_0^2\right)^{-1}-N^Q\,, 
\label{new_n_q_-3} 
\end{eqnarray} 
where $a=\frac{4\sqrt{3}}{3}|\lambda|=l^{-1}\sqrt{8/3}$, 
and $g(r)=1+O(r^2)$ for $r\rightarrow0$ and 
$g(r)=O^\infty(r^{-5})$ for $r\rightarrow\infty$. 
This new multiplier, function of the old multiplier, $\tilde{N}^M$, has 
as its properties for $r\rightarrow\infty$ 
\begin{eqnarray} 
\tilde{N}^M(t,r) &=& \tilde{N}_+(t) + O^\infty(r^{-2})\,, \\ 
\tilde{N}^Q(t,r) &=& \tilde{\Phi}_+(t) + O^\infty(r^{-1})\,, 
\end{eqnarray} 
and as its properties for $r\rightarrow0$ 
\begin{eqnarray} 
\tilde{N}^M(t,r) &=& \tilde{N}_0^M(t) + O(r^{2})\,, \\ 
\tilde{N}^M(t,r) &=& \tilde{\Phi}_0(t) + O(r^{2})\,, 
\end{eqnarray} 
where $\tilde{N}_0^M$ is given by 
\begin{eqnarray} 
\tilde{N}_0^M &=& N_1\Lambda_0R_2^{-1} 
\left(\frac38\,a^2\,R_0-\frac{3}{48}\,a\,Q_0^2\right)\\ 
&\stackrel{M'=0=Q'}{=}& N_1\Lambda_0^{-1}\,. 
\end{eqnarray} 
With this new constraint $\tilde{N}^M$, fixing $N_1 \Lambda_0^{-1}$ 
or fixing $\tilde{N}^M_0$ is equivalent, and the same happens with 
$\tilde{N}^Q$, where fixing $\tilde{\Phi}_0$ is fixing the multiplier 
$\tilde{N}^Q$ in the limit $r\to0$. 
There are no problems with 
$N^{\textrm{R}}$, which is left as determined in Eq. 
(\ref{mult_trans_2_-3}). 

The new action is then, summing both the bulk and the surface terms, 
\begin{eqnarray} 
S\left[M, \textrm{R},Q, P_M, P_{\textrm{R}},P_Q; \tilde{N}^M, 
  N^{\textrm{R}},N^Q\right] &=& \int \,dt\, \int_0^\infty \, dr \, 
\left\{P_M\dot{M}+P_\textrm{R}\dot{\textrm{R}}+P_Q\dot{Q}+ 
N^QQ'-N^{\textrm{R}}P_{\textrm{R}}\right.\nonumber \\ 
&&\left.+\tilde{N}^M 
\left[(1-g)M'+g\left(\frac34\,a\,(a\,R)^{\frac12}- 
\frac18a\,(a\,R_0)^{-\frac12}Q_0^2\right)^{-1}\right.\right. 
\nonumber \\ 
&&\left.\left.\left[(a\,R_0)^{-\frac12}M' 
+Q_0Q'\right]\right]\right\}+ 
\nonumber \\ 
&&  \int \, dt \, \left(2a^{-1} (a\,R_0)^{\frac12} \tilde{N}_0^M - 
  \tilde{N}_+ M_+ +\tilde{\Phi}_0Q_0- 
  \tilde{\Phi}_+Q_+\right)\,. \label{newaction_-3} 
\end{eqnarray} 
The new equations of motion are now 
\begin{eqnarray} \label{new_eom_1_-3} 
\dot{M} &=& 0\,, \\ 
\dot{\textrm{R}} &=& N^{\textrm{R}}\,, \\ 
\dot{Q} &=& 0\,, \\ 
\dot{P}_M &=& (N^M)'\,, \\ 
\dot{P}_{\textrm{R}} &=& 0\,, \\ 
\dot{P}_Q &=& (N^Q)'\,, \\ 
M' &=& 0\,, \\ 
P_{\textrm{R}} &=& 0\,,\\ 
Q' &=& 0\,. 
\label{new_eom_9_-3} 
\end{eqnarray} 
Here we understood $N^M$ to be a function of the new constraint, 
defined through Eq. (\ref{new_n_m_-3}). The resulting boundary terms 
of the variation of this new action, Eq. (\ref{newaction_-3}), 
are, first, terms proportional to 
$\delta M$, $\delta \textrm{R}$ and $\delta Q$ on the 
initial and final hypersurfaces, and, second, the term 
$\int \, dt \, \left(2a^{-1} (a\,R_0)^{\frac12} \delta\tilde{N}_0^M - 
M_+ \delta\tilde{N}_+ + Q_0\delta\tilde{\Phi}_0- 
  Q_+\delta\tilde{\Phi}_+\right)$. 
Here we used the expression in Eq. (\ref{deltar_0_-3}). The action in 
Eq. (\ref{newaction_-3}) yields the equations of motion, Eqs. 
(\ref{new_eom_1_-3})-(\ref{new_eom_9_-3}), provided that we fix 
the initial and final values of the new canonical variables and that 
we also fix both the values of $\tilde{N}^M_0$ and $\tilde{N}_+$, 
and $\tilde{\Phi}_0$ and $\tilde{\Phi}_+$. Thanks to 
the redefinition of the Lagrange multiplier, from $N^M$ to 
$\tilde{N}^M$, and from $N^Q$ to $\tilde{N}^Q$ 
the fixation of those quantities, $\tilde{N}^M_0$ and 
$\tilde{N}_+$, and $\tilde{\Phi}_0$ and $\tilde{\Phi}_+$, 
has the same meaning it had before the 
canonical transformations and the redefinition of $N^M$ and $N^Q$. 
This same meaning is guaranteed through the use of our gauge freedom to 
choose the multipliers, and at the same time not fixing the boundary 
variations independently of the choice of Lagrange multipliers, which 
in turn allow us to have a well defined variational principle for the 
action. 

\subsection{Hamiltonian reduction} \label{hr_-3} 

We now solve the constraints in order to reduce to the true dynamical 
degrees of freedom. The equations of motion, Eqs. 
(\ref{new_eom_1_-3})-(\ref{new_eom_9_-3}), allow us to write $M$ and $Q$ 
as independent functions of the radial coordinate $r$ 
\begin{eqnarray} 
M(t,r)&=&\textbf{m}(t)\,,\\ 
Q(t,r)&=&\textbf{q}(t)\,. 
\label{m_-3} 
\end{eqnarray} 
The reduced action, with the constraints and Eq (\ref{m_-3}) taken into 
account, is 
\begin{equation} \label{red_action_-3} 
S 
\left[\textbf{m},\textbf{q},\textbf{p}_{\textbf{m}},\textbf{p}_{\textbf{q}} 
;\tilde{N}_0^M,\tilde{N}_+,\tilde{\Phi}_0,\tilde{\Phi}_+\right] 
= \int dt\,\, 
\textbf{p}_{\textbf{m}}\dot{\bf{m}}+\textbf{p}_{\textbf{q}}\dot{\bf{q}} 
-\textbf{h}\,, 
\end{equation} 
where 
\begin{eqnarray} \label{new_p_m_-3} 
\textbf{p}_{\textbf{m}} = \int_0^\infty dr\,P_M\,, \\ 
\textbf{p}_{\textbf{q}} = \int_0^\infty dr\,P_Q\,, 
 \label{new_p_q_-3} 
\end{eqnarray} 
and the reduced Hamiltonian, $\textbf{h}$, is now written as 
\begin{equation} \label{red_Hamiltonian_-3} 
\textbf{h}(\textbf{m},\,\textbf{q};t) = -2a^{-1} 
(a\,R_{\textrm{h}})^{\frac12} \tilde{N}_0^M + 
\tilde{N}_+ \textbf{m}+\left(\tilde{\Phi}_+- 
\tilde{\Phi}_0\right)\textbf{q}\,, 
\end{equation} 
with $R_{\textrm{h}}$ being the horizon 
radius. We also have that $\textbf{m}> 
M_{\textrm{\tiny{crit}}}(\textbf{q})$. 
The equations of motion are then 
\begin{eqnarray} \label{red_eom_1_-3} 
\dot{\textbf{m}} &=& 0\,, \\ 
\dot{\textbf{q}} &=& 0\,, 
\label{red_eom_2_-3}\\ 
\dot{\textbf{p}}_{\textbf{m}} &=& 
(a\,R_{\textrm{h}})^{-\frac12} 
\left(\frac34\,a\,(a\,R_{\textrm{h}})^{\frac12}- 
\frac18a\,(a\,R_{\textrm{h}})^{-\frac12} 
\textbf{q}^2\right)^{-1}-\tilde{N}_+\,, 
\label{red_eom_3_-3}\\ 
\dot{\textbf{p}}_{\textbf{q}} &=&\frac12\textbf{q} 
\left(\frac34\,a\,(a\,R_{\textrm{h}})^{\frac12}- 
\frac18a\,(a\,R_{\textrm{h}})^{-\frac12} 
\textbf{q}^2\right)^{-1} 
+\tilde{\Phi}_0-\tilde{\Phi}_+\,. 
\label{red_eom_4_-3} 
\end{eqnarray} 
Here $\textbf{m}$ and $\textbf{q}$ are equal to the mass parameter $M$ 
and the charge parameter $Q$ of the 
classical solution in Eq. (\ref{metric_dilatonic}). 
The third equation of motion, Eq. (\ref{red_eom_3_-3}), 
describes the time evolution of the 
difference of the Killing times on the horizon and at infinity, due to 
$\textbf{p}_{\textbf{m}} = T_0 - T_+$ and Eq. (\ref{new_p_m_-3}). 
The last equation, Eq. (\ref{red_eom_4_-3}), 
gives the time evolution of $\textbf{p}_{\textbf{q}}$ 
as the difference between the time derivatives of 
the electromagnetic gauges $\xi(t,r)$ on the horizon and 
at infinity due to $\textbf{p}_{\textbf{m}} = \xi_0 - \xi_+$ 
and Eq. (\ref{new_p_q_-3}). 

\subsection{Quantum theory and partition function} 

The steps developed in Sections \ref{mtz} and \ref{zero} 
can be readily used here. So we do not spell out the 
corresponding calculations in detail. 
Thus, obtained as above, the time evolution operator is 
\begin{equation} 
K\left(\textbf{m},\textbf{q}; 
\mathcal{T},\Theta,\Xi_0,\Xi_+\right) = \exp 
\left[-i\textbf{m}\mathcal{T}+ 2\,i\,a^{-1}\, 
(a\,R_{\textrm{h}})^{\frac12}\Theta- 
i\textbf{q}\left(\Xi_+-\Xi_0\right)\right]\,. 
\label{k_op_-3} 
\end{equation} 
We can now develop the thermodynamic analysis as in Sections 
\ref{mtz} and \ref{zero}. 

\subsection{Thermodynamics} 

We can now build the partition function for this system, with the 
Hamiltonian given in Eq. (\ref{red_Hamiltonian_-3}). 
The steps are the same as was the case 
with $\omega=\infty$ and $\omega=0$. 
The thermodynamic ensemble is also the grand canonical ensemble. 
Again, with $\mathcal{T}=-i\beta$, $\Theta=-2\pi i$, $\Xi_0=0$, and 
$\Xi_+=i\beta\bar{\phi}$, we write the general form 
of the partition function as 
\begin{equation} 
\mathcal{Z} = Tr\left[ \hat{K}(-i\beta, -2\pi i, 
0,i\beta\bar{\phi}) \right]\,. 
\end{equation} 
This is realized as 
\begin{equation} \label{partition_function_ren_-3} 
\mathcal{Z}_{\textrm{ren}}(\beta,\bar{\phi}) = \mathcal{N} \int_{A'} 
\widetilde{\mu}\,dR_{\textrm{h}}d\textbf{q}\,\exp(-I_*)\,, 
\end{equation} 
where $\mathcal{N}$ is a normalization factor, 
$A'$ is the domain of integration given by $0\leq R_{\textrm{h}}$ 
and $\textbf{q}^2\leq6aR_{\textrm{h}}$, and the function 
$I_*(R_{\textrm{h}},\textbf{q})$, a kind of an 
effective action \cite{york1}. $I_*(R_{\textrm{h}},\textbf{q})$, 
is written as 
\begin{equation} \label{eff_action_-3} 
I_*(R_{\textrm{h}},\textbf{q}):= \frac{\beta}{2} 
(a\,R_{\textrm{h}})^{\frac32} 
-\left(\frac{\beta}{4}a^\frac12\textbf{q}+ 
4\pi a^{-\frac12}\right)R_{\textrm{h}}^\frac12- 
\textbf{q}\bar{\phi}\beta\,. 
\end{equation} 
Its critical points are at 
\begin{eqnarray} 
R_{\textrm{h}}^\pm &=& \frac43\pi\beta^{-1}a^{-2} 
\left(1\pm\sqrt{1+\frac38\beta^2a^2\pi^{-2}\bar{\phi}^2}\right)\,, 
\label{crit_r_-3}\\ 
\textbf{q}^\pm &=& -2\bar{\phi} (a\,R_{\textrm{h}}^\pm)^{-\frac12}\,. 
\label{crit_q_-3} 
\end{eqnarray} 
Given that $\bar{\phi}^2>-\frac83\beta^{-2}a^{-2}\pi^2$ there are always 
two critical points, but only $R_{\textrm{h}}^+$ is physical 
as it is positive, $R_{\textrm{h}}^+>0$. The other critical point is 
negative, $R_{\textrm{h}}^-<0$. 
Determining the zeros of the action (\ref{eff_action_-3}) 
when Eq. (\ref{crit_q_-3}) holds, we get 
\begin{eqnarray} 
    {R_{\textrm{h}}^0}^\pm &=& 4\pi\beta^{-1}a^{-2} 
    \left(1\pm\sqrt{1-\frac18\beta^2a^2\pi^{-2}\bar{\phi}^2} 
    \right)\,. 
    \label{zeros_eff_action_-3} 
\end{eqnarray} 
Now, for different values of $\bar{\phi}^2$ we may have different 
signs for the action (\ref{eff_action_-3}) evaluated at the 
physical critical point $R_{\textrm{h}}^+$. So, for 
$\bar{\phi}^2>8\beta^{-2}a^{-2}\pi^2$ there are no zeros and 
$I_*(R_{\textrm{h}}^+)>0$, where we take 
$\textbf{q}^+=\textbf{q}^+(R_{\textrm{h}}^+)$ to be a function 
of the critical point radius $R_{\textrm{h}}^+$. If 
$\bar{\phi}^2=8\beta^{-2}a^{-2}\pi^2$ then the effective action 
(\ref{eff_action_-3}) has one zero only, at $R_{\textrm{h}}^+$, 
i. e., $I_*(R_{\textrm{h}}^+)=0$. 
Finally, if $\bar{\phi}^2<8\beta^{-2}a^{-2}\pi^2$ then we have two zeros, 
given in Eq. (\ref{zeros_eff_action_-3}), 
and $I_*(R_{\textrm{h}}^+)<0$. Then, despite the variation of the 
sign of the effective action at the physical critical point 
$(R_{\textrm{h}}^+,\textbf{q}^+)$, there are no new minima when 
the order parameter $\bar{\phi}$ is changed across the border given 
by $\bar{\phi}^2=8\beta^{-2}a^{-2}\pi^2$, which means that there are 
no phase transitions, as the classical phase corresponds always 
to the global minimum. 
We can now derive the basic thermodynamic results, as long as 
the approximation for the saddle point is valid, which means that we 
have to work in the neighborhood of the classical solution 
(\ref{metric_dilatonic}), where the critical point dominates. 
The expected value of the energy $E$ is 
\begin{equation} 
\left\langle E \right\rangle = 
\left(-\frac{\partial}{\partial\beta}+\beta^{-1}\bar{\phi} 
\frac{\partial}{\partial\bar{\phi}}\right)\ln 
\mathcal{Z}_{\textrm{ren}} \approx \frac12 
(a\,R_{\textrm{h}}^+)^\frac32-\frac14(a\,R_{\textrm{h}}^+)^\frac12 
\textbf{q}^2 = \textbf{m}^+\,. 
\label{energy_-3} 
\end{equation} 
The charge expectation value is 
\begin{eqnarray} 
    \left\langle Q \right\rangle &=& \beta^{-1} 
    \frac{\partial}{\partial\bar{\phi}}\ln\mathcal{Z}_{\textrm{ren}} 
    \approx \textbf{q}^+\,. 
    \label{charge_-3} 
\end{eqnarray} 
Here we have the temperature 
$\textbf{T}\equiv\beta^{-1}$ given by 
\begin{eqnarray} 
\textbf{T}&=&\frac{1}{4\pi}\left(\frac32\,a^2\,R_{\textrm{h}}^+- 
\frac12\,a\,\left(\textbf{q}^+\right)^2\right)\,. 
\label{temperature_-3} 
\end{eqnarray} 
It can be shown that the derivative of 
$\textbf{m}^+(\beta)$ with respect to $\beta$ is negative, 
which means that the heat capacity is positive, i. e., 
$C_{\bar{\phi}}=-\beta^2\left(\frac{\partial\left\langle 
E\right\rangle}{\partial\beta}\right)>0$. 
The system is thus stable. Finally the entropy is given by 
\begin{equation} 
S = 
\left(1-\beta\frac{\partial}{\partial 
\beta}\right)(\ln\mathcal{Z}_{\textrm{ren}}(\beta)) 
\approx 4 \pi \sqrt{a^{-1}R_{\textrm{h}}^+}\,. 
\end{equation} 
The result recovers the entropy for the three-dimensional 
charged dilatonic black hole with $\omega=-3$ 
(see also \cite{ads3_bh}). 

\section{Conclusions} 
\label{conclusions} 

We have extended the work done in 
\cite{diaslemos} by adding a Maxwell term to the three-dimensional 
Brans-Dicke action, and thus, have continued Louko and collaborators' 
program \cite{louko1,louko2,louko3,louko4,louko5,bose} (see also 
\cite{kunstatter1,kunstatter2})) of studying the thermodynamic 
properties through Hamiltonian methods of black holes in several 
theories in different dimensions.  Specifically we have calculated the 
thermodynamic properties of black hole solutions with asymptotics that 
allow a well formulated Hamiltonian formalism in three-dimensional 
Brans-Dicke-Maxwell gravity. Only certain values of the Brans-Dicke 
parameter $\omega$ are allowed in this juncture.  The corresponding 
theories are general relativity, i.e., $\omega\to\pm\infty$, a 
dimensionally reduced cylindrical charged four-dimensional general 
relativity theory, i.e., $\omega=0$, and a theory representing a class 
of charged dilaton-gravity theories, with a typical $\omega$ given by 
$\omega=-3$, all of these coupled to a Maxwell field.  We have shown 
that for these theories, with well defined asymptotics for geometric 
and electromagnetic variables, the formalism fits well.  It was 
possible to have a well defined electric field for all the values of 
$\omega$ studied.  As in \cite{diaslemos} several modifications were 
needed. First, in the powers of the fall-off conditions, and second 
due to the presence of the scalar field, which was reflected in the 
fact that it changed the powers of the radial $R$ coordinate, to name 
a few.  Of course, the addition of electromagnetism also introduced a 
new pair of canonical variables, which added a good deal of complexity 
to the calculations.  Namely, the treatment of the mass term when 
electromagnetic interaction is included is not trivial at all in this 
Hamiltonian formalism in three dimensions, since there is a divergent 
asymptotic behavior of the mass $M$. Within the Hamiltonian formalism 
we solved this problem by introducing a renormalized mass.  This 
redefinition of the mass fits perfectly into the formalism for the 
$\omega\rightarrow\pm\infty$ (i.e., general relativity), it is not needed 
in the $\omega=0$ case, and conforms to the $\omega=-3$ case. 
Although, in order to build a three-dimensional Lorentzian Hamiltonian 
theory with electric charge, these modifications had to be made, a 
quantum theory and a statistical description of the systems in 
question in a grand canonical ensemble were derived, and the 
corresponding thermodynamics, with precise values for the entropy of 
the black holes studied, was established. 

\section*{Acknowledgments} 

GASD is supported by grant SFRH/BD/2003 from FCT.  This work was 
partially funded by Funda\c c\~ao para a Ci\^encia e Tecnologia (FCT) 
- Portugal, through project POCI/FP/63943/2005.

\end{document}